\documentclass[english,aps,prl,preprint, groupedaddress, amsmath]{revtex4-2}

\usepackage{fancyhdr,graphicx}
\usepackage[table,xcdraw]{xcolor}
\usepackage[normalem]{ulem}
\useunder{\uline}{\ul}{}
\usepackage[T1]{fontenc}
\usepackage{chemformula} % Formula
\usepackage[utf8x]{inputenc}
\usepackage{gensymb}
\usepackage{times}
\usepackage{color}
\usepackage[colorlinks,bookmarks=false,citecolor=blue,linkcolor=red,urlcolor=blue]{hyperref}
\input pdfcolor.tex
\usepackage{colortbl,amsthm,amsmath,amssymb,txfonts}
\usepackage{graphicx}
\usepackage{epsfig,color}
\usepackage{verbatim}
\usepackage{dcolumn}
\usepackage{longtable}
\usepackage[normalem]{ulem}
\usepackage{bm}
\usepackage{xcolor}
\usepackage{babel}

\begin{document}

\title{Universality of Quantum Phase Transitions in the Integer and Fractional Quantum Hall Regimes}
\author{Simrandeep Kaur$^\dagger$$^1$, Tanima Chanda$^\dagger$$^1$, Kazi Rafsanjani Amin$^\dagger$$^2$, Divya Sahani$^1$, Kenji Watanabe$^3$, Takashi Taniguchi$^4$, Unmesh Ghorai$^5$, Yuval Gefen$^6$, G. J. Sreejith$^{7}$}
\author{Aveek Bid$^{1*}$}
\affiliation{$^1$Department of Physics, Indian  Institute of Science, Bangalore 560012, India.
	\\ $^2$Department of Microtechnology and Nanoscience, Chalmers University of Technology, 412 96
	Gothenburg, Sweden.\\
	$^3$Research Center for Electronic and Optical Materials, National Institute for Materials Science, 1-1 Namiki, Tsukuba 305-0044, Japan.
	\\ $^4$Research Center for Materials Nanoarchitectonics, National Institute for Materials Science,  1-1 Namiki, Tsukuba 305-0044, Japan.
	\\$^5$Department of Theoretical Physics, Tata Institute of Fundamental Research, Homi Bhabha Road, Mumbai, 400005, India\\
	$^6$ Department of Condensed Matter Physics, Weizmann Institute of Science, Rehovot 76100, Israel.\\
	$^7$Indian Institute of Science Education and Research, Pune 411008, India.\\
	$^*$Email: aveek@iisc.ac.in}

	%%%%%%%%%%%%%%%%%%%%%%%%%%%%%%%%%%%%%%%%%%%%%%%%%%%%
	\begin{abstract}

\textbf{Fractional quantum Hall (FQH) phases emerge due to strong electronic interactions and are characterized by anyonic quasiparticles, each distinguished by unique topological parameters, fractional charge, and statistics. In contrast, the integer quantum Hall (IQH) effects can be understood from the band topology of non-interacting electrons. We report a surprising super-universality of the critical behavior across all FQH  and IQH transitions. Contrary to the anticipated state-dependent critical exponents, our findings reveal the same critical scaling exponent $\kappa = 0.41 \pm 0.02$ and localization length exponent $\gamma = 2.4 \pm 0.2$ for fractional and integer quantum Hall transitions. From these, we extract the value of the dynamical exponent $z\approx 1$. We have achieved this in ultra-high mobility trilayer graphene devices with a metallic screening layer close to the conduction channels. The observation of these global critical exponents across various quantum Hall phase transitions was masked in previous studies by significant sample-to-sample variation in the measured values of $\kappa$ in conventional semiconductor heterostructures, where long-range correlated disorder dominates. We show that the robust scaling exponents are valid in the limit of short-range disorder correlations. }
\end{abstract}

\maketitle
\section*{Introduction}

The Quantum Hall (QH) effect, observed in a two-dimensional electron gas subject to a perpendicular magnetic field, realizes multiple quantum phase transitions (QPT) between distinct insulating topological states~\cite{PhysRevLett.45.494}. The magnetic field $\mathbf{B}$ quenches the electronic kinetic energy into disorder-broadened discrete Landau energy levels (LL). All electronic single-particle states are localized, barring those at a specific critical energy $E_c$ near the center of each LL, which are extended~\cite{PhysRevLett.68.1375, PhysRevB.23.5632, PhysRevLett.54.831, PhysRevLett.61.593, Chalker1988, PhysRevLett.82.5100}.
When the Fermi energy lies between the extended states of two successive LLs, the system is in a distinct topological phase characterized by a quantized value of Hall resistance $R_{xy}$ and vanishingly small longitudinal resistance $R_{xx}$.
%At each distinct topological phase, the transverse resistance quantized to $R_{xy} = h/(Ne^2)$ accompanied by a vanishingly small longitudinal resistance $R_{xx} \approx 0$.
%This is the integer quantum Hall (IQH) regime.
As the Fermi energy approaches $E_c$, the localization length $\xi$ characterizing the single-particle states diverges as $\xi \sim |E-E_c|^{-\gamma}$ while the slowest time-scale diverges as $\tau \sim \xi^z \sim |E-E_c|^{-z\gamma}$~\cite{PhysRev.177.952, PhysRevLett.102.216801}.
The exponent $\gamma$ governs the critical divergence of the localization length as the filling fraction or magnetic field approach the critical values and $z$ governs the divergence of the coherence length with decreasing temperatures \cite{RevModPhys.69.315}.
From the finite-size scaling theory \cite{PhysRevLett.61.1297, RevModPhys.69.315},
\begin{equation}
  dR_{xy}/d\nu_{\nu=\nu_c} \propto T^{-1/z\gamma}
  \label{eq1}
\end{equation}
Here, $\nu=nh/eB$, $n$ is the areal charge-carrier density, $h$ is the Planck constant, $e$ is the electronic charge, and $T$ is the temperature. One additionally defines the scaling exponent $\kappa = 1/z\gamma$ \cite{Dodoo-Amoo_2014, RevModPhys.67.357, PhysRevLett.61.1297} that governs this temperature dependence of the slope of $R_{xy}$ as well as the width of the $R_{xx}$ peak at the transition.
The values of these three critical exponents (of which only two are independent) have been argued to be universal, with $\gamma \approx 2.3$, $\kappa \approx 0.42$, and $z =1$ for all IQH transitions~\cite{PhysRevLett.128.116801, PhysRevLett.61.1297,RevModPhys.69.315,PhysRevLett.54.831,PhysRevLett.64.1437}.

Low temperatures and high magnetic fields enhance the effective electron-electron interactions, producing a richer set of the fractional quantum Hall (FQH) phases at rational filling fractions~\cite{PhysRevLett.48.1559}.   The question then arises: Can IQH and FQH phase transitions be analyzed using a `unified' scaling framework~\cite{RevModPhys.49.435}? While the IQH phases originate from the topology of the single particle electronic Chern bands~\cite{TKNN}, the FQH phases are crucially underlain by strong electronic interactions. These are marked by distinct electronic correlations, topological order, ground state degeneracy, and topological entanglement. The transition between FQH plateaus is driven by a proliferation of anyonic quasiparticles (characterized by quasiparticle statistics and fractional charge). This picture may suggest that the critical behavior at the transitions depends on the specifics of the topological FQH states involved and is also different from the analogous transitions in the IQH regime.

Experimental investigations of scaling in the IQH regime have reported $\kappa$ varying between $0.16 \leq \kappa \leq 0.81$ (Supplementary Information, Supplementary Note 13). This wide variation has been attributed to varying disorder correlation lengths with a universal critical behavior seen only in samples with short-range disorder~\cite{Wei1992,li2005scaling}.
This lack of a tight constraint on $\kappa$ has hindered any claims of their universality.
Similar experimental investigations of scaling laws at transitions between FQH phases are scarce~\cite{engel1990critical, MACHIDA2001182, PhysRevLett.130.226503}. A recent experimental study on extremely high-mobility 2D electron gas confined to GaAs quantum wells found the value of $\kappa$ in the FQH regime to be non-universal, this observation being attributable to long-range disorder correlation~\cite{PhysRevLett.130.226503}. Thus, despite over three decades of study, the fundamental question of the values of the critical exponents across quantum Hall transitions (integer and fractional) remains unsettled~\cite{PhysRevLett.130.226503, PhysRevLett.128.116801, PhysRevResearch.4.033146}.

This article reports the experimental observation of a surprising super universality in the scaling exponents for transitions between various IQH and FQH phases in trilayer graphene. We measure both the scaling exponent $\kappa$ and the localization length exponent $\gamma$ independently over several integer-to-integer, integer-to-fractional, and fractional-to-fractional Quantum Hall transitions.  Contrary to the expected picture of multiple plateau-to-plateau quantum phase transitions, each with its own distinct critical properties, here we find that for all IQH and FQH plateau-to-plateau transitions (PT), $\kappa = 0.41 \pm 0.02$, $\gamma \approx 2.4 \pm 0.2$, and $z \approx 1$,  closely aligned with the predictions of the scaling theory of localization~\cite{RevModPhys.67.357}. Given the distinct origins of the two phenomena, this striking similarity of the critical exponents suggests a connection between the IQH and FQH effects that transcends the composite fermion (CF) framework.

We estimate the values of $\kappa$ near criticality ($\nu \approx \nu_c$) using three distinct approaches: (i) analyzing the critical divergence of $d R_{xy}/d\nu$, (ii) probing the critical divergence of the inverse width of $R_{xx}(T)$, and (iii) a scaling analysis of $R_{xy}$ near the critical point. The localization exponent $\gamma$ is obtained deep in the tails of the localized regime from the dependence of $G_{xx}$ on $\nu$. A scaling analysis of Quantum Hall transitions for fractional and integer states provides a second, independent way to extract $\gamma$.

The realizations of these quantum phase transitions in graphene-based systems are associated with a highly tunable set of parameters. These include the ability to alter electron density, which is typically unachievable in semiconductor heterostructures~\cite{PhysRevLett.124.156801}, the capability to manage screening, and the option to induce band mixing by applying a displacement field $\mathbf{D}$. This flexibility helps us establish that weak Landau level mixing does not significantly affect these critical exponents.

Graphene also provides a platform where the nature of disorder scattering can be controlled. This is because the electrical transport properties of high-mobility graphene devices are dominated by short-range impurity scattering, while those of low-mobility graphene devices are controlled by both short-ranged and long-ranged scattering potentials~\cite{Sarkar2015,Rhodes2019}. Thus, high-mobility graphene devices represent a natural candidate to investigate the universality of scaling exponents. Our comparative study between graphene devices of varying mobility shows that as long as long-range impurity scattering can be suppressed, the universality of scaling parameters persists, independent of the quantum Hall bulk phases involved.

\section*{Results}
Standard dry transfer technique is used for the fabrication of dual graphite-gated hexagonal-boron-nitride (hBN) encapsulated TLG devices [Fig.~\ref{fig:fig1}(a)] (for details, see Supplementary Information, Supplementary Note 1) \cite{pizzocchero2016hot}. Fig.~\ref{fig:fig1}(b) shows  measurements of the longitudinal resistance $R_{xx}$ and the transverse conductance $G_{xy}$ versus the Landau level  filling factor $\nu$; the measurements were performed at $\mathbf{B}= 13\mathrm{T}$, $T=20\mathrm{mK}$ and $\mathbf{D}=0 \mathrm{V/nm}$. We identify several major odd denominator FQH states by prominent dips in $R_{xx}$ and corresponding plateaus in $G_{xy}$. Indications of developing $\nu = 3+1/5$ and $3+2/7$ states are also seen. Several of these FQH states are resolved at $\mathbf{B} = 4.5\mathrm{T}$, attesting to the high quality of the device in terms of excellent homogeneity of number density and suppression of long-range scattering  (Supplementary Information, Supplementary Note 6).

The band structure of TLG is formed of monolayer-like and bilayer-like Landau levels (Fig.~\ref{fig:fig1}(c)) -- these are protected from mixing by the lattice mirror-symmetry~\cite{koshino2011landau}. The calculated LL spectrum as a function of $\mathbf{B}$ and energy $E$  is shown in Fig.~\ref{fig:fig1}(d), where blue (red) lines mark the monolayer-like (bilayer-like) LLs. For $\mathbf{B} > 8\mathrm{T}$, the $\nu=2$ and $\nu=3$ arise from the spin-split  $N_{M}=0^{-}\uparrow$ and $N_{M}=0^{-}\downarrow$ bands of the monolayer-like LLs.  Here, $(+,-)$ refers to the two valleys, and $(\uparrow, \downarrow)$ refers to electronic spins.  We confine our study to $8\mathrm{T} < \mathbf{B}  < 13\mathrm{T}$ to avoid Landau level-mixing at lower $\mathbf{B}$ and phase transitions between competing FQH states at higher $\mathbf{B}$ ~\cite{PhysRevB.84.241306, PhysRevLett.124.097604, PhysRevB.87.245425}.

\section*{Critical exponents near FQH plateau-to-plateau transitions:}

Fig.~\ref{fig:fig2}(a) shows the $T$-dependence of $R_{xy}$ between the IQH states $\nu =-2$ and $\nu = -1$. Similar data for transition between the FQH states $\nu = 2+2/3$ and  $\nu = 2+3/5$ are shown in Fig.~\ref{fig:fig2}(b). The critical points $\nu_c$ of the plateau-to-plateau transition  (identified as the crossing point of the $R_{xy}$ curves at different $T$) are indicated in the plots. The exponent $\kappa$ evaluated from the peak value of d$R_{xy}$/d$\nu$ versus $T$ near criticality (Figs.~\ref{fig:fig2}(c-d)) in both cases is $\kappa=0.41 \pm 0.01$. Analysis of the $T$-dependence of the inverse of the half-width of $R_{xx}$ as $\nu$ is varied between two consecutive FQH plateaus also yields $\kappa=0.41 \pm 0.02$ (Supplementary Information, Supplementary Note 2).

To demonstrate the scaling properties of $R_{xy}$ in the vicinity of $\nu_c$, we use the following form~\cite{RevModPhys.67.357}:
	\begin{equation}
		R_{xy}(\nu,T)=R_{xy}(\nu_c) f[\alpha(\nu-\nu_c)]
		\label{eq:fss1}
	\end{equation}
with $\alpha\propto T^{-\kappa}$. Here,  $f(0)=1$, and $f'(0)\neq 0$.  This gives us a third, independent method of estimating $\kappa$. Fig.~\ref{fig:fig2}(e) shows the plots of $R_{xy}/R_{xy}(\nu_c)$ at various temperatures as a function of $\alpha|\nu-\nu_c|$ for the $\nu = 2+1/3$ to $2+2/5$ transition. $\alpha(T)$ is optimized to collapse the various constant-temperature data onto a single curve (the upper branch of which is for $\nu < \nu_c$, and the lower branch is for $\nu > \nu_c$). From the plot of $\alpha$ versus $T$ (inset of Fig.~\ref{fig:fig2}(e)) we obtain $\kappa = 0.40 \pm 0.03$.

To check the validity of our scaling analysis, we perform the following error analysis:  The residue in the least square fit between the scaling curves (like those shown in Fig.~\ref{fig:fig2}(e)) for each assumed value of $\kappa$ is calculated. This quantity, which we call fit error', is presented in Supplementary Information Supplementary Figure 6 and Supplementary Figure 7 in a semi-$\mathrm{log}$ scale; we find that the fit error is indeed minimum for $\kappa = 0.41$.

Fig.~\ref{fig:fig3}(a) compiles our findings. These results indicate a \(\kappa\) value of \(0.41 \pm 0.03\) uniformly observed across all probed transitions between IQH and FQH states (compare with Supplementary Figure 14 of Supplementary Information). This consistency in scaling exponents spans various transition types, including (1) transitions from one IQH state to another, (2) transitions among different FQH states, and (3) transitions between an IQH state and a neighboring FQH state. It is important to emphasize that the observed universality of $\kappa$ goes beyond marking an experimental confirmation of a uniform scaling law across FQH transitions in any material. Given the distinct physics of IQH and FQH states, such constancy of the scaling exponent is remarkable and underscores the universal applicability of the scaling principle across QH transitions. This is the central result of this article.

\textbf{Locating the transition:}
The physics of the FQH effect of electrons at a filling factor $\nu$ can be mapped onto that of IQH of CF at a filling factor $\nu_{CF}$, with $\nu= \nu_{CF}/(2\nu_{CF}\pm1)$~\cite{PhysRevLett.63.199}. It follows that the critical points for the transition between successive FQH phases at $\nu= \nu_{CF}/(2\nu_{CF}\pm1)$ and $\nu= (\nu_{CF}+1)/(2(\nu_{CF}+1)\pm1)$ occurs at~\cite{PhysRevLett.128.116801, PhysRevLett.65.907}:
	\begin{equation}
		\nu_c=\frac{(\nu_{CF}+0.5)}{2(\nu_{CF}+0.5)\pm 1}.
		\label{Eqn:Sscalingnu}
	\end{equation}
	%where $\nu_{CF}$ is the LL index of composite fermions.
The experimentally obtained values of $\nu_c$, extracted either from the crossing point of the $R_{xy}$ isotherms or the maxima of $R_{xx}$, match exceptionally well with the theoretical predictions (Fig.~\ref{fig:fig3}(b)) (Supplementary Information, Supplementary Table 1).

\textbf{Robustness of the critical exponents against  LL mixing.} A non-zero vertical displacement field $\mathbf{D}$ gives rise to a complex phase diagram in TLG, with the Landau levels inter-crossing multiple times, resulting in significant LL mixing as either $\mathbf{D}$ or $\mathbf{B}$ is varied~\cite{zibrov2018emergent, rao2020gully, winterer2022spontaneous, PhysRevB.87.115422, wang2016first}. LL-mixing can change the effective interaction between the electrons~\cite{PhysRevB.87.245425}. However, as shown in Fig.~\ref{fig:fig3}(c), it does not significantly affect the universality of $\kappa$. This vital result suggests that as long as the anyons are weakly interacting, the critical behavior of the localization-delocalization transition remains unaltered.

\section*{Measurement of localization exponent $\gamma$:}

We now focus on the localized regime, far away from $E_c$, marked by the black rectangle in Fig.~\ref{fig:fig5}(a). Given the presence of strong interactions, it is reasonable to assume that transport in this localized part of the energy spectrum proceeds through  Efros–Shklovskii (ES) type hopping mechanism~\cite{efros1975coulomb}. The localization exponent $\gamma$ determines the $T$ dependence of longitudinal conductance $G_{xx}$~\cite{ono1982localization, efros1975coulomb}:
	 \begin{equation}
		 G_{xx}=G_0e^{-(T_0/T)^{1/2}}
		  \label{Eqn:scaling4}
		  \end{equation}
	 with
    \begin{equation}
		 k_BT_0 \propto |\delta\nu|^\gamma.
		  \label{Eqn:scaling5}
		  \end{equation}
The pre-factor $G_0 \propto 1/T$ and  $\delta\nu = (\nu-\nu_c)$.  Fig.~\ref{fig:fig5}(b) shows plots of \textrm{log}$(TG_{xx})$ versus $T^{-1/2}$ at different values of $\delta\nu$; the dotted lines are linear fits to the data.  The linearity of the data at low-$T$  is consistent with  transport by the ES hopping mechanism in the FQH regime (Eqn.~\ref{Eqn:scaling4}). At high-$T$ (in the region marked in Fig.~\ref{fig:fig5}(b) by a dotted ellipse), the values of $G_{xx}$ become relatively large, and the plots deviate from a straight line. In passing, we note that as we move progressively closer to the center of the plateau in $R_{xy}$, where the value of $G_{xx} \approx 0$ at low-$T$, the linearity of the plots persists to higher temperatures. Fitting $T_0$  (estimated from Eqn~\ref{Eqn:scaling4}) and $|\delta\nu_{CF}|$ to Eqn.~\ref{Eqn:scaling5} , we find the estimated $\gamma$ to lie in the range $2.3 - 2.6$ (Fig~\ref{fig:fig5}(c)) for FQH plateau-to-plateau transitions, very close to the predicted range of $\gamma=2.3-2.5$~\cite{PhysRevLett.128.116801}. The fact that the exponent controlling the divergence of the localization length at criticality is almost identical for both FQH and IQH states points to an effective model of localization that is universal across the different statistics of the quasiparticles in these QH phases. Furthermore, from  $\kappa = 1/z\gamma \approx 0.41 \pm 0.1$ and $\gamma \approx 2.3$, we get  $z \approx 1$, as expected for a strongly interacting system~\cite{PhysRevLett.89.276801, PhysRevLett.82.5100, PhysRevB.48.11167,PhysRevLett.102.216801}.

An independent estimate of $\gamma$ is obtained by casting Eq.~\ref{Eqn:scaling4} into a single-parameter scaling form~\cite{hohls2002hopping}:
	 \begin{equation}
		 G_{xx}\left(s\right)=\sigma^*se^{-(T^*s)^{1/2}},
		  \label{Eqn:scaling6}
		  \end{equation}
with the the scaling parameter $s=|\delta\nu_{CF}|^\gamma/T$. Fig.~\ref{fig:fig5}(d) shows the scaling plots of $G_{xx}/s$ versus $s^{1/2}$ for the PT  in ES regime from $\nu =3+2/5$  to $\nu = 3+3/7$. We find a near-perfect data collapse for all values of $\delta\nu_{CF}$ in the localized regime with $\gamma \approx 2.3$, providing an independent validation of the universality of $\gamma$.

\section*{Discussion}
We are now in a position to compare the universality of $\kappa$ seen in the FQH PT in our high-mobility TLG with non-universality of the same measured in the high-mobility 2D semiconductors~\cite{PhysRevLett.130.226503}.
The large spread in the observed values of $\kappa$ seen in the data in GaAs quantum wells was attributed to two main reasons~\cite{PhysRevLett.130.226503}. The first is the formation of numerous emerging FQH phases between $\nu=1/3$ and $2/5$, which limits the temperature range over which one observes the decrease of the width of $R_{xx}$ with $T$. Note that in Fig.~\ref{fig:fig1}(b), there are two incipient  FQH phases, $\nu = 3+1/5$ and $3+2/7$,  between the more robust phases $\nu = 3$ and $\nu = 3 + 1/3$. The incipient phases are weak enough not to affect the scaling of the transition region in $R_{xy}$ even at the lowest temperature employed here. As a result, we find $\kappa=0.42 \pm 0.01$ (Fig.~\ref{fig:fig3}(a)).

The second reason is related to the type of disorder in the sample~\cite{PhysRevLett.130.226503}. Universality in $\kappa$ is observed only when the effective disorder potential is short-ranged~\cite {li2005scaling}, as in our graphite-gated high-mobility graphene devices. This is not the case in GaAs/AlGaAs systems, where long-range scattering potential from the impurities cannot be ignored~\cite{PhysRevLett.130.226503}. We fabricated graphene devices without the graphite gate electrodes to probe the effect of long-range interactions on $\kappa$. The graphene channel was no longer screened from long-range Coulomb fluctuations arising from the \ch{SiO2} substrate; this was reflected in reduced mobility   $\sim 2 - 5~ \mathrm{m^2/Vs}$. While in these devices we do not find FQH states, the value of $\kappa$ for IQH transitions varied widely between $0.45-0.64$ (Supplementary Information, Supplementary Note 4), supporting the conclusions of Ref~\cite{PhysRevLett.130.226503}.

To summarize, our principal finding is that scaling properties for transitions involving Abelian FQH states and/or IQH phases are universal.
Specifically, we have demonstrated the scaling of the longitudinal conductance (with a scaling exponent $\kappa = 0.41 \pm 0.02$ and localization exponent $\gamma \approx 2.3$) in the IQH and FQH states in Bernal-stacked ABA trilayer graphene. This conclusion holds for plateau-to-plateau transitions between two consecutive  IQH states, two FQH states, and even between IQH and the adjoining FQH state,  underlining the universal character of the scaling. This universality of $\kappa$ persists even when an external displacement field hybridizes the Landau levels of Bernal-stacked TLG. In fact, we find deviations from universality in the value of $\kappa$ only in devices where long-range scattering dominates. To our knowledge, ours is the first definite scaling analysis of the QPT  over a series of fractional QH states.

FQH phases are underlined by strongly correlated and interacting electrons. Our results demonstrate a surprising correspondence between the FQH phase transitions and those of non-interacting electrons. The results indicate a super-universality in the localization-delocalization transitions across distinct anyonic species that represent the characteristic quasiparticles of the FQH phases. While much is known about the localization of electrons, the observed super universality motivates the study of localization in anyonic quasiparticles and the mechanism that drives their conduction in the presence of disorder and quasiparticle interactions. Our study raises the natural question of whether the universality observed in this context applies to transitions between other topological phases with fractional excitations, such as fractional Chern insulators~\cite{Han2023}.

\section*{Methods}
\subsection*{Device fabrication}

Devices of dual graphite gated ABA trilayer graphene (TLG) heterostructures were fabricated using a dry transfer technique (for details, see Supplementary Information Supplementary Note 1). Raman spectroscopy and optical contrast were used to determine the number of layers and the stacking sequence. The devices were patterned using electron beam lithography followed by reactive ion etching and thermal deposition of Cr/Pd/Au contacts. Dual electrostatic gates were used to simultaneously tune the areal number density $n = [(C_{tgVtg} + C_{bgVbg})/e + n_o]$ and the displacement field  $D=[(C_{bg}V_{bg}-C_{tg}V_{tg})/2\epsilon_0+D_{0}]$ across the device. Here $C_{bg} (C_{tg})$ is the capacitance of the back gate (top gate), and $V_{bg} (V_{tg})$ is the voltage of the back gate (top gate). The values of $C_{tg}$ and $C_{bg}$ are determined from quantum Hall measurements. $n_o$ and $D_o$ are the residual number density and electric field due to unavoidable impurities in the channel.
 \subsection*{Transport measurements}
The electrical transport measurements were performed in a dilution refrigerator (with a base temperature of $\mathrm{20~mK}$) at low frequency ($11.4~\mathrm{Hz}$)  using standard low-frequency measurement techniques, with a bias current of $10~\mathrm{nA}$.

	\section*{Data availability}
	The authors declare that the data supporting the findings of this study are available within the main text and its Supplementary Information. Other relevant data are available from the corresponding author upon request.

\section*{Code availability}
	The codes that support the findings of this study are available from the corresponding author upon request.

	\section*{Acknowledgements}
	We thank Jainendra K. Jain, Sankar Das Sarma, Rajdeep Sensarma, Nandini Trivedi, Ravindra Bhatt and Prasant Kumar for helpful discussions and clarifications. We acknowledge Ramya Nagarajan for data on low mobility samples. A.B. acknowledges funding from U.S. Army DEVCOM Indo-Pacific (Project number: FA5209   22P0166) and Department of Science and Technology, Govt of India (DST/SJF/PSA-01/2016-17). K.W. and T.T. acknowledge support from the JSPS KAKENHI (Grant Numbers 21H05233 and 23H02052) and World Premier International Research Center Initiative (WPI), MEXT, Japan. G.J.S. thanks Condensed Matter Theory Center and Joint Quantum Institute, University of Maryland College Park, for their hospitality during the preparation of this manuscript. Y.G. acknowledges the support by the Deutsche Forschungsgemeinschaft (DFG) through grant
	No. MI 658/10-2 and  RO 2247/11-1, the US-Israel Binational Science Foundation 2022391, and the Minerva Foundation. Y.G. is the incumbent of the InfoSys chair at IISc.

	\section*{Author contributions}
	S.K., T.C., K.R.A., D.S.,  and A.B. conceived the idea of the study,  conducted the measurements, and analyzed the results. T.T. and K.W. provided the hBN crystals. U.G., G.J.S., and Y.G. developed the theoretical model. All the authors contributed to preparing the manuscript.

	$^\dagger$ These authors contributed equally.

	\section*{Competing interests}
	The authors declare no competing interests.

	\clearpage

%\section{Figure Captions}

\begin{figure*}[t]
		\includegraphics[width=\columnwidth]{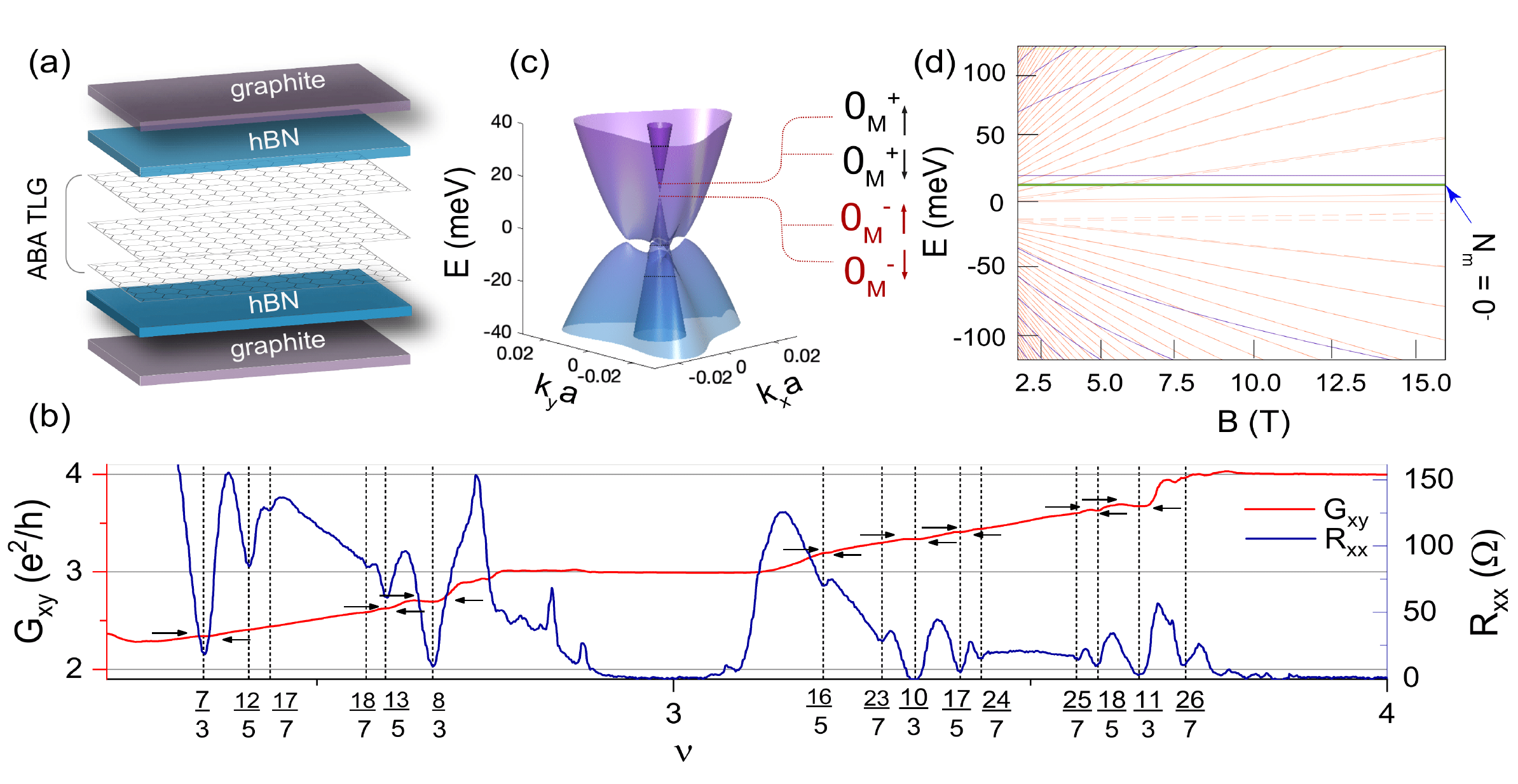}
		\small{\caption{\textbf{FQH in Bernal-stacked TLG} (a) Device schematic of TLG encapsulated between two hBN and few-layer graphite flakes. (b) Line plots of $G_{xy}$ (left-axis; solid red line) and $R_{xx}$ (right-axis; solid blue line)  versus $\nu$ measured at $\mathbf{B}=13\mathrm{T}$, $T=20$~mK, and $\mathbf{D}=0 \mathrm{V/nm}$. The dashed vertical lines mark the FQH states formed at corresponding $\nu$, and the arrows indicate corresponding plateaus in $G_{xy}$.  (c) Calculated band structure of Bernal stacked trilayer graphene for $\mathbf{D}=0\mathrm{V/nm}$. The four LLs of the $N_M = 0$ (The MLG LLs are marked by the subscripts $M$, and orbital contents are given by the numbers $0$) band are indicated schematically. (d) Calculated Landau levels as a function of energy $E$ and $\mathbf{B}$ for $\textbf{D}=0\mathrm{V/nm}$. The blue lines are the monolayer-like LLs, while the red lines are the bilayer-like LLs. The solid and dotted lines indicate the LLs from $K$ and $K'$-valley, respectively. The solid-green line is the spin-degenerate $N_{M}=0^{-}\uparrow$ and $N_{M}=0^{-}\downarrow$ monolayer-like LLs that host the FQH states probed in this article.}\label{fig:fig1}}
	\end{figure*}

	%\clearpage
	%%% Figure 2
	\begin{figure*}[t]
	\includegraphics[width=\columnwidth]{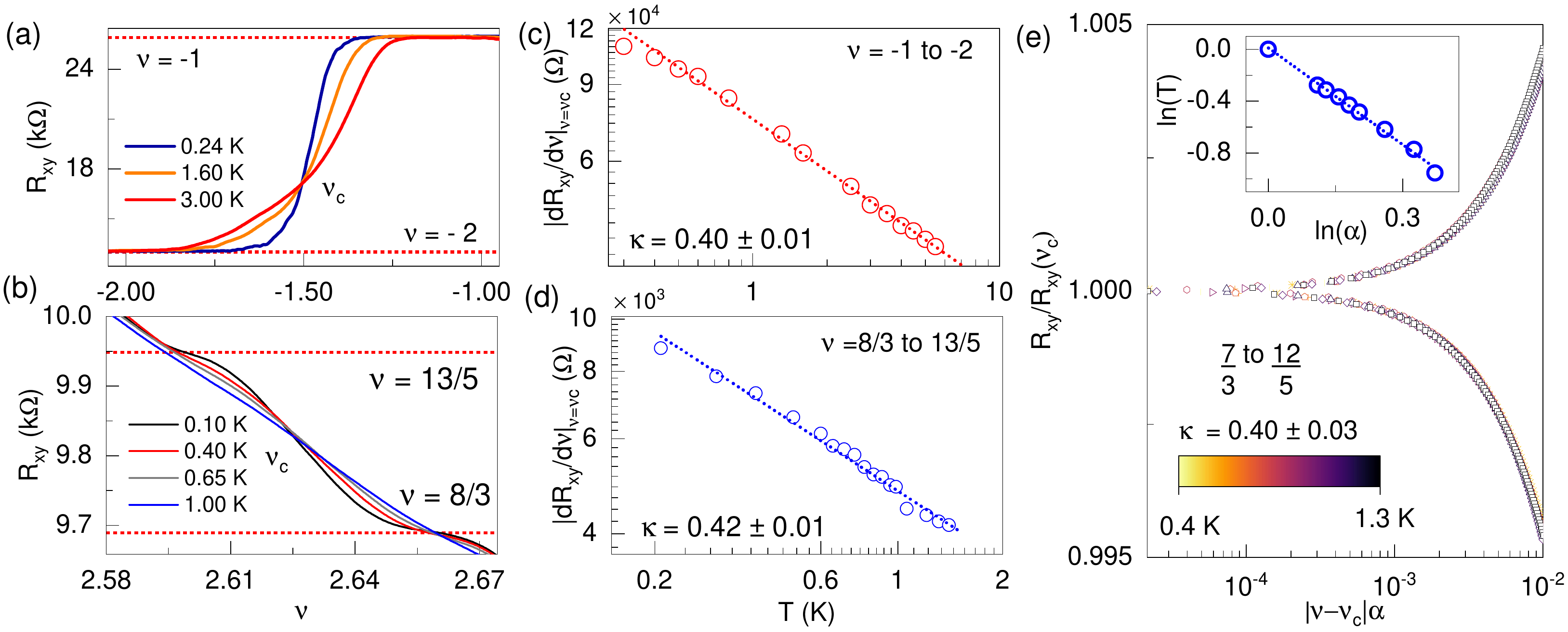}
		\small{\caption{\textbf{Scaling near $\nu=\nu_c$.}
				Plot of $R_{xy}$ versus $\nu$ for transition between the (a) IQH states $\nu = -2$ and $\nu = -1$ (the critical point $\nu_c = -1.5$), and (b) the  FQH states $2+2/3$ and $2+3/5$ ($\nu_c=2.625$). (c) Double logarithmic plot of $|\mathrm{d}R_{xy}/\mathrm{d}\nu|$ versus $T$ for the PT $\nu = -2$ and $\nu = -1$ at $\nu_c$. The dashed line is the fit to the data points using Eqn.~\ref{eq1}. (d) Same as in (c) for the  PT between FQH states $2+2/3$ and $2+3/5$. (e) Scaling analysis of $R_{xy}$ for the PT transition between $\nu = 2+1/3$ and $\nu = 2+2/5$. The inset is a plot of $T$ versus $\alpha$ in a double logarithmic scale (open circles); a linear fit to the data  (dotted line) yields $\kappa = 0.40 \pm 0.03$. (For an error analysis, see Supplementary Information, Supplementary Note 7.)}
			\label{fig:fig2}}

	\end{figure*}

	%\clearpage
	%%% Figure 3
	\begin{figure*}[t]
	\includegraphics[width=\columnwidth]{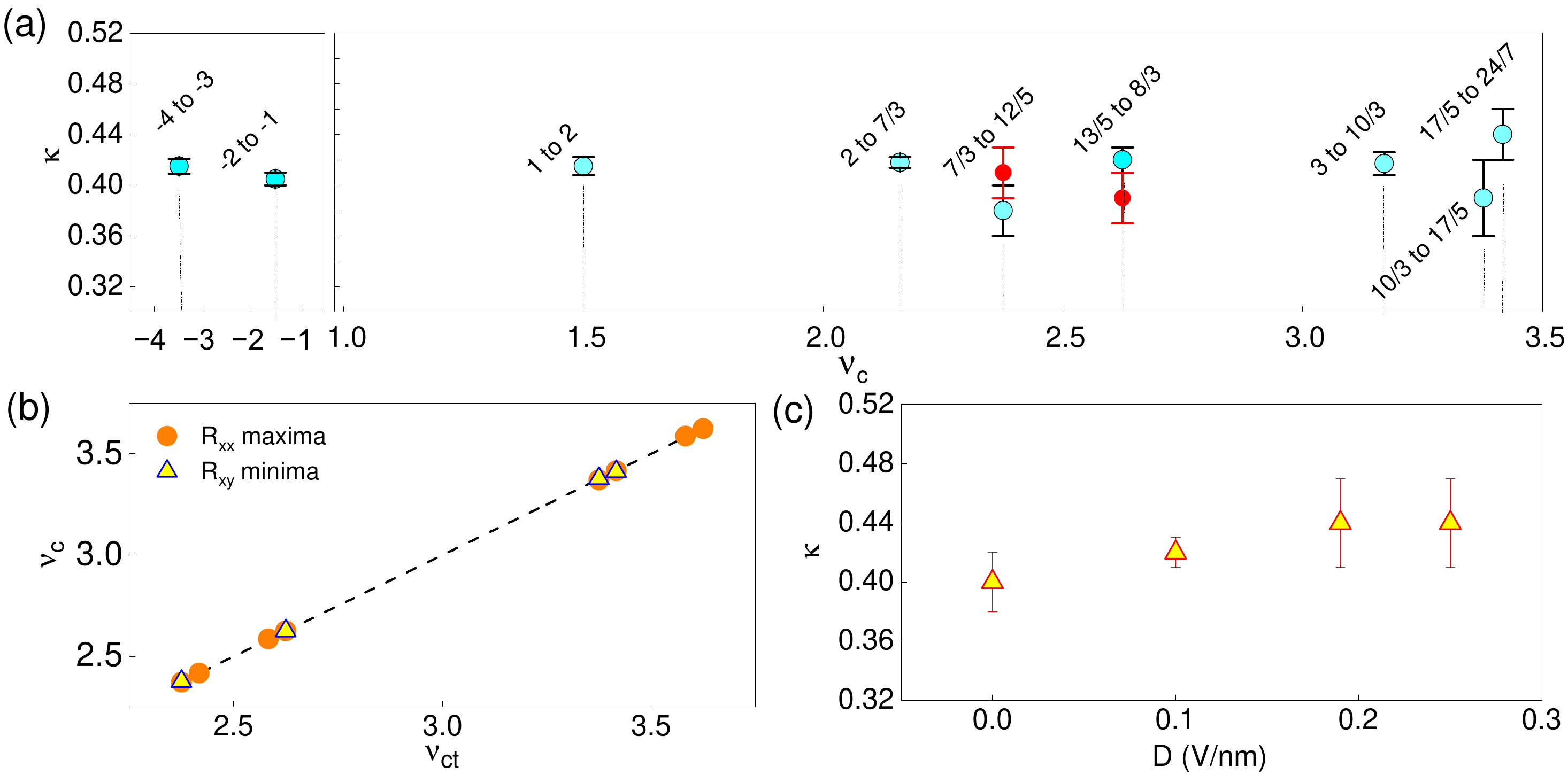}
		\small{\caption{ \textbf{Scaling exponents for different PT} (a) Plot of $\kappa$ as a function of $\nu_c$ corresponding to different PT evaluated from the maxima of derivative $(dR_{xy}/d\nu)^{max}$ near critical point. The dotted vertical lines mark the experimentally obtained $\nu_c$. The light blue symbols are for the $\kappa$ values obtained for trilayer graphene, and the red symbols are for the single-layer graphene. (b) Plot of experimentally obtained values of critical points, $\nu_c$ versus those theoretically calculated $\nu_{ct}$ ~\cite{PhysRevLett.128.116801}. The triangles are the values determined from crossing points of isotherms in $R_{xy}$ while the circles are determined from the $R_{xx}$ maxima. The black dashed line fits the data points with slope = $1.00 \pm 0.002$. (c) Plot of $\kappa$  versus $\mathbf{D}$ for the FQH transition from $\nu = 8/3$ to $\nu=13/5$ states evaluated from the maxima of derivative $(dR_{xy}/d\nu)^{max}$ near critical point. The error bars are determined from the least square fits to the data.}

			\label{fig:fig3}}

	\end{figure*}

		%%% Figure 4
	\begin{figure*}[t]
	\includegraphics[width=\columnwidth]{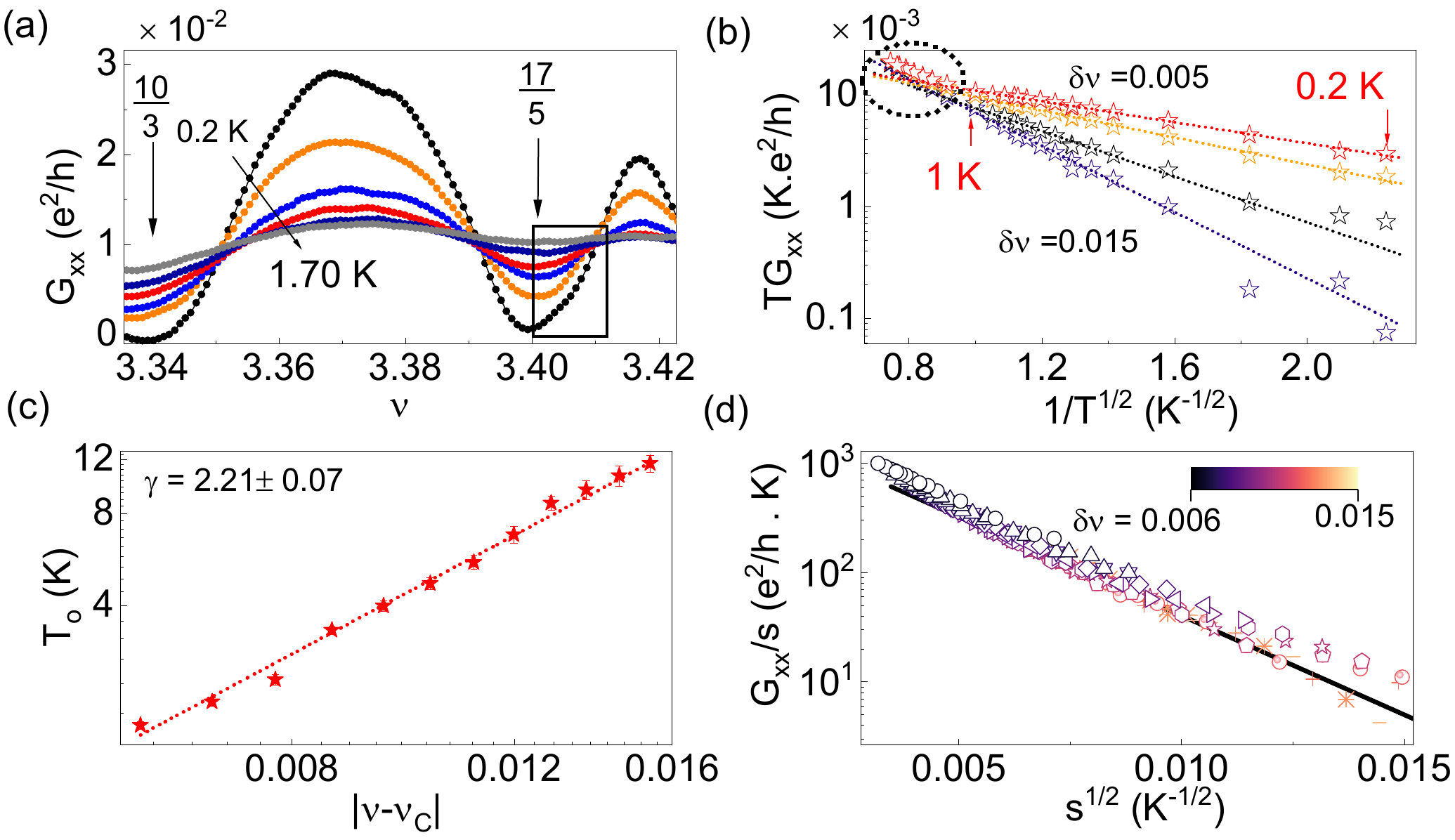}
		\small{\caption{\textbf{Scaling exponent in the ES regime for $\nu=3+2/5$ to $3+3/7$ transition} (a) Plots of the $T$-dependence of $G_{xx}$ versus filling factor $\nu$ for two FQH states between $\nu=3$ and $\nu = 4$.  The black box marks the region where the ES analysis was carried out. (b) Fit of ES Eqn.~\ref{Eqn:scaling4} (dotted lines) to the $G_{xx}$ data for the transition from $\nu = 3+2/5$ and $\nu = 3+3/7$. Each set of data points is for a given value of $\delta\nu$=|$\nu$-$\nu_c$| with $\nu_c=3.416$. The plots deviate from the expected ES behavior at high $T$ (the region is marked with an ellipse). (c) Plots of $T_0$ versus $\delta\nu_{CF}$. The dotted line is a linear fit to the data (see  Eqn.~\ref{Eqn:scaling5}). The slope yields the value of $\gamma$. The error bars are determined from the least square fits to the data in (b). (d) Plot of scaled longitudinal conductance $G_{xx}/s$ as a function of scaling parameter $s = |\delta\nu|^\gamma/T$ for PT between $\nu=17/5$ and $\nu=24/7$. The scatter points of different colors are for different values of $|\delta\nu|$, and the solid black line is fit to Eqn.~\ref{Eqn:scaling6}.
				\label{fig:fig5}}}
	\end{figure*}

	\clearpage

	\clearpage

	\section*{ Supplementary Information }
	\addto\captionsenglish{\renewcommand{\figurename}{}}
	\addto\captionsenglish{\renewcommand{\tablename}{}}
	\renewcommand{\theequation}{Supplementary Equation~\arabic{equation}}
	\renewcommand{\thesection}{S~\arabic{section}}
	\renewcommand{\thefigure}{\textbf{Supplementary Figure~\arabic{figure}}}
	\renewcommand{\thetable}{\textbf{Supplementary Table~\arabic{table}}}

	\setcounter{figure}{0}
	\setcounter{equation}{0}
	\setcounter{section}{0}
	\setcounter{table}{0}

	%%%%%%%%%%%%%%%%%%%%%%%%%%%%%%%%%%%%%%%%%%%%%%%%%%%%%%%%%%%%%%%%%%%%%%%%%%%%%%%%%%%%%%%%%%%%%%%%%%

	\section*{\textbf{Supplementary Note 1:
			Device fabrication, device schematics, and characterization}}
	%%% Figure 1

	\begin{figure}[h]
		\includegraphics[width=\columnwidth]{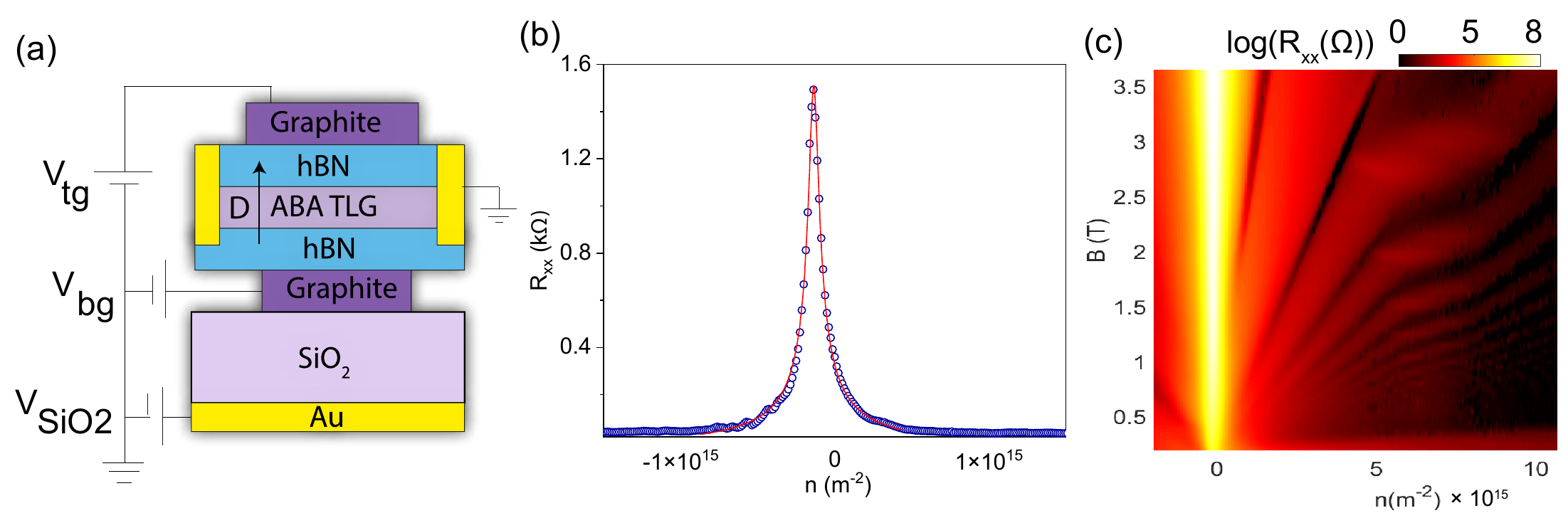}
		\caption{(a) Schematic of the device. Two gates $V_{bg}$ and $V_{tg}$ (with $\approx$ 40~nm thick bottom and $\approx$ 25~nm top hBN flakes as gate dielectrics and thin graphite as gate contacts) are used to tune the number density and displacement field across the flake. A silicon back-gate (with  \ch{SiO2} as the gate dielectric)  is used to dope the graphene contacts of the device to prevent the formation of p-n junctions. (b) Plot of resistance versus number density at $\mathbf{D} =0$~V/nm and $20$~mK. The red line is the fit to ~\ref{en. 0}. (c) Landau level fan diagram for TLG measured at $7$~K. Color map shows the $R_{xx}$ in logarithmic scale. }
		\label{fig:figS1}
	\end{figure}
	Bernal-stacked trilayer graphene (TLG), hBN, and graphite flakes are mechanically exfoliated on  Si substrates with a 300~nm thick top SiO$_2$ layer. TLG flakes are first identified through color contrast under an optical microscope and further confirmed using Raman spectroscopy \cite{cong2011raman,nguyen2014excitation}.  The standard dry pickup and transfer technique is used to fabricate the heterostructure. The flakes are picked up sequentially using polycarbonate (PC) film at $T = 120^\circ$~C in the following order: graphite/hBN/TLG/hBN/graphite. The entire stack, along with the PC film, is transferred on \ch{Si}/\ch{SiO2} substrate at $180^\circ$~C followed by cleaning in chloroform, acetone, and IPA solution to remove the PC residue. The heterostructure is then annealed in vacuum at $300^\circ$~C for $4$ hours. We employ electron beam lithography for defining the contacts on the heterostructure. This is followed by etching with a mixture of \ch{CHF3} and \ch{O2} gases and metal deposition with Cr/Pd/Au ($3$~nm/$12$~nm/$55$~nm) to create 1-D contacts~\cite{Wang614,PhysRevLett.126.096801}.

	Avoiding the formation of p-n junctions is absolutely essential if the devices are to be operated at high displacement fields~\cite{lui2011observation, wang2016first, datta2018landau}. We achieve this by doping the graphene contacts (that extend out of both the graphite gates) to high charge-carrier density. A schematic of the device is shown in ~\ref{fig:figS1}(a). Two common kinds of TLG flakes are typically obtained during mechanical exfoliation: ABA (or Bernal-stacked) and ABC. ABC, being a metastable stacking \cite{chen2019evidence, zhou2021superconductivity}, generally converts into ABA stacking during fabrication. These two phases are easily distinguishable by Raman spectroscopy and transport measurements -- displacement field opens up a band gap in ABC TLG ~\cite{zou2013transport, jhang2011stacking,PhysRevLett.70.3796}. In contrast, a band gap does not open in ABA TLG.

	To calculate the mobility of the sample, we have fitted the measured resistance $R$ as a function of number density at $\mathbf{D}$ $=0$ V/nm and $\mathbf{B}=0$ T with the following equation \cite{10.1063/1.3592338}:
	\begin{equation}
		R = R_{c}+\frac{L}{We\mu \sqrt{n^2+n_0^2}}
		\label{en. 0}
	\end{equation}
	where $R_c$, $L$, and $W$  are the contact resistance, length of the device, and width of the device, respectively. $\mu$ is the mobility of the device. From the fit ( \ref{fig:figS1}(b)), we extract $\mu = 40~\mathrm{m^2V^{-1}s^{-1}}$ and the intrinsic carrier concentration induced by charge impurity $n_0~\approx 3.32 \times 10^{13}$~$\mathrm{m^{-2}}$ reflecting the high quality and low impurity of the sample.

	The average distance between the impurities in the device is $l_i\approx 300$~nm.  For $\mathbf{B}=10$ T, the magnetic length is $l_B =\sqrt{\hbar/eB} \approx 8$~nm. Thus, $l_{B} \ll l_{i}$, implying that the charge impurities concentration is not large enough to produce a significant long-range potential. Also, $l_i$ is significantly larger than the distance between the graphene channel and the gates ($\sim 40$~nm). Thus, one can safely assume that the coulomb interactions due to the impurities are screened.

	~\ref{fig:figS1}(c) shows the Landau level fan diagram of the sample measured at 7~K. It matches pretty well with the simulated LL plot shown in Fig.~1(d) of the main manuscript with clear indications of monolayer-like Landau levels (LL) around a charge-carrier density $5\times10^{15}$~$\mathrm{m^{-2}}$ that cross the bilayer-like LLs confirming the system to be ABA trilayer graphene \cite{PhysRevLett.117.076807,taychatanapat2011quantum}.

	\section*{\textbf{Supplementary Note 2: Estimation of $\kappa$ from the temperature dependence of the width of $R_{xx}$. }}

	\begin{figure}[t]
		\includegraphics[width=\columnwidth]{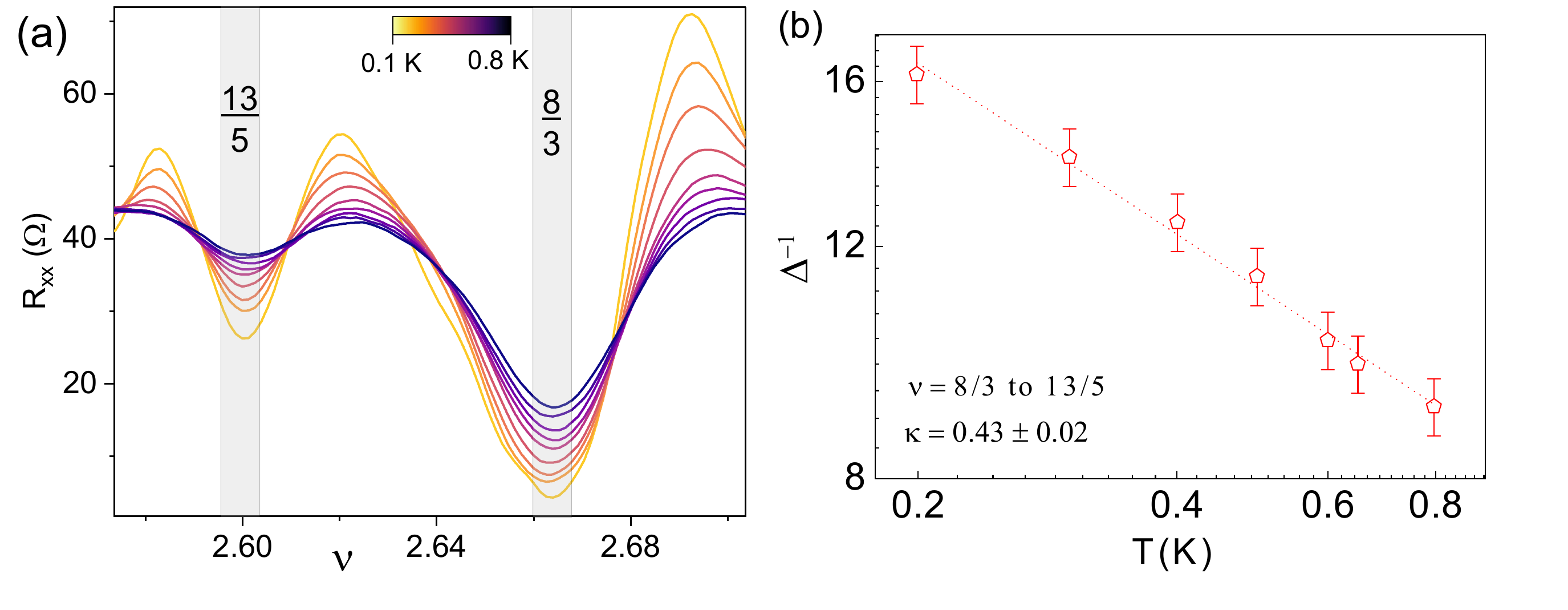}
		\caption{\textbf{Calculating $\kappa$ from width of $R_{xx}$}. (a) Longitudinal resistance as a function of filling factor at $\mathbf{B}=8.5 \mathrm{T}$. (b) Double logarithmic plots of the inverse of the half-width
			of longitudinal magnetoresistance $R_{xx}$ versus $T$ for PT between $\nu =2+2/3$ and $\nu=2+3/5$. Error bar is determined from least square fit to the data points.}

		\label{fig:figS2}
	\end{figure}
	At the critical point of the quantum Hall plateau-to-plateau transitions (PT), both d$R_{xy}$/d$\nu$ and the inverse of the half-width of $R_{xx}$ versus $\nu$ plot diverge according to power law $T^{-\kappa}$ \cite{RevModPhys.67.357}. In the main manuscript, we estimated the value of $\kappa$ by evaluating d$R_{xy}$/d$\nu$ close to the critical point. Here, we focus on the analysis of the width $\Delta$ of $R_{xx}$ (FWHM of $R_{xx}$ transition peak) versus $\nu$ \cite{Dodoo-Amoo_2014,engel1990critical}. At the critical point, $\Delta^{-1}$ diverges like $T^{-\kappa}$.
	The dependence of $\Delta^{-1}$ on $T$ for the transition between $\nu =2+2/3$ and $\nu=2+3/5$ is shown in ~\ref{fig:figS2}. The slope of linear fits to data yields $\kappa = 0.43 \pm 0.02$.

	\section*{\textbf{Supplementary Note  3: Critical behavior of various plateau-to-plateau transitions}}

	In ~\ref{table:kappa}, we compare our experimentally obtained values of $\nu_c$ with the theoretically predicted values~\cite{goldman1990nature,PhysRevLett.128.116801}:
	\begin{equation}
		\nu_c=\frac{(n+0.5)}{2\hspace{1mm}(n+0.5)\pm1};
		\label{Eqn:Sscalingnu}
	\end{equation}
	where $n$ is the LL index of composite Fermions.

	\begin{table}[h]
		\centering
		\resizebox{\columnwidth}{!}{
			%\begin{tabular}{|c|c|c|c|c|m{7cm}|} \hline
			\begin{tabular}{| p{.10\textwidth} | p{.10\textwidth} | p{.2\textwidth} | p{.20\textwidth} |p{.30\textwidth} |}
				\hline
				$\nu_1$&  $\nu_2$ &  $\nu_c^{xy}$ & $\nu_c^{xx}$ & $\nu_c$ (predicted)     \\
				\hline

				$\nu=\frac{7}{3}$ & $\nu=\frac{12}{5}$ & 2.375 $\pm$ 0.002 &  2.371 $\pm$ 0.003 & 2.375 \\
				\hline

				$\nu=\frac{12}{5}$ & $\nu=\frac{17}{7}$ & -- & 2.417 $\pm$ 0.003 & 2.417 \\ \hline

				$\nu=\frac{18}{7}$ & $\nu=\frac{13}{5}$  & --& 2.586 $\pm$ 0.002  & 2.583 \\ \hline

				$\nu=\frac{13}{5}$ & $\nu=\frac{8}{3}$ &  2.625 $\pm$0.003& 2.624 $\pm$ 0.002 & 2.625\\ \hline

				$\nu=\frac{10}{3}$ & $\nu=\frac{17}{5}$ & 3.377 $\pm$ 0.002 & 3.371 $\pm$ 0.003 & 3.375 \\ \hline

				$\nu=\frac{17}{5}$ & $\nu=\frac{24}{7}$ & 3.416 $\pm$ 0.003 & 3.417 $\pm$ 0.003 &3.417 \\ \hline

				$\nu=\frac{25}{7}$ & $\nu=\frac{18}{5}$ &-- & 3.588 $\pm$ 0.002 & 3.583 \\ \hline

				$\nu=\frac{18}{5}$ & $\nu=\frac{11}{3}$ &-- & 3.624 $\pm$ 0.004 & 3.625 \\ \hline

		\end{tabular}}

		\caption{Experimentally determined values of  $\nu_c$ for high-mobility Bernal-stacked trilayer graphene devices for  plateau-to-plateau transition between filling factors $\nu_1$ and $\nu_2$. $\nu_{c}^{xy}$ ($\nu_{c}^{xx}$)  is the value of the critical filling factor obtained from the crossing points of $R_{xy}$ (maxima of $R_{xx}$). Also tabulated are the theoretical predictions for $\nu_c$~\cite{goldman1990nature, PhysRevLett.128.116801}. Here, error bar in  $\nu_{c}^{xy}$ and $\nu_{c}^{xx}$ is range of $\nu$ where R$_{xy}$ intersects and R$_{xx}$ has maxima value.}
		\label{table:kappa}
	\end{table}

	%\clearpage %[KRA: Trying to make the table S1 appear after the equations]
	\section*{\textbf{Supplementary Note  4: Scaling in low-mobility devices}}
	\begin{figure*}[h]
		\includegraphics[width=0.8\columnwidth]{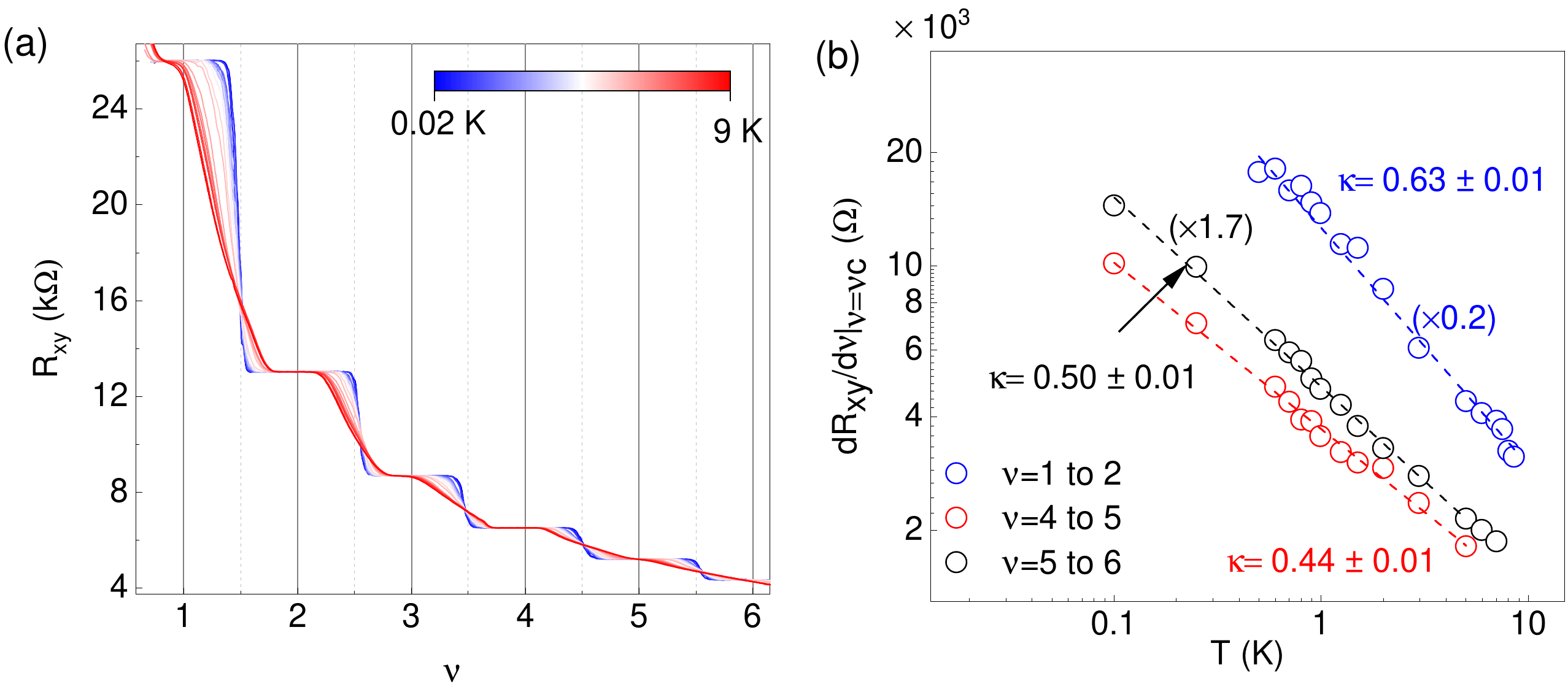}
		\caption{\textbf{Scaling exponents in low mobility graphene device}. (a) Plots of $R_{xy}$ versus filling factor at different temperatures. (b) Plots of  (d$R_{xy}/$d$\nu$)$^{max}$ at the critical point $\nu = \nu_C$ versus temperature in double logarithmic scale for various plateau-to-plateau transitions. The values of $\kappa$ extracted from the plots are mentioned in the plot. Here, error bar is calculated using least square fit to the data points.}

		\label{fig:figS3}
	\end{figure*}

	To compare the effect of long-range and short-range potential disorders~\cite{li2005scaling} on the scaling exponents, we fabricated hBN-encapsulated graphene heterostructures without the back graphite electrode. The number density across these devices is tuned using a Si/\ch{SiO2} gate. Despite being hBN encapsulated, effects of Coulomb impurities present at the \ch{SiO2} surface containing dangling bonds are not screened. These lead to long-range potential fluctuations across the device~\cite{Martin2008, Sarkar2015}. ~\ref{fig:figS3} shows the variation of d$R_{xy}/\mathrm{d}\nu$ at ${\nu=\nu_c}$  as a function of temperature for one such device for different plateau-to-plateau transitions. We observe a large spread in values of the scaling exponent $\kappa$, as opposed to the case of high-mobility devices discussed in the main manuscript, where the values of $\kappa$ were tightly clustered around the theoretically predicted value of $0.42$.  Our analysis supports the recent observations where the presence of long-range interactions made the scaling exponent non-universal~\cite{PhysRevLett.130.226503}.

	\section*{\textbf{Supplementary Note 5: Second derivative of $R_{xy}$ with temperature.}}

	As discussed in the main manuscript, a single parameter scaling function can be written down for the resistance tensor for plateau-to-plateau transitions~\cite{pruisken1988universal, RevModPhys.67.357,  WEI199034}:
	\begin{eqnarray}
		R_{xy}(\nu,T)=R_{xy}(\nu_c) f[T^{-\kappa}(\nu-\nu_c)]
		\label{Eqn:scaling2}
	\end{eqnarray}
	This immediately leads to
	\begin{equation}
		\frac{\mathrm{d}R_{xy}}{\mathrm{d}\nu}\propto T^{-\kappa}
	\end{equation}
	and
	\begin{equation}
		\frac{\mathrm{d}^2R_{xy}}{\mathrm{d}\nu^2}\propto T^{-2\kappa}
		\label{Eqn:Sscaling2}
	\end{equation}

\begin{figure}[h]
		\includegraphics[width=0.95\columnwidth]{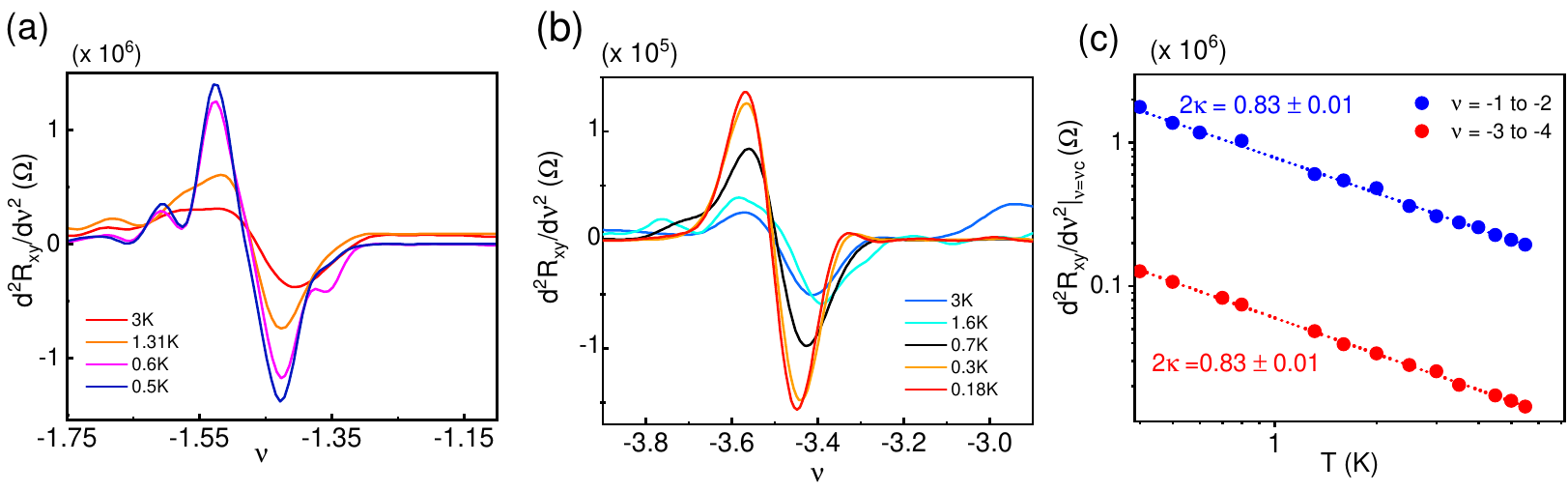}
		\caption{\textbf{The second derivative of $R_{xy}$ with temperature.} Plots of $\mathrm{d}^2R_{xy}/\mathrm{d}\nu^2$ vs $\nu$ at different temperatures for plateau-to-plateau transitions between between (a) $\nu=-1$ and $\nu=-2$ and (b) $\nu=-3$ and $\nu=-4$. (c)  $\mathrm{log}- \mathrm{log}$ plot of $\mathrm{d}^2R_{xy}/\mathrm{d}\nu^2$ vs $T$ for two different PT (open circles). The dotted lines are the linear fits to the data. Here, error bar is calculated using least square fit to the data points.}

		\label{fig:figS4}
	\end{figure}

	~\ref{fig:figS4} (a) and (b) show plots of d$^2R_{xy}/\mathrm{d}{\nu}^2$ as a function of temperature for two different plateau-to-plateau transitions. ~\ref{fig:figS4} (c) shows the variation of the d$^2R_{xy}/\mathrm{d}{\nu}^2$ at $\nu=\nu_c$ with temperature in $\mathrm{log}-\mathrm{log}$ scale. The slope yields $2\kappa \approx 0.83$, a value matching very closely with the prediction of ~\ref{Eqn:Sscaling2}.

	\section*{\textbf{Supplementary Note 6: Fractional Quantum Hall states at $\mathbf{B}= 4.5$~T.}}
\begin{figure*}[h]
		\includegraphics[width=0.9\columnwidth]{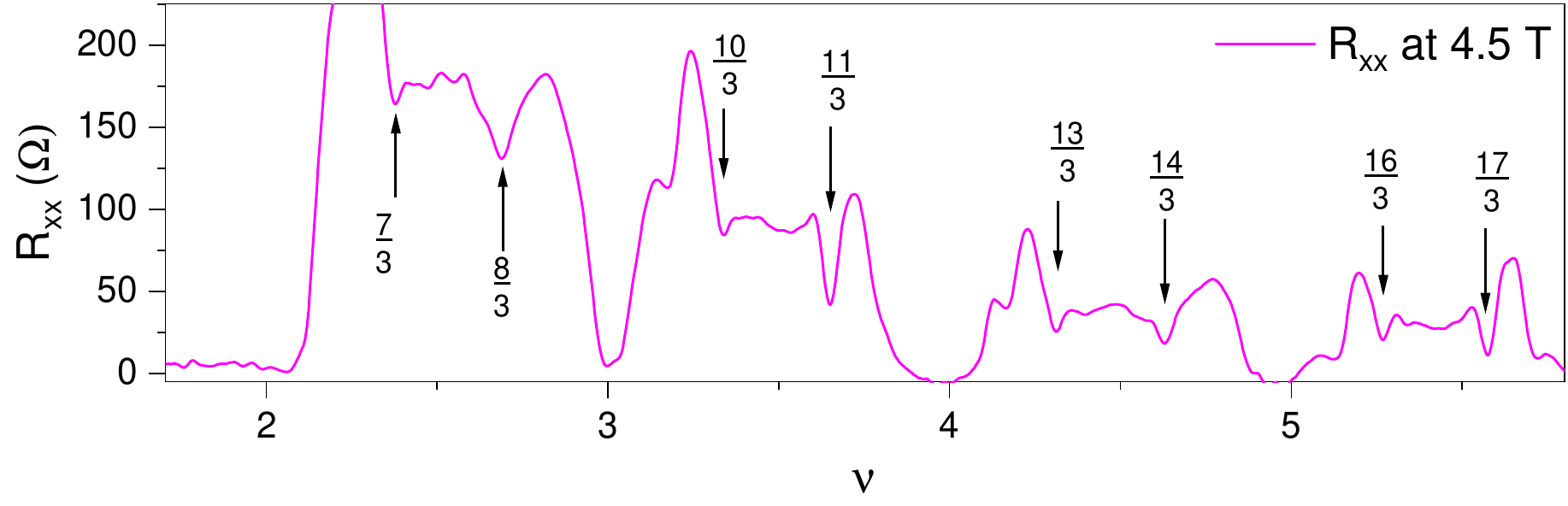}
		\caption{\textbf{Fractional Quantum Hall states at \textbf{B} $= 4.5$~T}.  Plot of $R_{xx}$ versus $\nu$ measured at \textbf{B}$=4.5$~T and $T=20$~mK. The major FQH that begin to form are marked by arrows.}

		\label{fig:figS5}
	\end{figure*}

	In ~\ref{fig:figS5} plots the longitudinal resistance $R_{xx}$ as a function of filling factor $\nu$. We can see the emergence of FQH states at $ \nu = N+1/3$ and $ \nu = N+2/3$ at \textbf{B}$= 4.5$~T.

	\section*{\textbf{Supplementary Note 7: Details of scaling analysis.}}

	\begin{figure*}[b]
		\includegraphics[width=0.9\columnwidth]{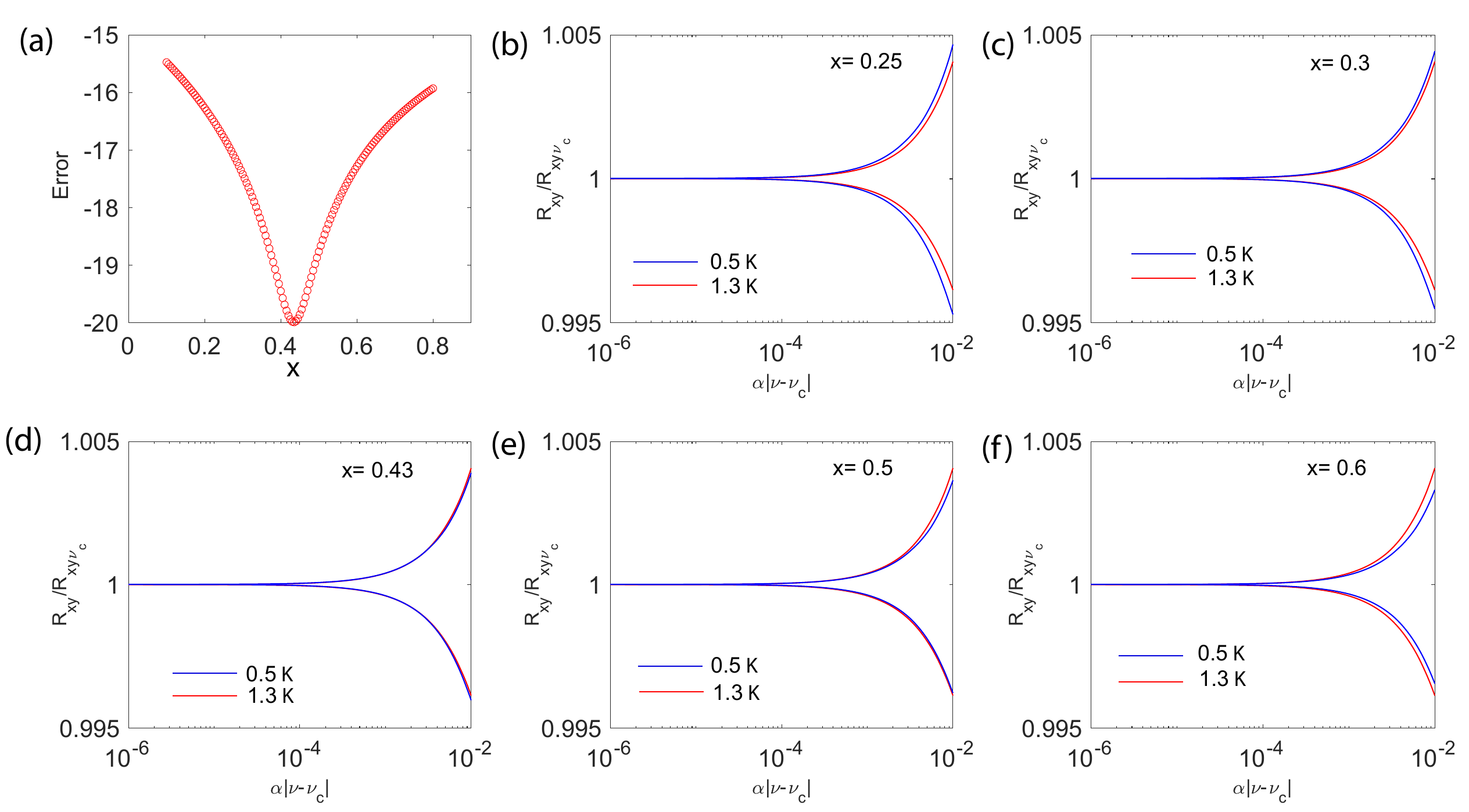}
		\caption{\textbf{Scaling analysis for transition between $\nu = 2 + 2/3$ and $\nu = 2 + 3/5$.} (a) Plot of the error in scaling versus $x$. (b-f) Scaling plot for different values of $x$ (the values of $x$ are marked inside the plot). }

		\label{fig:figS6}
	\end{figure*}
	\begin{figure*}[h]
		\includegraphics[width=\columnwidth]{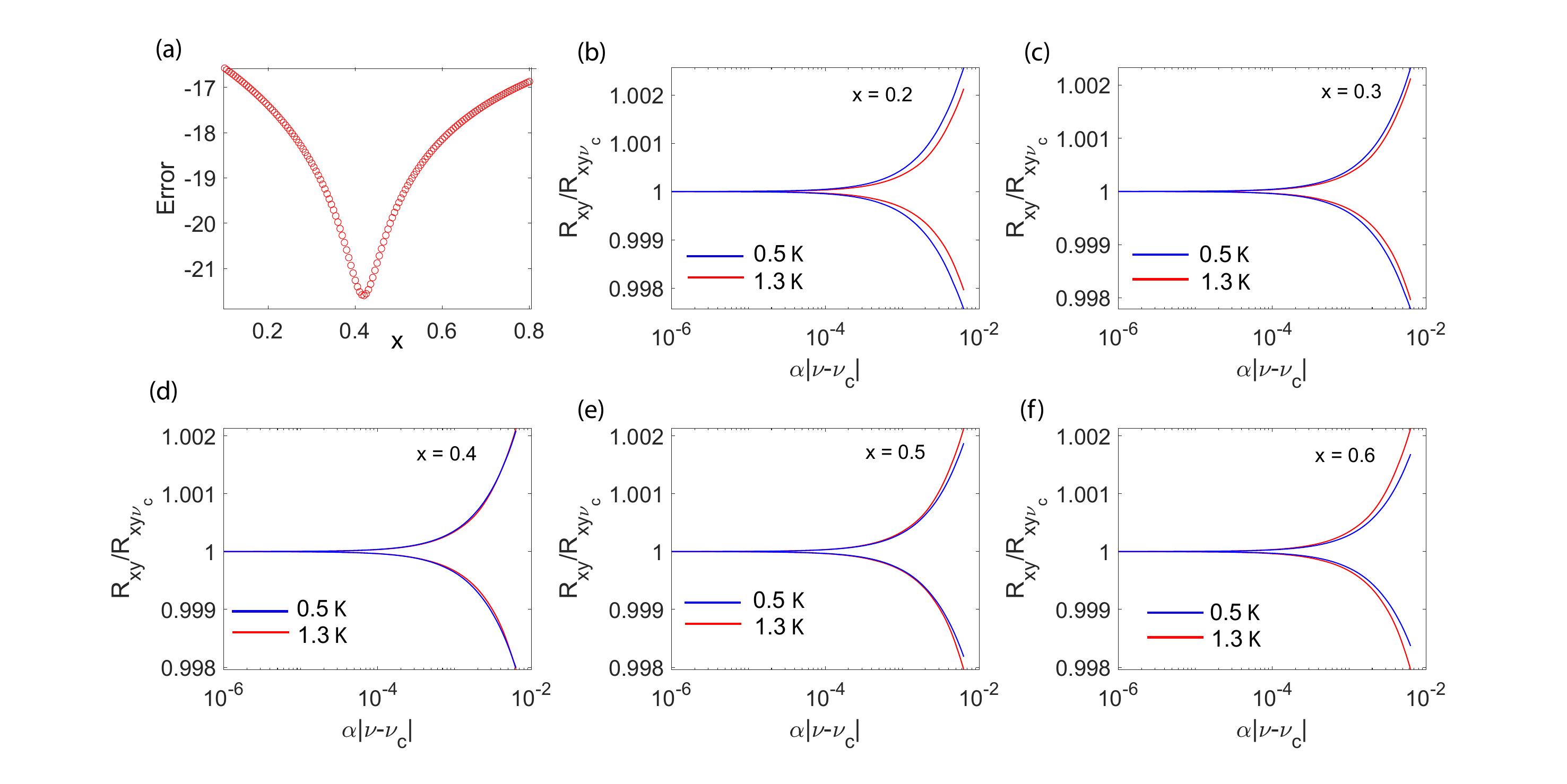}
		\caption{\textbf{Scaling analysis for transition between $\nu = 3 + 2/5$ and $\nu = 3 + 3/7$.} (a) Plot of the error in scaling versus $x$. (b-f) Scaling plot for different values of $x$ (the values of $x$ are marked inside the plot). }
		\label{fig:figS7_}
	\end{figure*}

	In this section, we describe the process followed to extract the value of $\kappa$. As discussed in the main manuscript, we use the following scaling equation~\cite{RevModPhys.67.357}:
	\begin{equation}
		R_{xy}(\nu,T)=R_{xy}(\nu_c) f[\alpha(\nu-\nu_c)]
		\label{eq:fss1}
	\end{equation}
	with
	\begin{equation}
		\alpha\propto T^{-x}
		\label{eq:fss2}
	\end{equation}
	~\ref{fig:figS6} (b-f) and \ref{fig:figS7_} (b-f) shows $R_{xy}/R_{xy}(\nu_c)$ at various temperatures as a function of $\alpha|\nu-\nu_c|$ for the $\nu = 2+2/3$ to $2+3/5$ and $3+ 2/5$ to $3+3/7$ transitions. The plots are for different values of $x$. The red line corresponds to $T = 1.3$~K, and the blue line corresponds to $T = 0.5$~K. For a perfect scaling, these two plots should collapse.  However, it is challenging to visually determine the value of $x$ that achieves the best scaling. To address this, the variance between the two plots is calculated as an `error' metric for the scaling accuracy. We identify $\kappa$ with the value of $x$  that minimizes this error. In this specific instance, the optimum value is $\kappa = 0.42$, as shown in ~\ref{fig:figS6}(a) and $\kappa = 0.40$ ( \ref{fig:figS7_} (a)) for $3+2/5$ to $3+3/7$.

	%\clearpage

	\section*{\textbf{Supplementary Note 8: Saturation of the derivative maxima at low temperature}}

\begin{figure*}[h]
		\includegraphics[width=0.5\columnwidth]{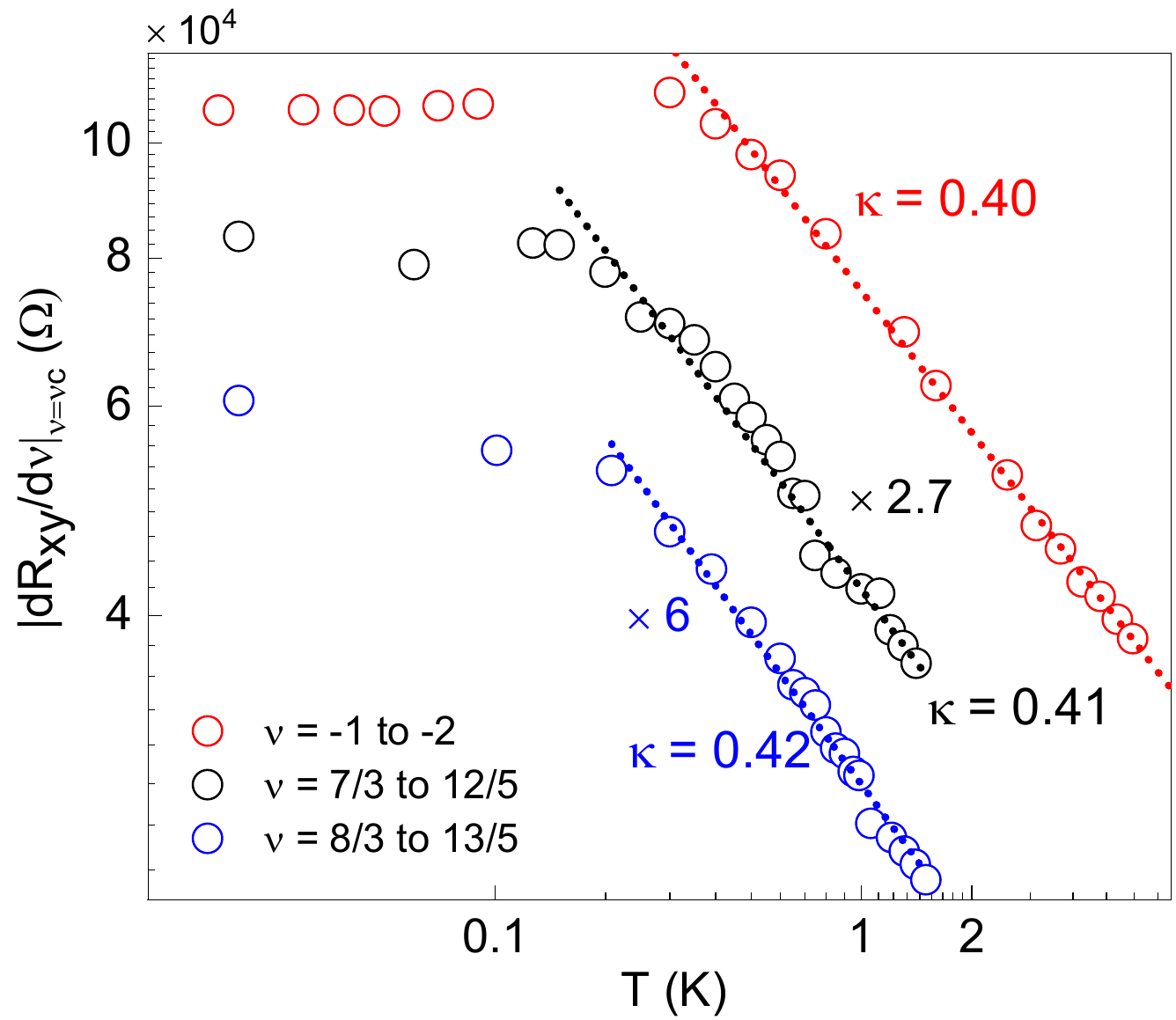}
		\caption{ Plot of the derivative of transverse resistance w.r.t to filling factor $\nu$ as a function of temperature for various plateau to plateau transition showing the saturation of derivative for low enough temperatures.}

		\label{fig:figS8_}
	\end{figure*}

	\ref{fig:figS8_} shows the plot of $(dR_{xy}/d\nu)$ at $\nu=\nu_c$ as a function of temperature for three representative plateau-to-plateau transitions in a high-mobility device. At low temperatures ($\mathrm{T>200~mK}$), the derivative maxima saturate. Similar saturation has been reported in previous studies on narrow devices~\cite{PhysRevLett.102.216801,PhysRevB.46.1596,PhysRevLett.67.883,PhysRevB.90.161408}. To understand this saturation, recall that the typical width of our devices is 3 $\mu$m. The phase coherence length $L_{\phi}$ exceeds the sample size ($L_\phi = T^{-p/2}$) at sufficiently low temperatures~\cite{Amin2018,PhysRevLett.70.3796,PhysRevB.46.1596}. As a result, at these low temperatures, the maxima of $dR_{xy}/d\nu$ is dictated solely by the device size, $L$. The condition $L<L_\phi$  leads to saturation of the derivative at low temperatures. We thus use only the data above $T = 200$~mK to extract the scaling exponent $\kappa$.

	\section*{\textbf{Supplementary Note 9: Scaling in  graphite-gated {h}BN encapsulated single layer  graphene }}

\begin{figure}[h]
		\includegraphics[width=0.9\columnwidth]{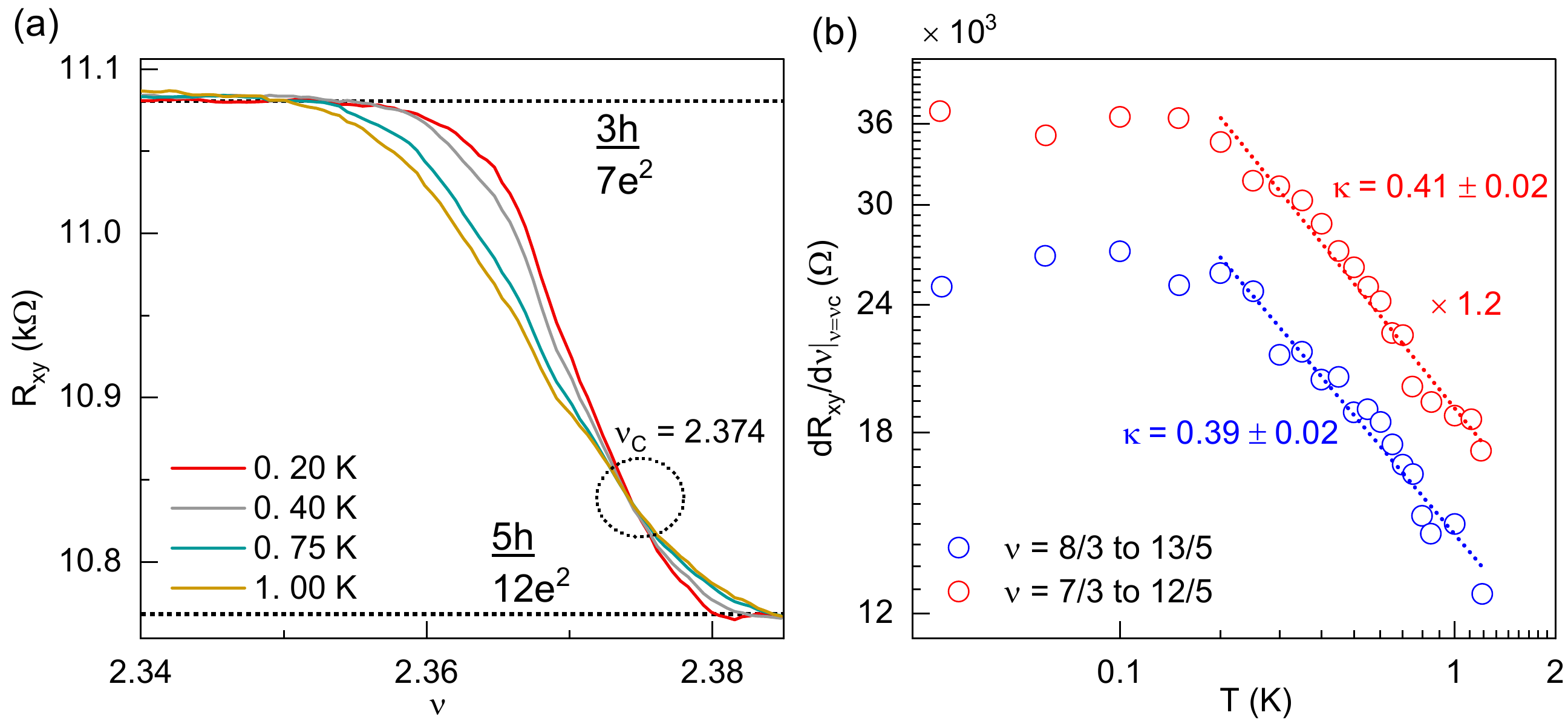}
		\caption{ (a) Plot of transverse resistance $R_{xy}$ as a function of filling factor for monolayer graphene. (b)  Plots of  d$R_{xy}/$d$\nu$ at the critical point $\nu = \nu_C$ versus temperature in double logarithmic  scale for FQHs plateau-to-plateau transitions. Here, error bar is calculated using least square fit to the data points. }
		\label{fig:figS9_}
	\end{figure}

	\ref{fig:figS9_} (a) shows the plot of transverse resistance $R_{xy}$ as a function of filling factor $\nu$ at different temperatures between the $2+1/3$ and $2+2/5$ plateau transition in graphite-gated hBN encapsulated single-layer high-mobility graphene.  \ref{fig:figS9_} (b) shows the plot of  $(dR_{xy}/d\nu)^{max}$ near the criticality as a function of temperature in a $\mathrm{log}-\mathrm{log}$ scale. From the slope of the data points, the obtained $\kappa$ is close to $0.41$. This further supports the observed universality in the FQH plateau-to-plateau transition.

	\section*{\textbf{Supplementary Note 10: Different configuration scaling analysis in high and low mobility device.}}

\begin{figure}[h]
		\includegraphics[width=0.8\columnwidth]{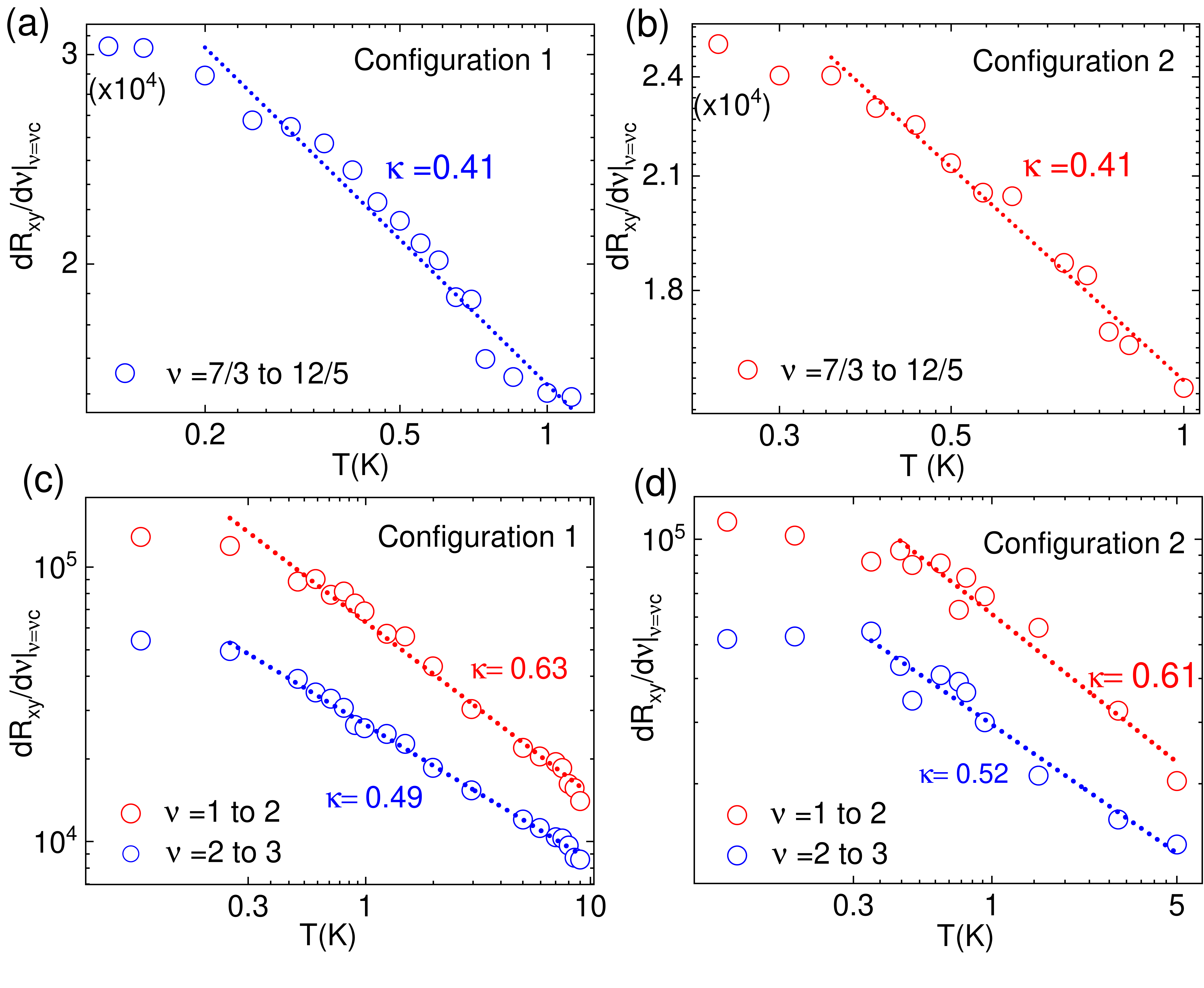}
		\caption{Double logarithmic plots of  d$R_{xy}/$d$\nu$ at the critical point $\nu = \nu_C$ versus $T$ (a), (b) in high-mobility graphene device (c), (d) in low mobility device measured in two different
			configurations. }

		\label{fig:figS10_}
	\end{figure}

	In \ref{fig:figS10_} shows the plot of evaluation $\kappa$ for different configurations measured both in high and low mobility devices. In case of high mobility device, $\kappa$ is 0.41 indicating uniformity across the device. In case of low mobility device, there is slight variation of $\kappa$ in different configuration. We conclude that critical exponents are uniform throughout the device for high mobility device and slight variation in low mobility devices.

	\section*{\textbf{Supplementary Note 11: Comparison of scaling analysis in the different low mobility devices.}}

\begin{figure*}[h]
		\includegraphics[width=0.8\columnwidth]{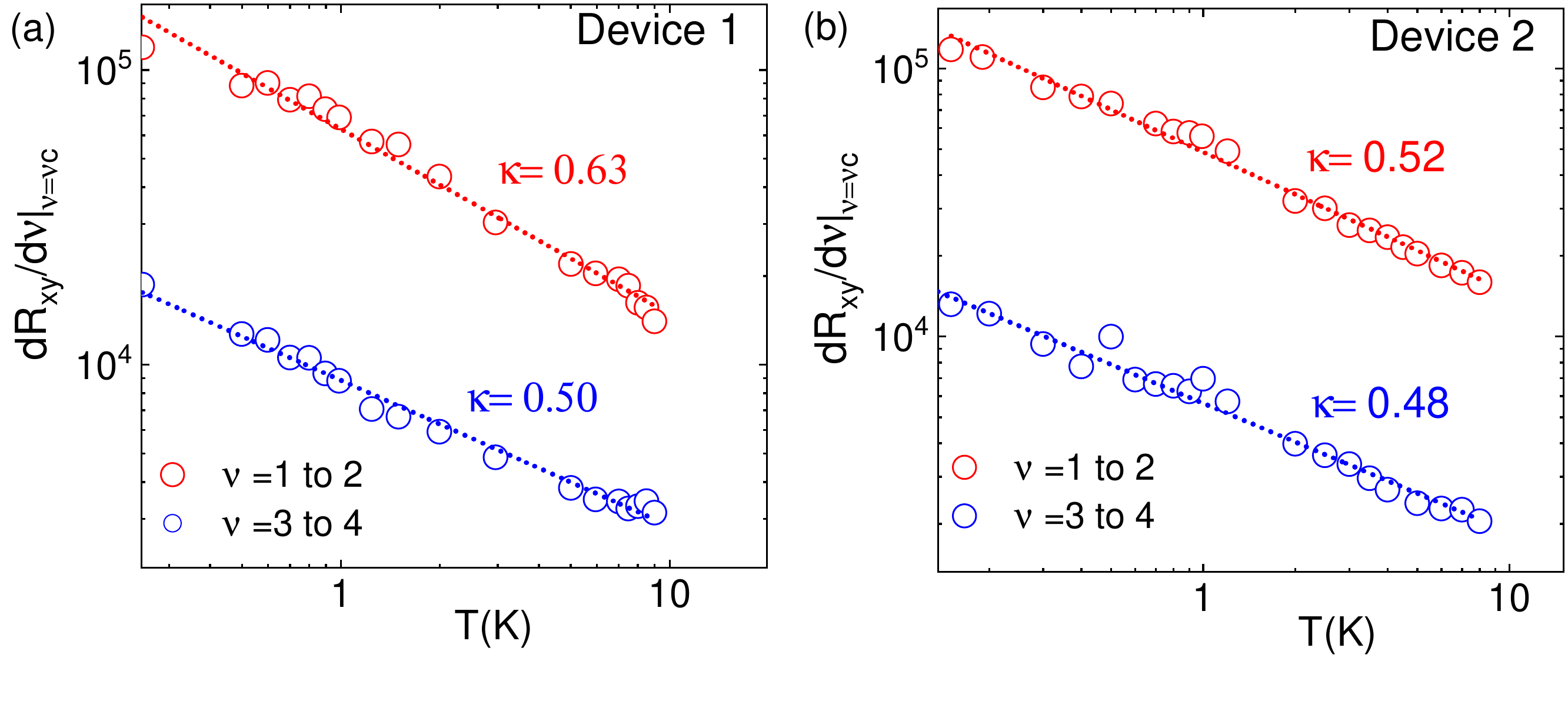}
		\caption{{Scaling exponents in two different low mobility graphene devices}. (a), (b) Double logarithmic plots of  d$R_{xy}/$d$\nu$ at the critical point $\nu = \nu_C$ versus temperature for two plateau-to-plateau transitions for two different low mobility devices.}

		\label{fig:figS11_}
	\end{figure*}

	In \ref{fig:figS11_} (a) and (b) shows the comparison plot of $(dR_{xy}/d\nu)^{max}$ as a function of temperature for two different low mobility hBN encapsulated samples. The evaluated $\kappa$ in both the samples deviates from the universal value.

	\section*{\textbf{Supplementary Note 12: Evaluation of $\kappa$ from three different analysis.}}

\begin{figure*}[h]
		\includegraphics[width=0.85\columnwidth]{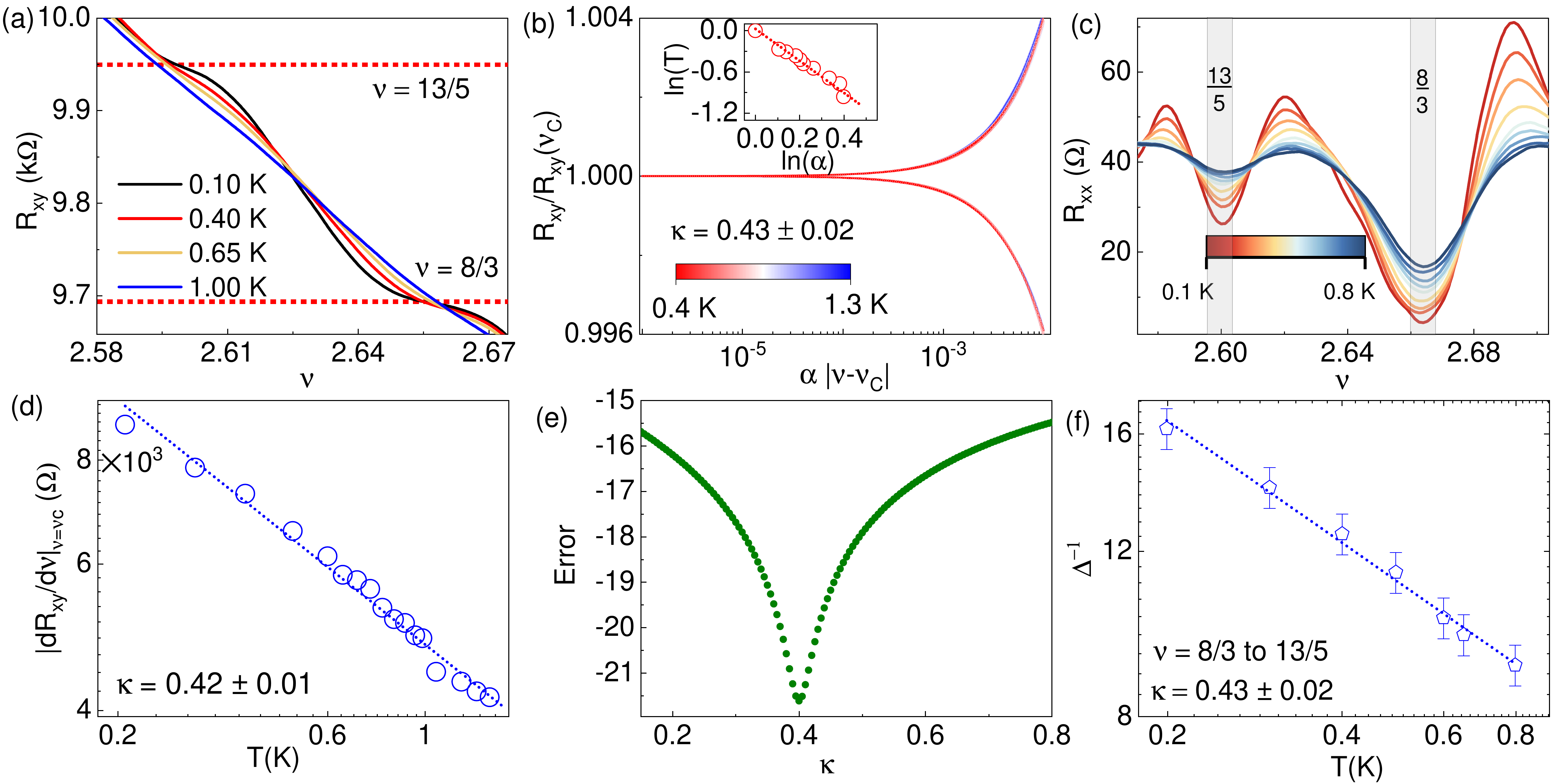}
		\caption{(a) Plot of $R_{xy}$ as a function of filling factor for the $\nu = 2+2/3$ to $2+3/5 $ transition. (b) Finite-size scaling analysis near the critical point $\nu_c$. (c) Plot of $R_{xx}$ as a function of filling factor $\nu$  for $2+2/3$ to $2+3/5$ plateau-to-plateau transition. (d) Evaluation of $\kappa$ from the derivative of $R_{xy}$ as a function of $\nu$. (e) Evaluation of $\kappa$ from finite-size scaling and error analysis between 0.4 K and 1.3 K. (f) Evaluation of $\kappa$ from full-width half maxima analysis.}

		\label{fig:figS12_}
	\end{figure*}

\begin{figure*}[h]
		\includegraphics[width=0.7\columnwidth]{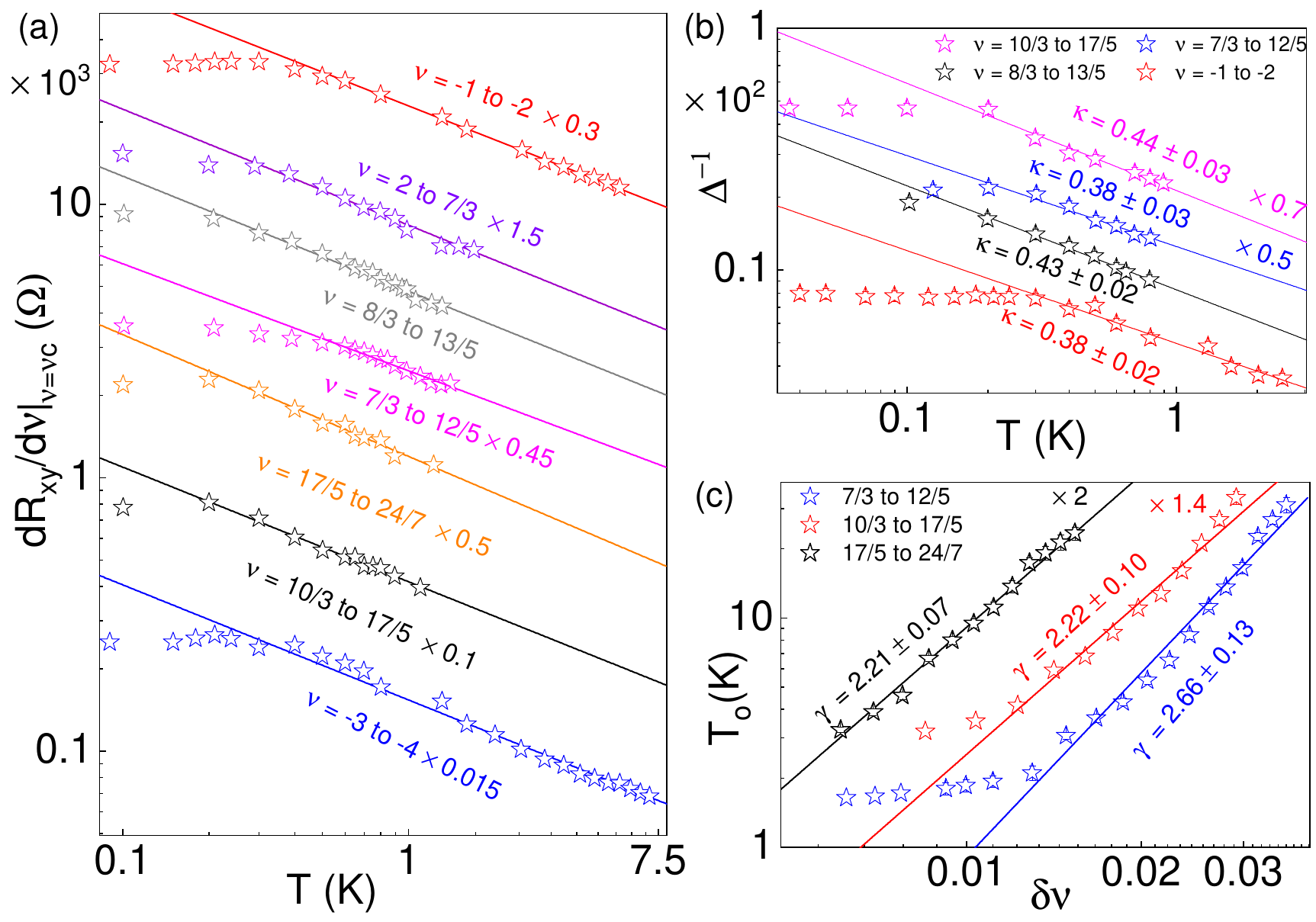}
		\caption{(a)  Plot of the derivative of transverse resistance with respect to $\nu$ as a function of temperature for different plateau-to-plateau transitions. The solid line is a linear fit to the data points. The slopes yield the value of $\kappa$. (b) Double logarithmic plot of the inverse of the half-width
			of longitudinal magnetoresistance $R_{xx}$ versus $T$ for several plateau to plateau transitions. The solid line is a linear fit to the data points. The slope yields the value of $\kappa$. (c) Plot of $T_0$ versus $\delta\nu$ for several PTs. The solid line is a linear fit to the data. The slope yields the value of $\gamma$. Here, error bar is calculated using least square fit to the data points.}

		\label{fig:figS13_}
	\end{figure*}

	\ref{fig:figS12_} shows the summary figure for the evaluation of $\kappa$ from three different analyses and \ref{fig:figS13_} shows the plot of evaluation of $\kappa$ and $\gamma$ for different plateau transitions.

	\section*{\textbf{Supplementary Note 13: Values of $\kappa$ from previous studies}}

	\begin{figure*}[h]

		\includegraphics[width=\columnwidth]{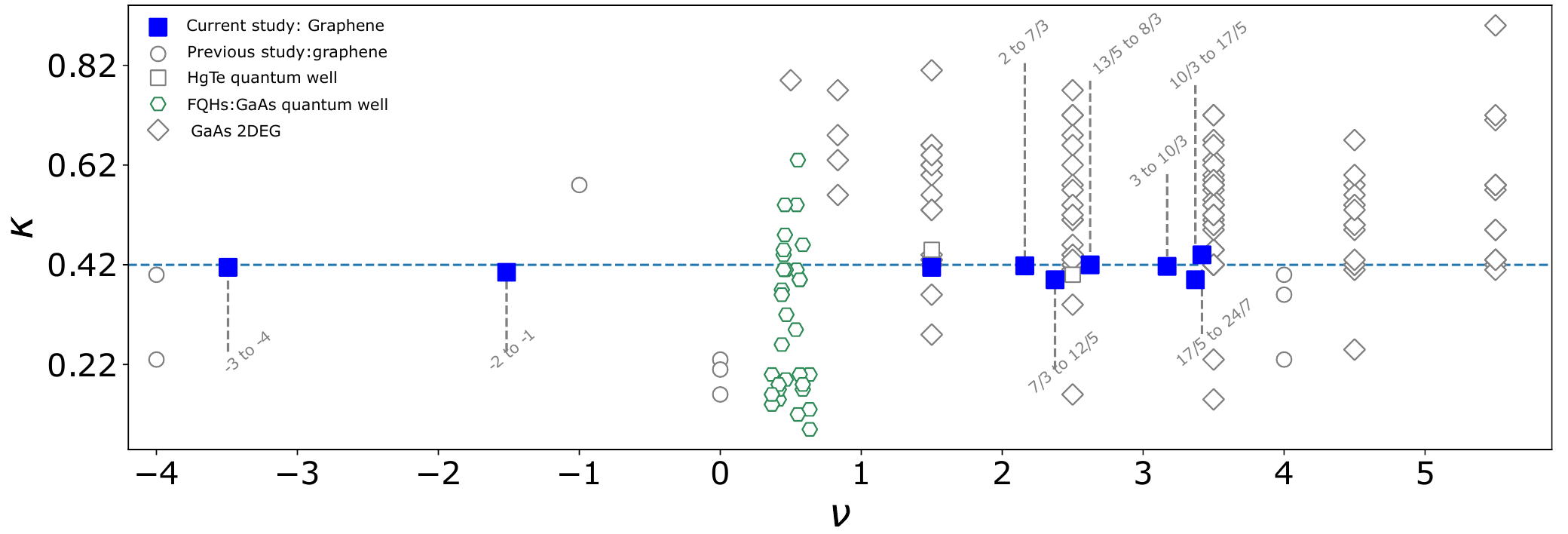}
		\caption{{A compilation of the values of $\kappa$ from previous studies~\cite{PhysRevLett.61.1294,  PhysRevB.43.6828,PhysRevLett.102.216801, PhysRevLett.61.1294, PhysRevB.80.241411, Shen2012, Pallecchi2013, Peters2014, PhysRevB.93.041421, Amado2012, PhysRevB.89.085422, Peters2014,PhysRevB.93.041421, Amado2012, PhysRevB.90.161408,DRESSELHAUS2021168676, ZIRNBAUER2019458}, represented by open symbols. The results of the current study are represented with filled squares and circles. Details of the data and the corresponding references are compiled in  ~\ref{tab:my_label} and ~\ref{tab:graphene}.}}

		\label{fig:FigS14}
	\end{figure*}
	\clearpage

	%\clearpage

	\begin{longtable}{| p{.10\textwidth} | p{.15\textwidth} | p{.420\textwidth} | p{.20\textwidth} |}

		\caption{{A compilation of the values of $\kappa$ obtained in 2D materials other than graphene by different groups.}\label{tab:my_label}  }\\

		\hline
		PPT	& $\kappa$	& Material &   Reference	       \\ \hline

		1$\rightarrow$2/3	&	0.77$\pm$0.02	&	 $\mathrm{Al_xGa_{1-x} As-Al_{0:33}Ga_{0:67}As}$	&	\cite{PhysRevB.43.6828}	\\ \hline
		1$\rightarrow$2/3	&	0.63$\pm$0.07	&	  $\mathrm{Al_xGa_{1-x} As-Al_{0:33}Ga_{0:67}As}$	&\cite{PhysRevB.43.6828}	\\ \hline
		1$\rightarrow$2/3	&	0.56$\pm$0.02	&	 $\mathrm{Al_xGa_{1-x} As-Al_{0:33}Ga_{0:67}As}$	&	\cite{PhysRevB.43.6828}	\\ \hline
		1$\rightarrow$2/3	&	0.68$\pm$0.05	&	  $\mathrm{Al_xGa_{1-x} As-Al_{0:33}Ga_{0:67}As}$	&	\cite{PhysRevB.43.6828}	\\ \hline
		1$\rightarrow$2	&	0.36$\pm$0.04	&	  $\mathrm{Al_xGa_{1-x} As-Al_{0:33}Ga_{0:67}As}$	&	\cite{PhysRevB.43.6828}	\\ \hline
		1$\rightarrow$2	&	0.56$\pm$0.05	&	 $\mathrm{Al_xGa_{1-x} As-Al_{0:33}Ga_{0:67}As}$	&	\cite{PhysRevB.43.6828}\\ \hline
		1$\rightarrow$2	&	0.81$\pm$0.04	&  $\mathrm{Al_xGa_{1-x} As-Al_{0:33}Ga_{0:67}As}$	&	\cite{PhysRevB.43.6828}	\\ \hline
		1$\rightarrow$2	&	0.44$\pm$0.02	&  $\mathrm{Al_xGa_{1-x} As-Al_{0:33}Ga_{0:67}As}$	&	\cite{PhysRevB.43.6828}	\\ \hline
		1$\rightarrow$2	&	0.53$\pm$0.07	& $\mathrm{Al_xGa_{1-x} As-Al_{0:33}Ga_{0:67}As}$	&	\cite{PhysRevB.43.6828}	\\ \hline
		1$\rightarrow$2	&	0.43$\pm$0.10	&  $\mathrm{Al_xGa_{1-x} As-Al_{0:33}Ga_{0:67}As}$	&	\cite{PhysRevB.43.6828}	\\ \hline
		1$\rightarrow$2	&	0.62$\pm$0.03	&  $\mathrm{Al_xGa_{1-x} As-Al_{0:33}Ga_{0:67}As}$&	\cite{PhysRevB.43.6828}	\\ \hline
		1$\rightarrow$2	&	0.28$\pm$0.06	&  $\mathrm{Al_xGa_{1-x} As-Al_{0:33}Ga_{0:67}As}$	&	\cite{PhysRevB.43.6828}	\\ \hline
		1$\rightarrow$2	&	0.53$\pm$0.06	& $\mathrm{Al_xGa_{(1-x)} As-Al_{0:33}Ga_{0:67}As}$	&	\cite{PhysRevB.43.6828}	\\ \hline
		1$\rightarrow$2	&	0.43$\pm$0.1	&  $\mathrm{Al_xGa_{1-x} As-Al_{0:33}Ga_{0:67}As}$	&	\cite{PhysRevB.43.6828}	\\ \hline
		2$\rightarrow$3	&	0.51$\pm$0.03	&	 $\mathrm{Al_xGa_{1-x} As-Al_{0:33}Ga_{0:67}As}$	&	\cite{PhysRevB.43.6828}	\\ \hline
		3$\rightarrow$4	&	0.51$\pm$0.03	&  $\mathrm{Al_xGa_{1-x} As-Al_{0:33}Ga_{0:67}As}$	&	\cite{PhysRevB.43.6828}\\ \hline
		3$\rightarrow$4	&	0.45$\pm$0.05	&	 $\mathrm{Al_xGa_{1-x} As-Al_{0:33}Ga_{0:67}As}$	&	\cite{PhysRevB.43.6828}	\\ \hline
		3$\rightarrow$4	&	0.45$\pm$0.05	&  $\mathrm{Al_xGa_{1-x} As-Al_{0:33}Ga_{0:67}As}$	&	\cite{PhysRevB.43.6828}	\\ \hline
		3$\rightarrow$4	&	0.52$\pm$0.03	&  $\mathrm{Al_xGa_{1-x} As-Al_{0:33}Ga_{0:67}As}$	&	\cite{PhysRevB.43.6828}	\\ \hline
		3$\rightarrow$4	&	0.63$\pm$0.03	&  $\mathrm{Al_xGa_{1-x} As-Al_{0:33}Ga_{0:67}As}$	&	\cite{PhysRevB.43.6828}	\\ \hline
		1$\rightarrow$2	&	0.42$\pm$0.04	&	$\mathrm{In_{0.53}Ga_{0.47}As/ InP}$	&	\cite{PhysRevLett.61.1294}\\ \hline
		2$\rightarrow$3, 3$\rightarrow$4	&	0.42$\pm$0.04	&	$\mathrm{In_{0.53}Ga_{0.47}As/ InP}$	&	\cite{PhysRevLett.61.1294}\\ \hline
		2$\rightarrow$3	&	0.72$\pm$0.05	&	GaAs/AlGaAs	&	\cite{PhysRevB.46.1596}	\\ \hline
		4$\rightarrow$5	&	0.15	&	Si-MOSFET	&	\cite{PhysRevB.46.1596}	\\ \hline
		3$\rightarrow$4	&	0.25	&	Si-MOSFET	&	\cite{PhysRevB.46.1596}	\\ \hline
		5$\rightarrow$6	&	0.15	&	Si-MOSFET	&	\cite{PhysRevB.46.1596}	\\ \hline
		2$\rightarrow$3,3$\rightarrow$4	&	0.90	&	Si-MOSFET	&	\cite{PhysRevB.46.1596}	\\ \hline
		1$\rightarrow$2,2$\rightarrow$3
		3$\rightarrow$4	&	0.62	&	Si-MOSFET	&	\cite{PhysRevB.46.1596}	\\ \hline
		6$\rightarrow$5	&	0.71	&	GaAs/AlGaAs	&	\cite{PhysRevB.78.233301}	\\ \hline
		7$\rightarrow$6	&	0.72	&	GaAs/AlGaAs	&	\cite{PhysRevB.78.233301}	\\ \hline
		6$\rightarrow$5	&	0.74	&	GaAs/AlGaAs	&	\cite{PhysRevB.78.233301}	\\ \hline
		7$\rightarrow$6	&	0.77	&	GaAs/AlGaAs	&	\cite{PhysRevB.78.233301}	\\ \hline
		8$\rightarrow$10	&	0.75$\pm$0.05	&	GaAs/AlGaAs	&	\cite{PhysRevB.78.233301}	\\ \hline
		1$\rightarrow$2	&	0.66$\pm$0.02	&	GaAs/AlGaAs	&	\cite{PhysRevLett.88.036802}	\\ \hline
		1$\rightarrow$2	&	0.6$\pm$0.02	&	GaAs/AlGaAs	&	\cite{PhysRevLett.88.036802}\\ \hline
		1$\rightarrow$2	&	0.62$\pm$0.03	&	GaAs/AlGaAs	&	\cite{PhysRevLett.88.036802}	\\ \hline
		6$\rightarrow$5	&	0.58	&	$\mathrm{\mathrm{Al_xGa_{1-x}As-Al_{0.33}Ga_{0.67}As ~(0\% Al)}}$&	\cite{li2005scaling}	\\ \hline
		5$\rightarrow$4	&	0.58	&	${\mathrm{Al_xGa_{1-x}As-Al_{0.33}Ga_{0.67}As ~(0\% Al)}}$&	\cite{li2005scaling}\\ \hline
		4$\rightarrow$3	&	0.57	&	$\mathrm{Al_xGa_{1-x}As-Al_{0.33}Ga_{0.67}As ~(0\% Al)}$	&	\cite{li2005scaling}	\\ \hline
		6$\rightarrow$5	&	0.57	&	$\mathrm{Al_xGa_{1-x}As-Al_{0.33}Ga_{0.67}As~ (0.21\% Al})$&\cite{li2005scaling}	\\ \hline
		5$\rightarrow$4	&	0.56	&	$\mathrm{Al_xGa_{1-x}As-Al_{0.33}Ga_{0.67}As~ (0.21\% Al)}$	&	\cite{li2005scaling}	\\ \hline
		4$\rightarrow$3	&	0.58	&	$\mathrm{Al_xGa_{1-x}As-Al_{0.33}Ga_{0.67}As ~ (0.21\% Al)}$	&	\cite{li2005scaling}	\\ \hline
		6$\rightarrow$5	&	0.49	&	$\mathrm{Al_xGa_{1-x}As-Al_{0.33}Ga_{0.67}As ~(0.33\% Al)}$	&\cite{li2005scaling}	\\ \hline
		5$\rightarrow$4	&	0.5	&	$\mathrm{Al_xGa_{1-x}As-Al_{0.33}Ga_{0.67}As~ (0.33\% Al)}$	&\cite{li2005scaling}	\\ \hline
		4$\rightarrow$3	&	0.49	&	$\mathrm{Al_xGa_{1-x}As-Al_{0.33}Ga_{0.67}As ~(0.33\% Al)}$	&\cite{li2005scaling}	\\ \hline
		6$\rightarrow$5	&	0.43 	&	$\mathrm{Al_xGa_{1-x}As-Al_{0.33}Ga_{0.67}As~ (0.85\% Al)}$	&\cite{li2005scaling}\\ \hline
		5$\rightarrow$4	&	0.42	&	$\mathrm{Al_xGa_{1-x}As-Al_{0.33}Ga_{0.67}As~ (0.85\% Al)}$ &\cite{li2005scaling}	\\ \hline
		4$\rightarrow$3	&	0.42	&	$\mathrm{Al_xGa_{1-x}As-Al_{0.33}Ga_{0.67}As ~(0.85\% Al)}$	&\cite{li2005scaling}	\\ \hline
		3$\rightarrow$2	&	0.41	&	$\mathrm{Al_xGa_{1-x}As-Al_{0.33}Ga_{0.67}As ~(0.85\% Al)}$	&	\cite{li2005scaling}	\\ \hline
		6$\rightarrow$5	&	0.42	&	$\mathrm{Al_xGa_{1-x}As-Al_{0.33}Ga_{0.67}As~ (0.85\% Al)}$	&	\cite{li2005scaling}	\\ \hline
		5$\rightarrow$4	&	0.41	&	$\mathrm{Al_xGa_{1-x}As-Al_{0.33}Ga_{0.67}As~ (0.85\% Al)}$	&\cite{li2005scaling}	\\ \hline
		4$\rightarrow$3	&	0.42	&	$\mathrm{Al_xGa_{1-x}As-Al_{0.33}Ga_{0.67}As~ (0.85\% Al)}$	&	\cite{li2005scaling} \\ \hline
		3$\rightarrow$2	&	0.42	&	$\mathrm{Al_xGa_{1-x}As-Al_{0.33}Ga_{0.67}As~ (0.85\% Al)}$	&\cite{li2005scaling}	\\ \hline
		6$\rightarrow$5	&	0.42	&	$\mathrm{Al_xGa_{1-x}As-Al_{0.33}Ga_{0.67}As~ (0.85\% Al)}$	&	\cite{li2005scaling}	\\ \hline
		5$\rightarrow$4	&	0.42	&	$\mathrm{Al_xGa_{1-x}As-Al_{0.33}Ga_{0.67}As~ (0.85\% Al)}$	&	\cite{li2005scaling}	\\ \hline
		4$\rightarrow$3	&	0.42	&	$\mathrm{Al_xGa_{1-x}As-Al_{0.33}Ga_{0.67}As ~(0.85\% Al)}$	&	\cite{li2005scaling}	\\ \hline
		3$\rightarrow$2	&	0.41	&	$\mathrm{Al_xGa_{1-x}As-Al_{0.33}Ga_{0.67}As ~(0.85\% Al)}$	&	\cite{li2005scaling}	\\ \hline
		6$\rightarrow$5	&	0.41	&	$\mathrm{Al_xGa_{1-x}As-Al_{0.33}Ga_{0.67}As ~(0.85\% Al)}$	&	\cite{li2005scaling}	\\ \hline
		5$\rightarrow$4	&	0.42	&	$\mathrm{Al_xGa_{1-x}As-Al_{0.33}Ga_{0.67}As ~(0.85\% Al)}$	&	\cite{li2005scaling}	\\ \hline
		4$\rightarrow$3	&	0.42	&	$\mathrm{Al_xGa_{1-x}As-Al_{0.33}Ga_{0.67}As ~(0.85\% Al)}$	&	\cite{li2005scaling}	\\ \hline
		3$\rightarrow$2	&	0.42	&	$\mathrm{Al_xGa_{1-x}As-Al_{0.33}Ga_{0.67}As~ (0.85\% Al)}$	&	\cite{li2005scaling}	\\ \hline
		6$\rightarrow$5	&	0.43	&	$\mathrm{Al_xGa_{1-x}As-Al_{0.33}Ga_{0.67}As ~(1.4\% Al)}$	&	\cite{li2005scaling}	\\ \hline
		5$\rightarrow$4	&	0.43	&	$\mathrm{Al_xGa_{1-x}As-Al_{0.33}Ga_{0.67}As~ (1.4\% Al)}$	&	\cite{li2005scaling}	\\ \hline
		4$\rightarrow$3	&	0.42	&	$\mathrm{Al_xGa_{1-x}As-Al_{0.33}Ga_{0.67}As ~(1.4\% Al)}$	&	\cite{li2005scaling}	\\ \hline
		3$\rightarrow$2	&	0.42	&	$\mathrm{Al_xGa_{1-x}As-Al_{0.33}Ga_{0.67}As~ (1.4\% Al)}$	&	\cite{li2005scaling}	\\ \hline
		\hline
		6$\rightarrow$5	&	0.49	&	$\mathrm{Al_xGa_{1-x}As-Al_{0.33}Ga_{0.67}As ~(1.9\% Al)}$	&	\cite{li2005scaling}	\\ \hline
		5$\rightarrow$4	&	0.49	&	$\mathrm{Al_xGa_{1-x}As-Al_{0.33}Ga_{0.67}As ~(1.9\% Al)}$	&	\cite{li2005scaling}	\\ \hline
		4$\rightarrow$3	&	0.5	&	$\mathrm{Al_xGa_{1-x}As-Al_{0.33}Ga_{0.67}As~ (1.9\% Al)}$	&	\cite{li2005scaling}	\\ \hline
		3$\rightarrow$2	&	0.51	&	$\mathrm{Al_xGa_{1-x}As-Al_{0.33}Ga_{0.67}As~ (1.9\% Al)}$	&	\cite{li2005scaling}	\\ \hline
		6$\rightarrow$5	&	0.58	&	$\mathrm{Al_xGa_{1-x}As-Al_{0.33}Ga_{0.67}As~ (2.6\% Al)}$	&\cite{li2005scaling}	\\ \hline
		5$\rightarrow$4	&	0.6	&	$\mathrm{Al_xGa_{1-x}As-Al_{0.33}Ga_{0.67}As~ (2.6\% Al)}$	&	\cite{li2005scaling}	\\ \hline
		4$\rightarrow$3	&	0.59	&	$\mathrm{Al_xGa_{1-x}As-Al_{0.33}Ga_{0.67}As ~(2.6\% Al)}$	&	\cite{li2005scaling}	\\ \hline
		3$\rightarrow$2	&	0.58	&	$\mathrm{Al_xGa_{1-x}As-Al_{0.33}Ga_{0.67}As~ (2.6\% Al)}$	&	\cite{li2005scaling}	\\ \hline
		4$\rightarrow$3	&	0.58	&	$\mathrm{Al_xGa_{1-x}As-Al_{0.33}Ga_{0.67}As~(4.1\% Al)}$	&	\cite{li2005scaling}	\\ \hline
		3$\rightarrow$2	&	0.57	&	$\mathrm{Al_xGa_{1-x}As-Al_{0.33}Ga_{0.67}As~(4.1\% Al)}$	&	\cite{li2005scaling}	\\ \hline
		4$\rightarrow$3	& 0.42$\pm$0.01	&	GaAs/AlGaAs 	&	\cite{Dodoo-Amoo_2014}	\\ \hline
		4$\rightarrow$3	&0.67$\pm$0.02	&	GaAs/AlGaAs	&	\cite{Dodoo-Amoo_2014}		\\ \hline
		4$\rightarrow$3	&0.55$\pm$0.04	&	GaAs/AlGaAs	&	\cite{Dodoo-Amoo_2014}		\\ \hline
		4$\rightarrow$3	&	0.54$\pm$0.02	&	GaAs/AlGaAs	&	\cite{Dodoo-Amoo_2014}		\\ \hline
		4$\rightarrow$3	&	0.23$\pm$0.02	&	GaAs/AlGaAs	&	\cite{Dodoo-Amoo_2014}		\\ \hline
		4$\rightarrow$3	&	0.66$\pm$0.03	&	GaAs/AlGaAs	&	\cite{Dodoo-Amoo_2014}		\\ \hline
		4$\rightarrow$3	&	0.60$\pm$0.02	&	GaAs/AlGaAs	&	\cite{Dodoo-Amoo_2014}		\\ \hline
		4$\rightarrow$3	&	 0.54$\pm$0.02	&	GaAs/AlGaAs	&	\cite{Dodoo-Amoo_2014}		\\ \hline
		3$\rightarrow$2	&	0.41$\pm$0.01	&	GaAs/AlGaAs&	\cite{Dodoo-Amoo_2014}		\\ \hline
		3$\rightarrow$2	&	0.44$\pm$0.02	&	GaAs/AlGaAs	&	\cite{Dodoo-Amoo_2014}		\\ \hline
		3$\rightarrow$2	&	0.46$\pm$0.02	&	GaAs/AlGaAs	&	\cite{Dodoo-Amoo_2014}		\\ \hline
		3$\rightarrow$2	&	0.34$\pm$0.01	&	GaAs/AlGaAs	&	\cite{Dodoo-Amoo_2014}		\\ \hline
		3$\rightarrow$2	&	0.44$\pm$0.02	&	GaAs/AlGaAs	&	\cite{Dodoo-Amoo_2014}		\\ \hline
		3$\rightarrow$2	&	0.42$\pm$0.03	&	GaAs/AlGaAs	&	\cite{Dodoo-Amoo_2014}		\\ \hline
		3$\rightarrow$2	&	0.43$\pm$0.03 &	GaAs/AlGaAs	&	\cite{Dodoo-Amoo_2014}	\\ \hline
		3$\rightarrow$2	&	0.16$\pm$0.02	&	GaAs/AlGaAs	&	\cite{Dodoo-Amoo_2014}		\\ \hline
		2/3$\rightarrow$3/5	&	0.09	&	GaAs quantum wells (50nm)	&	\cite{PhysRevLett.130.226503}	\\ \hline
		3/5$\rightarrow$4/7	&	0.46	&	GaAs quantum wells (50nm)	&	\cite{PhysRevLett.130.226503}	\\ \hline
		4/7$\rightarrow$5/9	&	0.39	&	GaAs quantum wells (50nm)	&	\cite{PhysRevLett.130.226503}	\\ \hline
		6/11$\rightarrow$5/9	&	0.41	&	GaAs quantum wells (50nm)	&	\cite{PhysRevLett.130.226503}	\\ \hline
		7/13$\rightarrow$8/15	&	0.29	&	GaAs quantum wells (50nm)	&	\cite{PhysRevLett.130.226503} \\ \hline
		7/15$\rightarrow$6/13	&	0.19	&	GaAs quantum wells (50nm)	&	\cite{PhysRevLett.130.226503}	\\ \hline
		6/13$\rightarrow$5/11	&	0.48	&	GaAs quantum wells (50nm)	&	\cite{PhysRevLett.130.226503}	\\ \hline
		5/11$\rightarrow$4/9	&	0.44	&	GaAs quantum wells (50nm)	&	\cite{PhysRevLett.130.226503}	\\ \hline
		4/9$\rightarrow$3/7	&	0.37	&	GaAs quantum wells (50nm)	&	\cite{PhysRevLett.130.226503}	\\ \hline
		3/7$\rightarrow$2/5	&	0.15	&	GaAs quantum wells (50nm)	&	\cite{PhysRevLett.130.226503}	\\ \hline
		2/5$\rightarrow$1/3	&	0.14	&	GaAs quantum wells (50nm)	&	\cite{PhysRevLett.130.226503}\\ \hline
		2/3$\rightarrow$3/5	&	0.20	&	GaAs quantum wells (30nm)	&	\cite{PhysRevLett.130.226503}	\\ \hline
		3/5$\rightarrow$4/7	&	0.17	&	GaAs quantum wells (30nm)	&	\cite{PhysRevLett.130.226503}	\\ \hline
		4/7$\rightarrow$5/9	&	0.20	&	GaAs quantum wells (30nm)	&	\cite{PhysRevLett.130.226503}	\\ \hline
		5/9$\rightarrow$6/11	&	0.63	&	GaAs quantum wells (30nm)	&	\cite{PhysRevLett.130.226503}	\\ \hline
		6/11$\rightarrow$7/13	&	0.54	&	GaAs quantum wells (30nm)	&	\cite{PhysRevLett.130.226503}	\\ \hline
		%7/13$\rightarrow$8/15	&	0.32	&	GaAs quantum wells (30nm)	&	\cite{PhysRevLett.130.226503} 	\\ \hline
		7/15$\rightarrow$8/17	&	0.32	&	GaAs quantum wells (30nm)	&	\cite{PhysRevLett.130.226503}	\\ \hline
		6/13$\rightarrow$7/15	&	0.41	&	GaAs quantum wells (30nm)	&	\cite{PhysRevLett.130.226503}	\\ \hline
		6/13$\rightarrow$5/11	&	0.54	&	GaAs quantum wells (30nm)	&	\cite{PhysRevLett.130.226503}	\\ \hline
		5/11$\rightarrow$4/9	&	0.41	&	GaAs quantum wells (30nm)	&	\cite{PhysRevLett.130.226503}	\\ \hline
		4/9$\rightarrow$3/7	&	0.26	&	GaAs quantum wells (30nm)	&	\cite{PhysRevLett.130.226503}\\ \hline
		3/7$\rightarrow$2/5	&	0.17	&	GaAs quantum wells (30nm)	&	\cite{PhysRevLett.130.226503}	\\ \hline
		2/5$\rightarrow$1/3	&	0.20	&	GaAs quantum wells (30nm)	&	\cite{PhysRevLett.130.226503}	\\ \hline
		2/3$\rightarrow$3/5	&	0.13	&	GaAs quantum wells (40nm)	&	\cite{PhysRevLett.130.226503}	\\ \hline
		3/5$\rightarrow$4/7	&	0.18	&	GaAs quantum wells (40nm)	&	\cite{PhysRevLett.130.226503}	\\ \hline
		4/7$\rightarrow$5/9	&	0.39	&	GaAs quantum wells (40nm)	&	\cite{PhysRevLett.130.226503}	\\ \hline
		5/9$\rightarrow$6/11	&	0.12	&	GaAs quantum wells (40nm)	&	\cite{PhysRevLett.130.226503}	\\ \hline
		5/11$\rightarrow$4/9	&	0.45	&	GaAs quantum wells (40nm)	&	\cite{PhysRevLett.130.226503}	\\ \hline
		4/9$\rightarrow$3/7	&	0.36	&	GaAs quantum wells (40nm)	&	\cite{PhysRevLett.130.226503}	\\ \hline
		3/7$\rightarrow$2/5	&	0.18	&	GaAs quantum wells (40nm)	&	\cite{PhysRevLett.130.226503}	\\ \hline
		2/5$\rightarrow$1/3	&	0.16	&	GaAs quantum wells (40nm)	&	\cite{PhysRevLett.130.226503}	\\ \hline
		2$\rightarrow$1	&	0.42	&	GaAs/AlGaAs	&	\cite{PhysRevB.45.3926}	\\ \hline

		3$\rightarrow$4	&	0.68$\pm$0.04	&	GaAs/AlGaAs	&	\cite{PhysRevLett.67.883}	\\ \hline

		4$\rightarrow$3	&	0.5 $\pm$0.03	&	GaAs/AlGaAs	&	\cite{YOO1994821}	\\ \hline
		5$\rightarrow$4	&	0.5$\pm$0.03	&	GaAs/AlGaAs	&	\cite{YOO1994821}	\\ \hline
		4$\rightarrow$3	&	0.62 $\pm$0.04	&	GaAs/AlGaAs	&	\cite{SKoch1995}	\\ \hline
		4$\rightarrow$3	&	0.59$\pm$0.04	&	GaAs/AlGaAs	&	\cite{SKoch1995}	\\ \hline
		2$\rightarrow$1	&	0.66$\pm$0.02	&	GaAs/AlGaAs	&	\cite{PhysRevLett.88.036802}	\\ \hline
		2$\rightarrow$1	&	0.60$\pm$0.0	&	GaAs/AlGaAs	&	\cite{PhysRevLett.88.036802}	\\ \hline
		2$\rightarrow$1	&	0.62$\pm$0.02	&	GaAs/AlGaAs&	\cite{PhysRevLett.88.036802}	\\ \hline
		2$\rightarrow$1	&	0.64 $\pm$0.09	&	GaAs/AlGaAs	&	\cite{HUANG2004232}	\\ \hline
		3$\rightarrow$4 &	0.66 - 0.77	&	GaAs/AlGaAs	&	\cite{TU2007108}	\\ \hline

		1$\rightarrow$0	&	0.79	&	GaAs/AlGaAs	&	\cite{doi:10.1143/JPSJ.76.094703}	\\ \hline
		3$\rightarrow$2	&	0.54	&	GaAs/AlGaAs	&	\cite{doi:10.1143/JPSJ.76.094703}	\\ \hline
		4$\rightarrow$3	&	0.42	&	GaAs/AlGaAs	&	\cite{PhysRevB.81.033305}	\\ \hline
		4$\rightarrow$3	&	0.58	&	GaAs/AlGaAs&	\cite{PhysRevB.81.033305}	\\ \hline
		3$\rightarrow$2	&	0.52$\pm$0.01	&	GaAs/AlGaAs	&	\cite{PhysRevB.93.075307}	\\ \hline
		4$\rightarrow$3	&	0.52$\pm$0.02	&	GaAs/AlGaAs	&	\cite{PhysRevB.93.075307}	\\ \hline
		5$\rightarrow$4	&	0.53$\pm$0.02	&	GaAs/AlGaAs	&	\cite{PhysRevB.93.075307}	\\ \hline
		1$\rightarrow$2	&	0.45$\pm$0.04	&	HgTe Quantum wells (5.9 nm)	&	\cite{PhysRevB.93.125308}	\\ \hline
		2$\rightarrow$3	&	0.40$\pm$0.02	&	HgTe Quantum wells (5.9 nm)	&	\cite{PhysRevB.93.125308}	\\ \hline

	\end{longtable}
	\clearpage

	\begin{longtable}{| p{.15\textwidth} | p{.15\textwidth} | p{.40\textwidth} | p{.20\textwidth} |}
		\caption{{Values of $\kappa$ obtained in graphene from previous studies. The results of our present study are also included.} \label{tab:graphene} } \\
		\hline
		PPT	&	               $\kappa$	&	Material &  References 	     \\ \hline
		2     $\rightarrow$  6	&	0.23$\pm$0.02	&	Graphene on \ch{SiO_2}	&	\cite{PhysRevB.86.085433}	\\ \hline
		-2    $\rightarrow$  -6	&	0.23$\pm$0.02	&	Graphene on \ch{SiO_2}	&	\cite{PhysRevB.86.085433}	\\ \hline
		-10  $\rightarrow$  -6	&	0.23$\pm$0.02	&	Graphene on \ch{SiO_2}	&	\cite{PhysRevB.86.085433}	\\ \hline
		10   $\rightarrow$   6	&	0.23$\pm$0.02	&	Graphene on \ch{SiO_2}	&	\cite{PhysRevB.86.085433}	\\ \hline
		-2  $\rightarrow$  2	&	0.23$\pm$0.02	&	Graphene on \ch{SiO_2}	&	\cite{PhysRevB.86.085433}	\\ \hline
		6 $\rightarrow$ 10	&	0.40$\pm$0.04	&	Graphene on \ch{SiO_2}	&	\cite{PhysRevB.80.241411} \\ \hline
		2  $\rightarrow$ 6	&	0.40$\pm$0.04	&	Graphene on \ch{SiO_2}	&	\cite{PhysRevB.80.241411} 	\\ \hline
		-2  $\rightarrow$ -6	&	0.40$\pm$0.03	&	Graphene on \ch{SiO_2}	&	\cite{PhysRevB.80.241411} 	\\ \hline
		-6  $\rightarrow$ -10	&	0.40$\pm$0.03	&	Graphene on \ch{SiO_2}	&	\cite{PhysRevB.80.241411} 	\\ \hline
		6  $\rightarrow$ 10	&	0.41$\pm$0.03	&	Graphene on \ch{SiO_2}	&	\cite{PhysRevB.80.241411}	\\
		\hline
		-2  $\rightarrow$ 2	&	0.16$\pm$0.05	&	Graphene on \ch{SiO_2} Corbino geometry	&	\cite{Peters2014}	\\ \hline
		-2  $\rightarrow$ 0	&	0.58 $\pm$ 0.03	&	Graphene on \ch{SiO_2} (hall bar)	& \cite{amado2010plateau}	 \\ \hline
		-2  $\rightarrow$ 2	&	0.21$\pm$0.01	&	Graphene (pnp junction)	&	\cite{PhysRevB.93.041421}	\\ \hline
		2  $\rightarrow$ 6	&	0.36$\pm$0.01	&	Graphene (pnp junction)	&	\cite{PhysRevB.93.041421}	\\ \hline
		6  $\rightarrow$ 10	&	0.35$\pm$0.01	&	Graphene (pnp junction)	&	\cite{PhysRevB.93.041421}	\\ \hline
		16  $\rightarrow$ 12	&	0.27$\pm$0.01	&	Encapsulated BLG	&	\cite{PhysRevB.90.161408}	\\ \hline
		12  $\rightarrow$ 8	&	0.32$\pm$0.01	&	Encapsulated BLG	&	\cite{PhysRevB.90.161408}	\\ \hline
		16  $\rightarrow$ 12	&	0.30$\pm$0.01	&	Encapsulated BLG	&	\cite{PhysRevB.90.161408}	\\ \hline
		12  $\rightarrow$ 8	&	0.32$\pm$0.01	&	Encapsulated BLG	&	\cite{PhysRevB.90.161408}	\\ \hline
		-8  $\rightarrow$ -4	&	0.30$\pm$0.02	&	Encapsulated BLG	&	\cite{PhysRevB.90.161408}	\\ \hline
		-8  $\rightarrow$ -4	&	0.29$\pm$0.02	&	Encapsulated BLG	&	\cite{PhysRevB.90.161408}	\\ \hline
		-16 $\rightarrow$-12	&	0.32$\pm$0.02	&	Encapsulated BLG	&	\cite{PhysRevB.90.161408}	\\
		\hline
		-4 $\rightarrow$-3	&	0.41$\pm$0.01	&	high mobility device	&	current study	\\  \hline
		-2 $\rightarrow$-1	&	0.40$\pm$0.01	&	high mobility device	&	current study	\\
		\hline
		2$\rightarrow$7/3	&	0.42$\pm$0.01	&	high mobility device	&	current study	\\  \hline
		7/3$\rightarrow$12/5	&	0.38$\pm$0.02	&  high mobility device	&	current study	\\  \hline
		10/3$\rightarrow$17/5	&	0.39$\pm$0.03	&  high mobility device	&	current study	\\  \hline
		13/5$\rightarrow$8/3 &	0.42$\pm$0.01	&	 high mobility device	&	current study	\\  \hline
		3$\rightarrow$10/3 &	0.42$\pm$0.01	&	 high mobility device	&	current study	\\  \hline
		17/5$\rightarrow$24/7 &	0.44$\pm$0.02	&	 high mobility device	&	current study	\\  \hline
		1$\rightarrow$2 &	0.41$\pm$0.01	&	 high mobility device	&	current study	\\  \hline
		1$\rightarrow$2 &	0.63$\pm$0.01	&	 low mobility device	&	current study	\\  \hline
		2$\rightarrow$3 &	0.49$\pm$0.01	&	 low mobility device	&	current study	\\  \hline
		3$\rightarrow$4 &	0.50$\pm$0.01	&	 low mobility device	&	current study	\\  \hline
		4$\rightarrow$5 &	0.44$\pm$0.01	&	low mobility device	&	current study	\\  \hline
		5$\rightarrow$6 &	0.50$\pm$0.01	&	low mobility device	&	current study	\\  \hline
	\end{longtable}

	\section*{\textbf{Supplementary Note 14: Landau levels in ABA trilayer graphene }}

\begin{figure*}[h]
		\includegraphics[width=0.5\columnwidth]{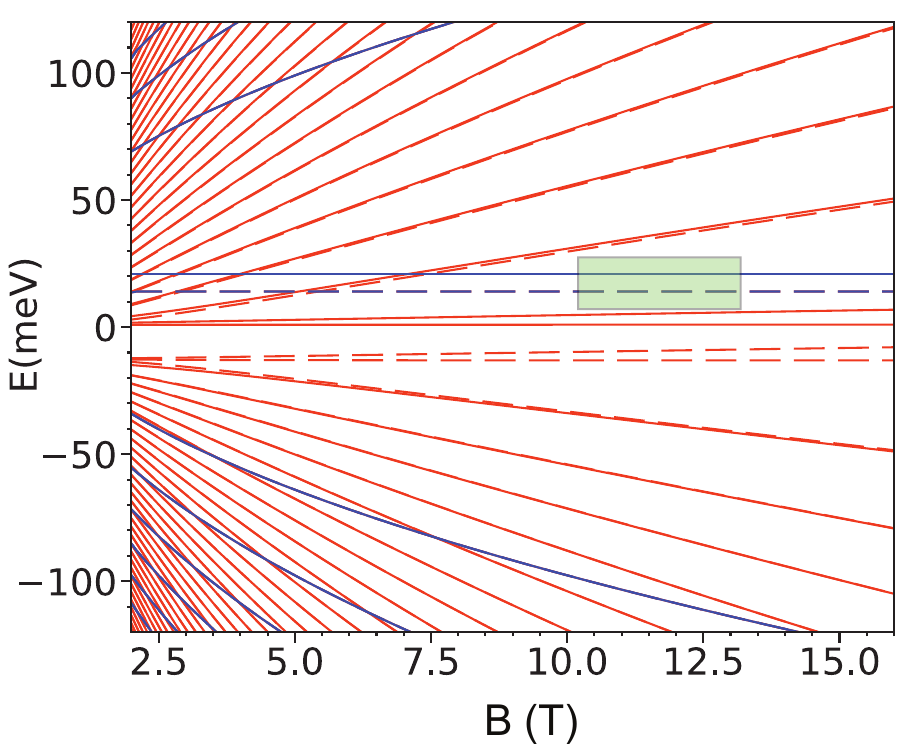}
		\caption{\textbf{Simulated Landau level plot of ABA trilayer graphene.} Red lines (Blue lines) mark the LLs originating from the bilayer-like band (monolayer-like band). All the analysis was conducted in the shaded region.}

		\label{fig:figS15_}
	\end{figure*}

	ABA trilayer graphene is a multiband system consisting of a monolayer-like band and a bilayer-like band (Fig.~1(c) of the main text) \cite{PhysRevB.87.115422,Taychatanapat2011,Datta2017}. In the presence of a magnetic field, Landau levels (LLs) originating from these two bands disperse differently ($\sqrt{B}$ for ML band and $B$ for BL band) \cite{doi:10.1126/sciadv.aax6550}. This difference leads to multiple crossings between the LLs of two bands. \ref{fig:figS15_} shows the simulated Landau level plot calculated at $\mathbf{D} =0$ V/nm. Here, red lines (blue lines) represent the LLs originating from the bilayer-like (monolayer-like) bands. To ensure that the localization physics is unaffected by landau level mixing effects, we performed all the experiments above 8  T where no such phase transitions are present. Under these conditions, the system remains in the $N_M = 0$ LL of the monolayer-like band (shown by the shaded region).

	The inherent mirror-symmetry of the system about the middle graphene layer precludes mixing between the monolayer-like and bilayer-like bands. The application of a finite-$\mathbf{D}$ field breaks this symmetry, allowing the mixing between the monolayer-like and bilayer-like bands.\cite{Winterer2022,PhysRevLett.121.056801,PhysRevB.87.115422,PhysRevLett.117.066601}

	\clearpage
	\def\bibsection{\section*{\refname}}
	\bibliography{arXiv}

\begin{thebibliography}{97}%
\makeatletter
\providecommand \@ifxundefined [1]{%
 \@ifx{#1\undefined}
}%
\providecommand \@ifnum [1]{%
 \ifnum #1\expandafter \@firstoftwo
 \else \expandafter \@secondoftwo
 \fi
}%
\providecommand \@ifx [1]{%
 \ifx #1\expandafter \@firstoftwo
 \else \expandafter \@secondoftwo
 \fi
}%
\providecommand \natexlab [1]{#1}%
\providecommand \enquote  [1]{``#1''}%
\providecommand \bibnamefont  [1]{#1}%
\providecommand \bibfnamefont [1]{#1}%
\providecommand \citenamefont [1]{#1}%
\providecommand \href@noop [0]{\@secondoftwo}%
\providecommand \href [0]{\begingroup \@sanitize@url \@href}%
\providecommand \@href[1]{\@@startlink{#1}\@@href}%
\providecommand \@@href[1]{\endgroup#1\@@endlink}%
\providecommand \@sanitize@url [0]{\catcode `\\12\catcode `\$12\catcode
  `\&12\catcode `\#12\catcode `\^12\catcode `\_12\catcode `\%12\relax}%
\providecommand \@@startlink[1]{}%
\providecommand \@@endlink[0]{}%
\providecommand \url  [0]{\begingroup\@sanitize@url \@url }%
\providecommand \@url [1]{\endgroup\@href {#1}{\urlprefix }}%
\providecommand \urlprefix  [0]{URL }%
\providecommand \Eprint [0]{\href }%
\providecommand \doibase [0]{https://doi.org/}%
\providecommand \selectlanguage [0]{\@gobble}%
\providecommand \bibinfo  [0]{\@secondoftwo}%
\providecommand \bibfield  [0]{\@secondoftwo}%
\providecommand \translation [1]{[#1]}%
\providecommand \BibitemOpen [0]{}%
\providecommand \bibitemStop [0]{}%
\providecommand \bibitemNoStop [0]{.\EOS\space}%
\providecommand \EOS [0]{\spacefactor3000\relax}%
\providecommand \BibitemShut  [1]{\csname bibitem#1\endcsname}%
\let\auto@bib@innerbib\@empty
%</preamble>
\bibitem [{\citenamefont {Klitzing}\ \emph {et~al.}(1980)\citenamefont
  {Klitzing}, \citenamefont {Dorda},\ and\ \citenamefont
  {Pepper}}]{PhysRevLett.45.494}%
  \BibitemOpen
  \bibfield  {author} {\bibinfo {author} {\bibfnamefont {K.~v.}\ \bibnamefont
  {Klitzing}}, \bibinfo {author} {\bibfnamefont {G.}~\bibnamefont {Dorda}},\
  and\ \bibinfo {author} {\bibfnamefont {M.}~\bibnamefont {Pepper}},\ }\href
  {https://doi.org/10.1103/PhysRevLett.45.494} {\bibfield  {journal} {\bibinfo
  {journal} {Phys. Rev. Lett.}\ }\textbf {\bibinfo {volume} {45}},\ \bibinfo
  {pages} {494} (\bibinfo {year} {1980})}\BibitemShut {NoStop}%
\bibitem [{\citenamefont {Huo}\ and\ \citenamefont
  {Bhatt}(1992)}]{PhysRevLett.68.1375}%
  \BibitemOpen
  \bibfield  {author} {\bibinfo {author} {\bibfnamefont {Y.}~\bibnamefont
  {Huo}}\ and\ \bibinfo {author} {\bibfnamefont {R.~N.}\ \bibnamefont
  {Bhatt}},\ }\href {https://doi.org/10.1103/PhysRevLett.68.1375} {\bibfield
  {journal} {\bibinfo  {journal} {Phys. Rev. Lett.}\ }\textbf {\bibinfo
  {volume} {68}},\ \bibinfo {pages} {1375} (\bibinfo {year}
  {1992})}\BibitemShut {NoStop}%
\bibitem [{\citenamefont {Laughlin}(1981)}]{PhysRevB.23.5632}%
  \BibitemOpen
  \bibfield  {author} {\bibinfo {author} {\bibfnamefont {R.~B.}\ \bibnamefont
  {Laughlin}},\ }\href {https://doi.org/10.1103/PhysRevB.23.5632} {\bibfield
  {journal} {\bibinfo  {journal} {Phys. Rev. B}\ }\textbf {\bibinfo {volume}
  {23}},\ \bibinfo {pages} {5632} (\bibinfo {year} {1981})}\BibitemShut
  {NoStop}%
\bibitem [{\citenamefont {Aoki}\ and\ \citenamefont
  {Ando}(1985)}]{PhysRevLett.54.831}%
  \BibitemOpen
  \bibfield  {author} {\bibinfo {author} {\bibfnamefont {H.}~\bibnamefont
  {Aoki}}\ and\ \bibinfo {author} {\bibfnamefont {T.}~\bibnamefont {Ando}},\
  }\href {https://doi.org/10.1103/PhysRevLett.54.831} {\bibfield  {journal}
  {\bibinfo  {journal} {Phys. Rev. Lett.}\ }\textbf {\bibinfo {volume} {54}},\
  \bibinfo {pages} {831} (\bibinfo {year} {1985})}\BibitemShut {NoStop}%
\bibitem [{\citenamefont {Chalker}\ and\ \citenamefont
  {Daniell}(1988)}]{PhysRevLett.61.593}%
  \BibitemOpen
  \bibfield  {author} {\bibinfo {author} {\bibfnamefont {J.~T.}\ \bibnamefont
  {Chalker}}\ and\ \bibinfo {author} {\bibfnamefont {G.~J.}\ \bibnamefont
  {Daniell}},\ }\href {https://doi.org/10.1103/PhysRevLett.61.593} {\bibfield
  {journal} {\bibinfo  {journal} {Phys. Rev. Lett.}\ }\textbf {\bibinfo
  {volume} {61}},\ \bibinfo {pages} {593} (\bibinfo {year} {1988})}\BibitemShut
  {NoStop}%
\bibitem [{\citenamefont {Chalker}\ and\ \citenamefont
  {Coddington}(1988)}]{Chalker1988}%
  \BibitemOpen
  \bibfield  {author} {\bibinfo {author} {\bibfnamefont {J.~T.}\ \bibnamefont
  {Chalker}}\ and\ \bibinfo {author} {\bibfnamefont {P.~D.}\ \bibnamefont
  {Coddington}},\ }\href {https://doi.org/10.1088/0022-3719/21/14/008}
  {\bibfield  {journal} {\bibinfo  {journal} {Journal of Physics C: Solid State
  Physics}\ }\textbf {\bibinfo {volume} {21}},\ \bibinfo {pages} {2665}
  (\bibinfo {year} {1988})}\BibitemShut {NoStop}%
\bibitem [{\citenamefont {Huckestein}\ and\ \citenamefont
  {Backhaus}(1999)}]{PhysRevLett.82.5100}%
  \BibitemOpen
  \bibfield  {author} {\bibinfo {author} {\bibfnamefont {B.}~\bibnamefont
  {Huckestein}}\ and\ \bibinfo {author} {\bibfnamefont {M.}~\bibnamefont
  {Backhaus}},\ }\href {https://doi.org/10.1103/PhysRevLett.82.5100} {\bibfield
   {journal} {\bibinfo  {journal} {Phys. Rev. Lett.}\ }\textbf {\bibinfo
  {volume} {82}},\ \bibinfo {pages} {5100} (\bibinfo {year}
  {1999})}\BibitemShut {NoStop}%
\bibitem [{\citenamefont {Halperin}\ and\ \citenamefont
  {Hohenberg}(1969)}]{PhysRev.177.952}%
  \BibitemOpen
  \bibfield  {author} {\bibinfo {author} {\bibfnamefont {B.~I.}\ \bibnamefont
  {Halperin}}\ and\ \bibinfo {author} {\bibfnamefont {P.~C.}\ \bibnamefont
  {Hohenberg}},\ }\href {https://doi.org/10.1103/PhysRev.177.952} {\bibfield
  {journal} {\bibinfo  {journal} {Phys. Rev.}\ }\textbf {\bibinfo {volume}
  {177}},\ \bibinfo {pages} {952} (\bibinfo {year} {1969})}\BibitemShut
  {NoStop}%
\bibitem [{\citenamefont {Li}\ \emph {et~al.}(2009)\citenamefont {Li},
  \citenamefont {Vicente}, \citenamefont {Xia}, \citenamefont {Pan},
  \citenamefont {Tsui}, \citenamefont {Pfeiffer},\ and\ \citenamefont
  {West}}]{PhysRevLett.102.216801}%
  \BibitemOpen
  \bibfield  {author} {\bibinfo {author} {\bibfnamefont {W.}~\bibnamefont
  {Li}}, \bibinfo {author} {\bibfnamefont {C.~L.}\ \bibnamefont {Vicente}},
  \bibinfo {author} {\bibfnamefont {J.~S.}\ \bibnamefont {Xia}}, \bibinfo
  {author} {\bibfnamefont {W.}~\bibnamefont {Pan}}, \bibinfo {author}
  {\bibfnamefont {D.~C.}\ \bibnamefont {Tsui}}, \bibinfo {author}
  {\bibfnamefont {L.~N.}\ \bibnamefont {Pfeiffer}},\ and\ \bibinfo {author}
  {\bibfnamefont {K.~W.}\ \bibnamefont {West}},\ }\href
  {https://doi.org/10.1103/PhysRevLett.102.216801} {\bibfield  {journal}
  {\bibinfo  {journal} {Phys. Rev. Lett.}\ }\textbf {\bibinfo {volume} {102}},\
  \bibinfo {pages} {216801} (\bibinfo {year} {2009})}\BibitemShut {NoStop}%
\bibitem [{\citenamefont {Sondhi}\ \emph {et~al.}(1997)\citenamefont {Sondhi},
  \citenamefont {Girvin}, \citenamefont {Carini},\ and\ \citenamefont
  {Shahar}}]{RevModPhys.69.315}%
  \BibitemOpen
  \bibfield  {author} {\bibinfo {author} {\bibfnamefont {S.~L.}\ \bibnamefont
  {Sondhi}}, \bibinfo {author} {\bibfnamefont {S.~M.}\ \bibnamefont {Girvin}},
  \bibinfo {author} {\bibfnamefont {J.~P.}\ \bibnamefont {Carini}},\ and\
  \bibinfo {author} {\bibfnamefont {D.}~\bibnamefont {Shahar}},\ }\href
  {https://doi.org/10.1103/RevModPhys.69.315} {\bibfield  {journal} {\bibinfo
  {journal} {Rev. Mod. Phys.}\ }\textbf {\bibinfo {volume} {69}},\ \bibinfo
  {pages} {315} (\bibinfo {year} {1997})}\BibitemShut {NoStop}%
\bibitem [{\citenamefont {Pruisken}(1988{\natexlab{a}})}]{PhysRevLett.61.1297}%
  \BibitemOpen
  \bibfield  {author} {\bibinfo {author} {\bibfnamefont {A.~M.~M.}\
  \bibnamefont {Pruisken}},\ }\href
  {https://doi.org/10.1103/PhysRevLett.61.1297} {\bibfield  {journal} {\bibinfo
   {journal} {Phys. Rev. Lett.}\ }\textbf {\bibinfo {volume} {61}},\ \bibinfo
  {pages} {1297} (\bibinfo {year} {1988}{\natexlab{a}})}\BibitemShut {NoStop}%
\bibitem [{\citenamefont {Dodoo-Amoo}\ \emph {et~al.}(2014)\citenamefont
  {Dodoo-Amoo}, \citenamefont {Saeed}, \citenamefont {Mistry}, \citenamefont
  {Khanna}, \citenamefont {Li}, \citenamefont {Linfield}, \citenamefont
  {Davies},\ and\ \citenamefont {Cunningham}}]{Dodoo-Amoo_2014}%
  \BibitemOpen
  \bibfield  {author} {\bibinfo {author} {\bibfnamefont {N.~A.}\ \bibnamefont
  {Dodoo-Amoo}}, \bibinfo {author} {\bibfnamefont {K.}~\bibnamefont {Saeed}},
  \bibinfo {author} {\bibfnamefont {D.}~\bibnamefont {Mistry}}, \bibinfo
  {author} {\bibfnamefont {S.~P.}\ \bibnamefont {Khanna}}, \bibinfo {author}
  {\bibfnamefont {L.}~\bibnamefont {Li}}, \bibinfo {author} {\bibfnamefont
  {E.~H.}\ \bibnamefont {Linfield}}, \bibinfo {author} {\bibfnamefont {A.~G.}\
  \bibnamefont {Davies}},\ and\ \bibinfo {author} {\bibfnamefont {J.~E.}\
  \bibnamefont {Cunningham}},\ }\href
  {https://doi.org/10.1088/0953-8984/26/47/475801} {\bibfield  {journal}
  {\bibinfo  {journal} {Journal of Physics: Condensed Matter}\ }\textbf
  {\bibinfo {volume} {26}},\ \bibinfo {pages} {475801} (\bibinfo {year}
  {2014})}\BibitemShut {NoStop}%
\bibitem [{\citenamefont {Huckestein}(1995)}]{RevModPhys.67.357}%
  \BibitemOpen
  \bibfield  {author} {\bibinfo {author} {\bibfnamefont {B.}~\bibnamefont
  {Huckestein}},\ }\href {https://doi.org/10.1103/RevModPhys.67.357} {\bibfield
   {journal} {\bibinfo  {journal} {Rev. Mod. Phys.}\ }\textbf {\bibinfo
  {volume} {67}},\ \bibinfo {pages} {357} (\bibinfo {year} {1995})}\BibitemShut
  {NoStop}%
\bibitem [{\citenamefont {Pu}\ \emph {et~al.}(2022)\citenamefont {Pu},
  \citenamefont {Sreejith},\ and\ \citenamefont
  {Jain}}]{PhysRevLett.128.116801}%
  \BibitemOpen
  \bibfield  {author} {\bibinfo {author} {\bibfnamefont {S.}~\bibnamefont
  {Pu}}, \bibinfo {author} {\bibfnamefont {G.~J.}\ \bibnamefont {Sreejith}},\
  and\ \bibinfo {author} {\bibfnamefont {J.~K.}\ \bibnamefont {Jain}},\ }\href
  {https://doi.org/10.1103/PhysRevLett.128.116801} {\bibfield  {journal}
  {\bibinfo  {journal} {Phys. Rev. Lett.}\ }\textbf {\bibinfo {volume} {128}},\
  \bibinfo {pages} {116801} (\bibinfo {year} {2022})}\BibitemShut {NoStop}%
\bibitem [{\citenamefont {Huckestein}\ and\ \citenamefont
  {Kramer}(1990)}]{PhysRevLett.64.1437}%
  \BibitemOpen
  \bibfield  {author} {\bibinfo {author} {\bibfnamefont {B.}~\bibnamefont
  {Huckestein}}\ and\ \bibinfo {author} {\bibfnamefont {B.}~\bibnamefont
  {Kramer}},\ }\href {https://doi.org/10.1103/PhysRevLett.64.1437} {\bibfield
  {journal} {\bibinfo  {journal} {Phys. Rev. Lett.}\ }\textbf {\bibinfo
  {volume} {64}},\ \bibinfo {pages} {1437} (\bibinfo {year}
  {1990})}\BibitemShut {NoStop}%
\bibitem [{\citenamefont {Tsui}\ \emph {et~al.}(1982)\citenamefont {Tsui},
  \citenamefont {Stormer},\ and\ \citenamefont
  {Gossard}}]{PhysRevLett.48.1559}%
  \BibitemOpen
  \bibfield  {author} {\bibinfo {author} {\bibfnamefont {D.~C.}\ \bibnamefont
  {Tsui}}, \bibinfo {author} {\bibfnamefont {H.~L.}\ \bibnamefont {Stormer}},\
  and\ \bibinfo {author} {\bibfnamefont {A.~C.}\ \bibnamefont {Gossard}},\
  }\href {https://doi.org/10.1103/PhysRevLett.48.1559} {\bibfield  {journal}
  {\bibinfo  {journal} {Phys. Rev. Lett.}\ }\textbf {\bibinfo {volume} {48}},\
  \bibinfo {pages} {1559} (\bibinfo {year} {1982})}\BibitemShut {NoStop}%
\bibitem [{\citenamefont {Hohenberg}\ and\ \citenamefont
  {Halperin}(1977)}]{RevModPhys.49.435}%
  \BibitemOpen
  \bibfield  {author} {\bibinfo {author} {\bibfnamefont {P.~C.}\ \bibnamefont
  {Hohenberg}}\ and\ \bibinfo {author} {\bibfnamefont {B.~I.}\ \bibnamefont
  {Halperin}},\ }\href {https://doi.org/10.1103/RevModPhys.49.435} {\bibfield
  {journal} {\bibinfo  {journal} {Rev. Mod. Phys.}\ }\textbf {\bibinfo {volume}
  {49}},\ \bibinfo {pages} {435} (\bibinfo {year} {1977})}\BibitemShut
  {NoStop}%
\bibitem [{\citenamefont {Thouless}\ \emph {et~al.}(1982)\citenamefont
  {Thouless}, \citenamefont {Kohmoto}, \citenamefont {Nightingale},\ and\
  \citenamefont {den Nijs}}]{TKNN}%
  \BibitemOpen
  \bibfield  {author} {\bibinfo {author} {\bibfnamefont {D.~J.}\ \bibnamefont
  {Thouless}}, \bibinfo {author} {\bibfnamefont {M.}~\bibnamefont {Kohmoto}},
  \bibinfo {author} {\bibfnamefont {M.~P.}\ \bibnamefont {Nightingale}},\ and\
  \bibinfo {author} {\bibfnamefont {M.}~\bibnamefont {den Nijs}},\ }\href
  {https://doi.org/10.1103/PhysRevLett.49.405} {\bibfield  {journal} {\bibinfo
  {journal} {Phys. Rev. Lett.}\ }\textbf {\bibinfo {volume} {49}},\ \bibinfo
  {pages} {405} (\bibinfo {year} {1982})}\BibitemShut {NoStop}%
\bibitem [{\citenamefont {Wei}\ \emph {et~al.}(1992{\natexlab{a}})\citenamefont
  {Wei}, \citenamefont {Lin}, \citenamefont {Tsui},\ and\ \citenamefont
  {Pruisken}}]{Wei1992}%
  \BibitemOpen
  \bibfield  {author} {\bibinfo {author} {\bibfnamefont {H.~P.}\ \bibnamefont
  {Wei}}, \bibinfo {author} {\bibfnamefont {S.~Y.}\ \bibnamefont {Lin}},
  \bibinfo {author} {\bibfnamefont {D.~C.}\ \bibnamefont {Tsui}},\ and\
  \bibinfo {author} {\bibfnamefont {A.~M.~M.}\ \bibnamefont {Pruisken}},\
  }\href {https://doi.org/10.1103/PhysRevB.45.3926} {\bibfield  {journal}
  {\bibinfo  {journal} {Phys. Rev. B}\ }\textbf {\bibinfo {volume} {45}},\
  \bibinfo {pages} {3926} (\bibinfo {year} {1992}{\natexlab{a}})}\BibitemShut
  {NoStop}%
\bibitem [{\citenamefont {Li}\ \emph {et~al.}(2005)\citenamefont {Li},
  \citenamefont {Cs\'athy}, \citenamefont {Tsui}, \citenamefont {Pfeiffer},\
  and\ \citenamefont {West}}]{li2005scaling}%
  \BibitemOpen
  \bibfield  {author} {\bibinfo {author} {\bibfnamefont {W.}~\bibnamefont
  {Li}}, \bibinfo {author} {\bibfnamefont {G.~A.}\ \bibnamefont {Cs\'athy}},
  \bibinfo {author} {\bibfnamefont {D.~C.}\ \bibnamefont {Tsui}}, \bibinfo
  {author} {\bibfnamefont {L.~N.}\ \bibnamefont {Pfeiffer}},\ and\ \bibinfo
  {author} {\bibfnamefont {K.~W.}\ \bibnamefont {West}},\ }\href
  {https://doi.org/10.1103/PhysRevLett.94.206807} {\bibfield  {journal}
  {\bibinfo  {journal} {Phys. Rev. Lett.}\ }\textbf {\bibinfo {volume} {94}},\
  \bibinfo {pages} {206807} (\bibinfo {year} {2005})}\BibitemShut {NoStop}%
\bibitem [{\citenamefont {Engel}\ \emph {et~al.}(1990)\citenamefont {Engel},
  \citenamefont {Wei}, \citenamefont {Tsui},\ and\ \citenamefont
  {Shayegan}}]{engel1990critical}%
  \BibitemOpen
  \bibfield  {author} {\bibinfo {author} {\bibfnamefont {L.}~\bibnamefont
  {Engel}}, \bibinfo {author} {\bibfnamefont {H.}~\bibnamefont {Wei}}, \bibinfo
  {author} {\bibfnamefont {D.}~\bibnamefont {Tsui}},\ and\ \bibinfo {author}
  {\bibfnamefont {M.}~\bibnamefont {Shayegan}},\ }\href
  {https://www.sciencedirect.com/science/article/pii/003960289090820X}
  {\bibfield  {journal} {\bibinfo  {journal} {Surface science}\ }\textbf
  {\bibinfo {volume} {229}},\ \bibinfo {pages} {13} (\bibinfo {year}
  {1990})}\BibitemShut {NoStop}%
\bibitem [{\citenamefont {Machida}\ \emph {et~al.}(2001)\citenamefont
  {Machida}, \citenamefont {Ishizuka}, \citenamefont {Komiyama}, \citenamefont
  {Muraki},\ and\ \citenamefont {Hirayama}}]{MACHIDA2001182}%
  \BibitemOpen
  \bibfield  {author} {\bibinfo {author} {\bibfnamefont {T.}~\bibnamefont
  {Machida}}, \bibinfo {author} {\bibfnamefont {S.}~\bibnamefont {Ishizuka}},
  \bibinfo {author} {\bibfnamefont {S.}~\bibnamefont {Komiyama}}, \bibinfo
  {author} {\bibfnamefont {K.}~\bibnamefont {Muraki}},\ and\ \bibinfo {author}
  {\bibfnamefont {Y.}~\bibnamefont {Hirayama}},\ }\href
  {https://doi.org/https://doi.org/10.1016/S0921-4526(01)00297-6} {\bibfield
  {journal} {\bibinfo  {journal} {Physica B: Condensed Matter}\ }\textbf
  {\bibinfo {volume} {298}},\ \bibinfo {pages} {182} (\bibinfo {year}
  {2001})},\ \bibinfo {note} {international Conference on High Magnetic Fields
  in Semiconductors}\BibitemShut {NoStop}%
\bibitem [{\citenamefont {Madathil}\ \emph {et~al.}(2023)\citenamefont
  {Madathil}, \citenamefont {Villegas~Rosales}, \citenamefont {Tai},
  \citenamefont {Chung}, \citenamefont {Pfeiffer}, \citenamefont {West},
  \citenamefont {Baldwin},\ and\ \citenamefont
  {Shayegan}}]{PhysRevLett.130.226503}%
  \BibitemOpen
  \bibfield  {author} {\bibinfo {author} {\bibfnamefont {P.~T.}\ \bibnamefont
  {Madathil}}, \bibinfo {author} {\bibfnamefont {K.~A.}\ \bibnamefont
  {Villegas~Rosales}}, \bibinfo {author} {\bibfnamefont {C.~T.}\ \bibnamefont
  {Tai}}, \bibinfo {author} {\bibfnamefont {Y.~J.}\ \bibnamefont {Chung}},
  \bibinfo {author} {\bibfnamefont {L.~N.}\ \bibnamefont {Pfeiffer}}, \bibinfo
  {author} {\bibfnamefont {K.~W.}\ \bibnamefont {West}}, \bibinfo {author}
  {\bibfnamefont {K.~W.}\ \bibnamefont {Baldwin}},\ and\ \bibinfo {author}
  {\bibfnamefont {M.}~\bibnamefont {Shayegan}},\ }\href
  {https://doi.org/10.1103/PhysRevLett.130.226503} {\bibfield  {journal}
  {\bibinfo  {journal} {Phys. Rev. Lett.}\ }\textbf {\bibinfo {volume} {130}},\
  \bibinfo {pages} {226503} (\bibinfo {year} {2023})}\BibitemShut {NoStop}%
\bibitem [{\citenamefont {Kumar}\ \emph {et~al.}(2022)\citenamefont {Kumar},
  \citenamefont {Nosov},\ and\ \citenamefont
  {Raghu}}]{PhysRevResearch.4.033146}%
  \BibitemOpen
  \bibfield  {author} {\bibinfo {author} {\bibfnamefont {P.}~\bibnamefont
  {Kumar}}, \bibinfo {author} {\bibfnamefont {P.~A.}\ \bibnamefont {Nosov}},\
  and\ \bibinfo {author} {\bibfnamefont {S.}~\bibnamefont {Raghu}},\ }\href
  {https://doi.org/10.1103/PhysRevResearch.4.033146} {\bibfield  {journal}
  {\bibinfo  {journal} {Phys. Rev. Res.}\ }\textbf {\bibinfo {volume} {4}},\
  \bibinfo {pages} {033146} (\bibinfo {year} {2022})}\BibitemShut {NoStop}%
\bibitem [{\citenamefont {Pan}\ \emph {et~al.}(2020)\citenamefont {Pan},
  \citenamefont {Kang}, \citenamefont {Lilly}, \citenamefont {Reno},
  \citenamefont {Baldwin}, \citenamefont {West}, \citenamefont {Pfeiffer},\
  and\ \citenamefont {Tsui}}]{PhysRevLett.124.156801}%
  \BibitemOpen
  \bibfield  {author} {\bibinfo {author} {\bibfnamefont {W.}~\bibnamefont
  {Pan}}, \bibinfo {author} {\bibfnamefont {W.}~\bibnamefont {Kang}}, \bibinfo
  {author} {\bibfnamefont {M.~P.}\ \bibnamefont {Lilly}}, \bibinfo {author}
  {\bibfnamefont {J.~L.}\ \bibnamefont {Reno}}, \bibinfo {author}
  {\bibfnamefont {K.~W.}\ \bibnamefont {Baldwin}}, \bibinfo {author}
  {\bibfnamefont {K.~W.}\ \bibnamefont {West}}, \bibinfo {author}
  {\bibfnamefont {L.~N.}\ \bibnamefont {Pfeiffer}},\ and\ \bibinfo {author}
  {\bibfnamefont {D.~C.}\ \bibnamefont {Tsui}},\ }\href
  {https://doi.org/10.1103/PhysRevLett.124.156801} {\bibfield  {journal}
  {\bibinfo  {journal} {Phys. Rev. Lett.}\ }\textbf {\bibinfo {volume} {124}},\
  \bibinfo {pages} {156801} (\bibinfo {year} {2020})}\BibitemShut {NoStop}%
\bibitem [{\citenamefont {Sarkar}\ \emph {et~al.}(2015)\citenamefont {Sarkar},
  \citenamefont {Amin}, \citenamefont {Modak}, \citenamefont {Singh},
  \citenamefont {Mukerjee},\ and\ \citenamefont {Bid}}]{Sarkar2015}%
  \BibitemOpen
  \bibfield  {author} {\bibinfo {author} {\bibfnamefont {S.}~\bibnamefont
  {Sarkar}}, \bibinfo {author} {\bibfnamefont {K.~R.}\ \bibnamefont {Amin}},
  \bibinfo {author} {\bibfnamefont {R.}~\bibnamefont {Modak}}, \bibinfo
  {author} {\bibfnamefont {A.}~\bibnamefont {Singh}}, \bibinfo {author}
  {\bibfnamefont {S.}~\bibnamefont {Mukerjee}},\ and\ \bibinfo {author}
  {\bibfnamefont {A.}~\bibnamefont {Bid}},\ }\href
  {http://dx.doi.org/10.1038/srep16772} {\bibfield  {journal} {\bibinfo
  {journal} {Scientific Reports}\ }\textbf {\bibinfo {volume} {5}},\ \bibinfo
  {pages} {16772} (\bibinfo {year} {2015})}\BibitemShut {NoStop}%
\bibitem [{\citenamefont {Rhodes}\ \emph {et~al.}(2019)\citenamefont {Rhodes},
  \citenamefont {Chae}, \citenamefont {Ribeiro-Palau},\ and\ \citenamefont
  {Hone}}]{Rhodes2019}%
  \BibitemOpen
  \bibfield  {author} {\bibinfo {author} {\bibfnamefont {D.}~\bibnamefont
  {Rhodes}}, \bibinfo {author} {\bibfnamefont {S.~H.}\ \bibnamefont {Chae}},
  \bibinfo {author} {\bibfnamefont {R.}~\bibnamefont {Ribeiro-Palau}},\ and\
  \bibinfo {author} {\bibfnamefont {J.}~\bibnamefont {Hone}},\ }\href
  {https://doi.org/10.1038/s41563-019-0366-8} {\bibfield  {journal} {\bibinfo
  {journal} {Nature Materials}\ }\textbf {\bibinfo {volume} {18}},\ \bibinfo
  {pages} {541} (\bibinfo {year} {2019})}\BibitemShut {NoStop}%
\bibitem [{\citenamefont {Pizzocchero}\ \emph {et~al.}(2016)\citenamefont
  {Pizzocchero}, \citenamefont {Gammelgaard}, \citenamefont {Jessen},
  \citenamefont {Caridad}, \citenamefont {Wang}, \citenamefont {Hone},
  \citenamefont {B{\o}ggild},\ and\ \citenamefont
  {Booth}}]{pizzocchero2016hot}%
  \BibitemOpen
  \bibfield  {author} {\bibinfo {author} {\bibfnamefont {F.}~\bibnamefont
  {Pizzocchero}}, \bibinfo {author} {\bibfnamefont {L.}~\bibnamefont
  {Gammelgaard}}, \bibinfo {author} {\bibfnamefont {B.~S.}\ \bibnamefont
  {Jessen}}, \bibinfo {author} {\bibfnamefont {J.~M.}\ \bibnamefont {Caridad}},
  \bibinfo {author} {\bibfnamefont {L.}~\bibnamefont {Wang}}, \bibinfo {author}
  {\bibfnamefont {J.}~\bibnamefont {Hone}}, \bibinfo {author} {\bibfnamefont
  {P.}~\bibnamefont {B{\o}ggild}},\ and\ \bibinfo {author} {\bibfnamefont
  {T.~J.}\ \bibnamefont {Booth}},\ }\href@noop {} {\bibfield  {journal}
  {\bibinfo  {journal} {Nature communications}\ }\textbf {\bibinfo {volume}
  {7}},\ \bibinfo {pages} {11894} (\bibinfo {year} {2016})}\BibitemShut
  {NoStop}%
\bibitem [{\citenamefont {Koshino}\ and\ \citenamefont
  {McCann}(2011)}]{koshino2011landau}%
  \BibitemOpen
  \bibfield  {author} {\bibinfo {author} {\bibfnamefont {M.}~\bibnamefont
  {Koshino}}\ and\ \bibinfo {author} {\bibfnamefont {E.}~\bibnamefont
  {McCann}},\ }\href {https://link.aps.org/doi/10.1103/PhysRevB.83.165443}
  {\bibfield  {journal} {\bibinfo  {journal} {Physical Review B}\ }\textbf
  {\bibinfo {volume} {83}},\ \bibinfo {pages} {165443} (\bibinfo {year}
  {2011})}\BibitemShut {NoStop}%
\bibitem [{\citenamefont {Papi\ifmmode~\acute{c}\else \'{c}\fi{}}\ \emph
  {et~al.}(2011)\citenamefont {Papi\ifmmode~\acute{c}\else \'{c}\fi{}},
  \citenamefont {Abanin}, \citenamefont {Barlas},\ and\ \citenamefont
  {Bhatt}}]{PhysRevB.84.241306}%
  \BibitemOpen
  \bibfield  {author} {\bibinfo {author} {\bibfnamefont {Z.}~\bibnamefont
  {Papi\ifmmode~\acute{c}\else \'{c}\fi{}}}, \bibinfo {author} {\bibfnamefont
  {D.~A.}\ \bibnamefont {Abanin}}, \bibinfo {author} {\bibfnamefont
  {Y.}~\bibnamefont {Barlas}},\ and\ \bibinfo {author} {\bibfnamefont {R.~N.}\
  \bibnamefont {Bhatt}},\ }\href {https://doi.org/10.1103/PhysRevB.84.241306}
  {\bibfield  {journal} {\bibinfo  {journal} {Phys. Rev. B}\ }\textbf {\bibinfo
  {volume} {84}},\ \bibinfo {pages} {241306} (\bibinfo {year}
  {2011})}\BibitemShut {NoStop}%
\bibitem [{\citenamefont {Zhu}\ \emph {et~al.}(2020)\citenamefont {Zhu},
  \citenamefont {Sheng},\ and\ \citenamefont
  {Sodemann}}]{PhysRevLett.124.097604}%
  \BibitemOpen
  \bibfield  {author} {\bibinfo {author} {\bibfnamefont {Z.}~\bibnamefont
  {Zhu}}, \bibinfo {author} {\bibfnamefont {D.~N.}\ \bibnamefont {Sheng}},\
  and\ \bibinfo {author} {\bibfnamefont {I.}~\bibnamefont {Sodemann}},\ }\href
  {https://doi.org/10.1103/PhysRevLett.124.097604} {\bibfield  {journal}
  {\bibinfo  {journal} {Phys. Rev. Lett.}\ }\textbf {\bibinfo {volume} {124}},\
  \bibinfo {pages} {097604} (\bibinfo {year} {2020})}\BibitemShut {NoStop}%
\bibitem [{\citenamefont {Sodemann}\ and\ \citenamefont
  {MacDonald}(2013)}]{PhysRevB.87.245425}%
  \BibitemOpen
  \bibfield  {author} {\bibinfo {author} {\bibfnamefont {I.}~\bibnamefont
  {Sodemann}}\ and\ \bibinfo {author} {\bibfnamefont {A.~H.}\ \bibnamefont
  {MacDonald}},\ }\href {https://doi.org/10.1103/PhysRevB.87.245425} {\bibfield
   {journal} {\bibinfo  {journal} {Phys. Rev. B}\ }\textbf {\bibinfo {volume}
  {87}},\ \bibinfo {pages} {245425} (\bibinfo {year} {2013})}\BibitemShut
  {NoStop}%
\bibitem [{\citenamefont {Jain}(1989)}]{PhysRevLett.63.199}%
  \BibitemOpen
  \bibfield  {author} {\bibinfo {author} {\bibfnamefont {J.~K.}\ \bibnamefont
  {Jain}},\ }\href {https://doi.org/10.1103/PhysRevLett.63.199} {\bibfield
  {journal} {\bibinfo  {journal} {Phys. Rev. Lett.}\ }\textbf {\bibinfo
  {volume} {63}},\ \bibinfo {pages} {199} (\bibinfo {year} {1989})}\BibitemShut
  {NoStop}%
\bibitem [{\citenamefont {Goldman}\ \emph
  {et~al.}(1990{\natexlab{a}})\citenamefont {Goldman}, \citenamefont {Jain},\
  and\ \citenamefont {Shayegan}}]{PhysRevLett.65.907}%
  \BibitemOpen
  \bibfield  {author} {\bibinfo {author} {\bibfnamefont {V.~J.}\ \bibnamefont
  {Goldman}}, \bibinfo {author} {\bibfnamefont {J.~K.}\ \bibnamefont {Jain}},\
  and\ \bibinfo {author} {\bibfnamefont {M.}~\bibnamefont {Shayegan}},\ }\href
  {https://doi.org/10.1103/PhysRevLett.65.907} {\bibfield  {journal} {\bibinfo
  {journal} {Phys. Rev. Lett.}\ }\textbf {\bibinfo {volume} {65}},\ \bibinfo
  {pages} {907} (\bibinfo {year} {1990}{\natexlab{a}})}\BibitemShut {NoStop}%
\bibitem [{\citenamefont {Zibrov}\ \emph {et~al.}(2018)\citenamefont {Zibrov},
  \citenamefont {Rao}, \citenamefont {Kometter}, \citenamefont {Spanton},
  \citenamefont {Li}, \citenamefont {Dean}, \citenamefont {Taniguchi},
  \citenamefont {Watanabe}, \citenamefont {Serbyn},\ and\ \citenamefont
  {Young}}]{zibrov2018emergent}%
  \BibitemOpen
  \bibfield  {author} {\bibinfo {author} {\bibfnamefont {A.~A.}\ \bibnamefont
  {Zibrov}}, \bibinfo {author} {\bibfnamefont {P.}~\bibnamefont {Rao}},
  \bibinfo {author} {\bibfnamefont {C.}~\bibnamefont {Kometter}}, \bibinfo
  {author} {\bibfnamefont {E.~M.}\ \bibnamefont {Spanton}}, \bibinfo {author}
  {\bibfnamefont {J.}~\bibnamefont {Li}}, \bibinfo {author} {\bibfnamefont
  {C.~R.}\ \bibnamefont {Dean}}, \bibinfo {author} {\bibfnamefont
  {T.}~\bibnamefont {Taniguchi}}, \bibinfo {author} {\bibfnamefont
  {K.}~\bibnamefont {Watanabe}}, \bibinfo {author} {\bibfnamefont
  {M.}~\bibnamefont {Serbyn}},\ and\ \bibinfo {author} {\bibfnamefont {A.~F.}\
  \bibnamefont {Young}},\ }\href
  {https://link.aps.org/doi/10.1103/PhysRevLett.121.167601} {\bibfield
  {journal} {\bibinfo  {journal} {Physical Review Letters}\ }\textbf {\bibinfo
  {volume} {121}},\ \bibinfo {pages} {167601} (\bibinfo {year}
  {2018})}\BibitemShut {NoStop}%
\bibitem [{\citenamefont {Rao}\ and\ \citenamefont
  {Serbyn}(2020)}]{rao2020gully}%
  \BibitemOpen
  \bibfield  {author} {\bibinfo {author} {\bibfnamefont {P.}~\bibnamefont
  {Rao}}\ and\ \bibinfo {author} {\bibfnamefont {M.}~\bibnamefont {Serbyn}},\
  }\href {https://link.aps.org/doi/10.1103/PhysRevB.101.245411} {\bibfield
  {journal} {\bibinfo  {journal} {Physical Review B}\ }\textbf {\bibinfo
  {volume} {101}},\ \bibinfo {pages} {245411} (\bibinfo {year}
  {2020})}\BibitemShut {NoStop}%
\bibitem [{\citenamefont {Winterer}\ \emph
  {et~al.}(2022{\natexlab{a}})\citenamefont {Winterer}, \citenamefont {Seiler},
  \citenamefont {Ghazaryan}, \citenamefont {Geisenhof}, \citenamefont
  {Watanabe}, \citenamefont {Taniguchi}, \citenamefont {Serbyn},\ and\
  \citenamefont {Weitz}}]{winterer2022spontaneous}%
  \BibitemOpen
  \bibfield  {author} {\bibinfo {author} {\bibfnamefont {F.}~\bibnamefont
  {Winterer}}, \bibinfo {author} {\bibfnamefont {A.~M.}\ \bibnamefont
  {Seiler}}, \bibinfo {author} {\bibfnamefont {A.}~\bibnamefont {Ghazaryan}},
  \bibinfo {author} {\bibfnamefont {F.~R.}\ \bibnamefont {Geisenhof}}, \bibinfo
  {author} {\bibfnamefont {K.}~\bibnamefont {Watanabe}}, \bibinfo {author}
  {\bibfnamefont {T.}~\bibnamefont {Taniguchi}}, \bibinfo {author}
  {\bibfnamefont {M.}~\bibnamefont {Serbyn}},\ and\ \bibinfo {author}
  {\bibfnamefont {R.~T.}\ \bibnamefont {Weitz}},\ }\href
  {https://doi.org/10.1021/acs.nanolett.2c00435} {\bibfield  {journal}
  {\bibinfo  {journal} {Nano Letters}\ }\textbf {\bibinfo {volume} {22}},\
  \bibinfo {pages} {3317} (\bibinfo {year} {2022}{\natexlab{a}})}\BibitemShut
  {NoStop}%
\bibitem [{\citenamefont {Serbyn}\ and\ \citenamefont
  {Abanin}(2013)}]{PhysRevB.87.115422}%
  \BibitemOpen
  \bibfield  {author} {\bibinfo {author} {\bibfnamefont {M.}~\bibnamefont
  {Serbyn}}\ and\ \bibinfo {author} {\bibfnamefont {D.~A.}\ \bibnamefont
  {Abanin}},\ }\href {https://doi.org/10.1103/PhysRevB.87.115422} {\bibfield
  {journal} {\bibinfo  {journal} {Phys. Rev. B}\ }\textbf {\bibinfo {volume}
  {87}},\ \bibinfo {pages} {115422} (\bibinfo {year} {2013})}\BibitemShut
  {NoStop}%
\bibitem [{\citenamefont {Wang}\ \emph
  {et~al.}(2016{\natexlab{a}})\citenamefont {Wang}, \citenamefont {Li},
  \citenamefont {Fry},\ and\ \citenamefont {Cheng}}]{wang2016first}%
  \BibitemOpen
  \bibfield  {author} {\bibinfo {author} {\bibfnamefont {Y.-P.}\ \bibnamefont
  {Wang}}, \bibinfo {author} {\bibfnamefont {X.-G.}\ \bibnamefont {Li}},
  \bibinfo {author} {\bibfnamefont {J.~N.}\ \bibnamefont {Fry}},\ and\ \bibinfo
  {author} {\bibfnamefont {H.-P.}\ \bibnamefont {Cheng}},\ }\href
  {https://link.aps.org/doi/10.1103/PhysRevB.94.165428} {\bibfield  {journal}
  {\bibinfo  {journal} {Physical Review B}\ }\textbf {\bibinfo {volume} {94}},\
  \bibinfo {pages} {165428} (\bibinfo {year} {2016}{\natexlab{a}})}\BibitemShut
  {NoStop}%
\bibitem [{\citenamefont {Efros}\ and\ \citenamefont
  {Shklovskii}(1975)}]{efros1975coulomb}%
  \BibitemOpen
  \bibfield  {author} {\bibinfo {author} {\bibfnamefont {A.~L.}\ \bibnamefont
  {Efros}}\ and\ \bibinfo {author} {\bibfnamefont {B.~I.}\ \bibnamefont
  {Shklovskii}},\ }\href {https://doi.org/10.1088/0022-3719/8/4/003} {\bibfield
   {journal} {\bibinfo  {journal} {Journal of Physics C: Solid State Physics}\
  }\textbf {\bibinfo {volume} {8}},\ \bibinfo {pages} {L49} (\bibinfo {year}
  {1975})}\BibitemShut {NoStop}%
\bibitem [{\citenamefont {Ono}(1982)}]{ono1982localization}%
  \BibitemOpen
  \bibfield  {author} {\bibinfo {author} {\bibfnamefont {Y.}~\bibnamefont
  {Ono}},\ }\href@noop {} {\bibfield  {journal} {\bibinfo  {journal} {Journal
  of the Physical Society of Japan}\ }\textbf {\bibinfo {volume} {51}},\
  \bibinfo {pages} {237} (\bibinfo {year} {1982})}\BibitemShut {NoStop}%
\bibitem [{\citenamefont {Hohls}\ \emph
  {et~al.}(2002{\natexlab{a}})\citenamefont {Hohls}, \citenamefont {Zeitler},
  \citenamefont {Haug}, \citenamefont {Meisels}, \citenamefont {Dybko},\ and\
  \citenamefont {Kuchar}}]{PhysRevLett.89.276801}%
  \BibitemOpen
  \bibfield  {author} {\bibinfo {author} {\bibfnamefont {F.}~\bibnamefont
  {Hohls}}, \bibinfo {author} {\bibfnamefont {U.}~\bibnamefont {Zeitler}},
  \bibinfo {author} {\bibfnamefont {R.~J.}\ \bibnamefont {Haug}}, \bibinfo
  {author} {\bibfnamefont {R.}~\bibnamefont {Meisels}}, \bibinfo {author}
  {\bibfnamefont {K.}~\bibnamefont {Dybko}},\ and\ \bibinfo {author}
  {\bibfnamefont {F.}~\bibnamefont {Kuchar}},\ }\href
  {https://doi.org/10.1103/PhysRevLett.89.276801} {\bibfield  {journal}
  {\bibinfo  {journal} {Phys. Rev. Lett.}\ }\textbf {\bibinfo {volume} {89}},\
  \bibinfo {pages} {276801} (\bibinfo {year} {2002}{\natexlab{a}})}\BibitemShut
  {NoStop}%
\bibitem [{\citenamefont {Polyakov}\ and\ \citenamefont
  {Shklovskii}(1993{\natexlab{a}})}]{PhysRevB.48.11167}%
  \BibitemOpen
  \bibfield  {author} {\bibinfo {author} {\bibfnamefont {D.~G.}\ \bibnamefont
  {Polyakov}}\ and\ \bibinfo {author} {\bibfnamefont {B.~I.}\ \bibnamefont
  {Shklovskii}},\ }\href {https://doi.org/10.1103/PhysRevB.48.11167} {\bibfield
   {journal} {\bibinfo  {journal} {Phys. Rev. B}\ }\textbf {\bibinfo {volume}
  {48}},\ \bibinfo {pages} {11167} (\bibinfo {year}
  {1993}{\natexlab{a}})}\BibitemShut {NoStop}%
\bibitem [{\citenamefont {Hohls}\ \emph
  {et~al.}(2002{\natexlab{b}})\citenamefont {Hohls}, \citenamefont {Zeitler},\
  and\ \citenamefont {Haug}}]{hohls2002hopping}%
  \BibitemOpen
  \bibfield  {author} {\bibinfo {author} {\bibfnamefont {F.}~\bibnamefont
  {Hohls}}, \bibinfo {author} {\bibfnamefont {U.}~\bibnamefont {Zeitler}},\
  and\ \bibinfo {author} {\bibfnamefont {R.~J.}\ \bibnamefont {Haug}},\ }\href
  {https://link.aps.org/doi/10.1103/PhysRevLett.88.036802} {\bibfield
  {journal} {\bibinfo  {journal} {Physical review letters}\ }\textbf {\bibinfo
  {volume} {88}},\ \bibinfo {pages} {036802} (\bibinfo {year}
  {2002}{\natexlab{b}})}\BibitemShut {NoStop}%
\bibitem [{\citenamefont {Han}\ \emph {et~al.}(2023)\citenamefont {Han},
  \citenamefont {Lu}, \citenamefont {Scuri}, \citenamefont {Sung},
  \citenamefont {Wang}, \citenamefont {Han}, \citenamefont {Watanabe},
  \citenamefont {Taniguchi}, \citenamefont {Park},\ and\ \citenamefont
  {Ju}}]{Han2023}%
  \BibitemOpen
  \bibfield  {author} {\bibinfo {author} {\bibfnamefont {T.}~\bibnamefont
  {Han}}, \bibinfo {author} {\bibfnamefont {Z.}~\bibnamefont {Lu}}, \bibinfo
  {author} {\bibfnamefont {G.}~\bibnamefont {Scuri}}, \bibinfo {author}
  {\bibfnamefont {J.}~\bibnamefont {Sung}}, \bibinfo {author} {\bibfnamefont
  {J.}~\bibnamefont {Wang}}, \bibinfo {author} {\bibfnamefont {T.}~\bibnamefont
  {Han}}, \bibinfo {author} {\bibfnamefont {K.}~\bibnamefont {Watanabe}},
  \bibinfo {author} {\bibfnamefont {T.}~\bibnamefont {Taniguchi}}, \bibinfo
  {author} {\bibfnamefont {H.}~\bibnamefont {Park}},\ and\ \bibinfo {author}
  {\bibfnamefont {L.}~\bibnamefont {Ju}},\ }\bibfield  {journal} {\bibinfo
  {journal} {Nature Nanotechnology}\ }\href
  {https://doi.org/10.1038/s41565-023-01520-1} {10.1038/s41565-023-01520-1}
  (\bibinfo {year} {2023})\BibitemShut {NoStop}%
\bibitem [{\citenamefont {Cong}\ \emph {et~al.}(2011)\citenamefont {Cong},
  \citenamefont {Yu}, \citenamefont {Sato}, \citenamefont {Shang},
  \citenamefont {Saito}, \citenamefont {Dresselhaus},\ and\ \citenamefont
  {Dresselhaus}}]{cong2011raman}%
  \BibitemOpen
  \bibfield  {author} {\bibinfo {author} {\bibfnamefont {C.}~\bibnamefont
  {Cong}}, \bibinfo {author} {\bibfnamefont {T.}~\bibnamefont {Yu}}, \bibinfo
  {author} {\bibfnamefont {K.}~\bibnamefont {Sato}}, \bibinfo {author}
  {\bibfnamefont {J.}~\bibnamefont {Shang}}, \bibinfo {author} {\bibfnamefont
  {R.}~\bibnamefont {Saito}}, \bibinfo {author} {\bibfnamefont {G.~F.}\
  \bibnamefont {Dresselhaus}},\ and\ \bibinfo {author} {\bibfnamefont {M.~S.}\
  \bibnamefont {Dresselhaus}},\ }\href {https://doi.org/10.1021/nn203472f}
  {\bibfield  {journal} {\bibinfo  {journal} {ACS Nano}\ }\textbf {\bibinfo
  {volume} {5}},\ \bibinfo {pages} {8760} (\bibinfo {year} {2011})},\ \Eprint
  {https://arxiv.org/abs/https://doi.org/10.1021/nn203472f}
  {https://doi.org/10.1021/nn203472f} \BibitemShut {NoStop}%
\bibitem [{\citenamefont {Nguyen}\ \emph {et~al.}(2014)\citenamefont {Nguyen},
  \citenamefont {Lee}, \citenamefont {Yoon},\ and\ \citenamefont
  {Cheong}}]{nguyen2014excitation}%
  \BibitemOpen
  \bibfield  {author} {\bibinfo {author} {\bibfnamefont {T.~A.}\ \bibnamefont
  {Nguyen}}, \bibinfo {author} {\bibfnamefont {J.-U.}\ \bibnamefont {Lee}},
  \bibinfo {author} {\bibfnamefont {D.}~\bibnamefont {Yoon}},\ and\ \bibinfo
  {author} {\bibfnamefont {H.}~\bibnamefont {Cheong}},\ }\href
  {https://www.nature.com/articles/srep04630} {\bibfield  {journal} {\bibinfo
  {journal} {Scientific reports}\ }\textbf {\bibinfo {volume} {4}},\ \bibinfo
  {pages} {4630} (\bibinfo {year} {2014})}\BibitemShut {NoStop}%
\bibitem [{\citenamefont {Wang}\ \emph {et~al.}(2013)\citenamefont {Wang},
  \citenamefont {Meric}, \citenamefont {Huang}, \citenamefont {Gao},
  \citenamefont {Gao}, \citenamefont {Tran}, \citenamefont {Taniguchi},
  \citenamefont {Watanabe}, \citenamefont {Campos}, \citenamefont {Muller},
  \citenamefont {Guo}, \citenamefont {Kim}, \citenamefont {Hone}, \citenamefont
  {Shepard},\ and\ \citenamefont {Dean}}]{Wang614}%
  \BibitemOpen
  \bibfield  {author} {\bibinfo {author} {\bibfnamefont {L.}~\bibnamefont
  {Wang}}, \bibinfo {author} {\bibfnamefont {I.}~\bibnamefont {Meric}},
  \bibinfo {author} {\bibfnamefont {P.~Y.}\ \bibnamefont {Huang}}, \bibinfo
  {author} {\bibfnamefont {Q.}~\bibnamefont {Gao}}, \bibinfo {author}
  {\bibfnamefont {Y.}~\bibnamefont {Gao}}, \bibinfo {author} {\bibfnamefont
  {H.}~\bibnamefont {Tran}}, \bibinfo {author} {\bibfnamefont {T.}~\bibnamefont
  {Taniguchi}}, \bibinfo {author} {\bibfnamefont {K.}~\bibnamefont {Watanabe}},
  \bibinfo {author} {\bibfnamefont {L.~M.}\ \bibnamefont {Campos}}, \bibinfo
  {author} {\bibfnamefont {D.~A.}\ \bibnamefont {Muller}}, \bibinfo {author}
  {\bibfnamefont {J.}~\bibnamefont {Guo}}, \bibinfo {author} {\bibfnamefont
  {P.}~\bibnamefont {Kim}}, \bibinfo {author} {\bibfnamefont {J.}~\bibnamefont
  {Hone}}, \bibinfo {author} {\bibfnamefont {K.~L.}\ \bibnamefont {Shepard}},\
  and\ \bibinfo {author} {\bibfnamefont {C.~R.}\ \bibnamefont {Dean}},\ }\href
  {https://doi.org/10.1126/science.1244358} {\bibfield  {journal} {\bibinfo
  {journal} {Science}\ }\textbf {\bibinfo {volume} {342}},\ \bibinfo {pages}
  {614} (\bibinfo {year} {2013})},\ \Eprint
  {https://arxiv.org/abs/http://science.sciencemag.org/content/342/6158/614.full.pdf}
  {http://science.sciencemag.org/content/342/6158/614.full.pdf} \BibitemShut
  {NoStop}%
\bibitem [{\citenamefont {Tiwari}\ \emph {et~al.}(2021)\citenamefont {Tiwari},
  \citenamefont {Srivastav},\ and\ \citenamefont
  {Bid}}]{PhysRevLett.126.096801}%
  \BibitemOpen
  \bibfield  {author} {\bibinfo {author} {\bibfnamefont {P.}~\bibnamefont
  {Tiwari}}, \bibinfo {author} {\bibfnamefont {S.~P.}\ \bibnamefont
  {Srivastav}},\ and\ \bibinfo {author} {\bibfnamefont {A.}~\bibnamefont
  {Bid}},\ }\href {https://doi.org/10.1103/PhysRevLett.126.096801} {\bibfield
  {journal} {\bibinfo  {journal} {Phys. Rev. Lett.}\ }\textbf {\bibinfo
  {volume} {126}},\ \bibinfo {pages} {096801} (\bibinfo {year}
  {2021})}\BibitemShut {NoStop}%
\bibitem [{\citenamefont {Lui}\ \emph {et~al.}(2011)\citenamefont {Lui},
  \citenamefont {Li}, \citenamefont {Mak}, \citenamefont {Cappelluti},\ and\
  \citenamefont {Heinz}}]{lui2011observation}%
  \BibitemOpen
  \bibfield  {author} {\bibinfo {author} {\bibfnamefont {C.~H.}\ \bibnamefont
  {Lui}}, \bibinfo {author} {\bibfnamefont {Z.}~\bibnamefont {Li}}, \bibinfo
  {author} {\bibfnamefont {K.~F.}\ \bibnamefont {Mak}}, \bibinfo {author}
  {\bibfnamefont {E.}~\bibnamefont {Cappelluti}},\ and\ \bibinfo {author}
  {\bibfnamefont {T.~F.}\ \bibnamefont {Heinz}},\ }\href
  {https://www.nature.com/articles/nphys2102} {\bibfield  {journal} {\bibinfo
  {journal} {Nature Physics}\ }\textbf {\bibinfo {volume} {7}},\ \bibinfo
  {pages} {944} (\bibinfo {year} {2011})}\BibitemShut {NoStop}%
\bibitem [{\citenamefont {Datta}\ \emph
  {et~al.}(2018{\natexlab{a}})\citenamefont {Datta}, \citenamefont {Agarwal},
  \citenamefont {Samanta}, \citenamefont {Ratnakar}, \citenamefont {Watanabe},
  \citenamefont {Taniguchi}, \citenamefont {Sensarma},\ and\ \citenamefont
  {Deshmukh}}]{datta2018landau}%
  \BibitemOpen
  \bibfield  {author} {\bibinfo {author} {\bibfnamefont {B.}~\bibnamefont
  {Datta}}, \bibinfo {author} {\bibfnamefont {H.}~\bibnamefont {Agarwal}},
  \bibinfo {author} {\bibfnamefont {A.}~\bibnamefont {Samanta}}, \bibinfo
  {author} {\bibfnamefont {A.}~\bibnamefont {Ratnakar}}, \bibinfo {author}
  {\bibfnamefont {K.}~\bibnamefont {Watanabe}}, \bibinfo {author}
  {\bibfnamefont {T.}~\bibnamefont {Taniguchi}}, \bibinfo {author}
  {\bibfnamefont {R.}~\bibnamefont {Sensarma}},\ and\ \bibinfo {author}
  {\bibfnamefont {M.~M.}\ \bibnamefont {Deshmukh}},\ }\href
  {https://link.aps.org/doi/10.1103/PhysRevLett.121.056801} {\bibfield
  {journal} {\bibinfo  {journal} {Physical Review Letters}\ }\textbf {\bibinfo
  {volume} {121}},\ \bibinfo {pages} {056801} (\bibinfo {year}
  {2018}{\natexlab{a}})}\BibitemShut {NoStop}%
\bibitem [{\citenamefont {Chen}\ \emph {et~al.}(2019)\citenamefont {Chen},
  \citenamefont {Jiang}, \citenamefont {Wu}, \citenamefont {Lyu}, \citenamefont
  {Li}, \citenamefont {Chittari}, \citenamefont {Watanabe}, \citenamefont
  {Taniguchi}, \citenamefont {Shi}, \citenamefont {Jung} \emph
  {et~al.}}]{chen2019evidence}%
  \BibitemOpen
  \bibfield  {author} {\bibinfo {author} {\bibfnamefont {G.}~\bibnamefont
  {Chen}}, \bibinfo {author} {\bibfnamefont {L.}~\bibnamefont {Jiang}},
  \bibinfo {author} {\bibfnamefont {S.}~\bibnamefont {Wu}}, \bibinfo {author}
  {\bibfnamefont {B.}~\bibnamefont {Lyu}}, \bibinfo {author} {\bibfnamefont
  {H.}~\bibnamefont {Li}}, \bibinfo {author} {\bibfnamefont {B.~L.}\
  \bibnamefont {Chittari}}, \bibinfo {author} {\bibfnamefont {K.}~\bibnamefont
  {Watanabe}}, \bibinfo {author} {\bibfnamefont {T.}~\bibnamefont {Taniguchi}},
  \bibinfo {author} {\bibfnamefont {Z.}~\bibnamefont {Shi}}, \bibinfo {author}
  {\bibfnamefont {J.}~\bibnamefont {Jung}}, \emph {et~al.},\ }\href
  {https://www.nature.com/articles/s41567-018-0387-2} {\bibfield  {journal}
  {\bibinfo  {journal} {Nature Physics}\ }\textbf {\bibinfo {volume} {15}},\
  \bibinfo {pages} {237} (\bibinfo {year} {2019})}\BibitemShut {NoStop}%
\bibitem [{\citenamefont {Zhou}\ \emph {et~al.}(2021)\citenamefont {Zhou},
  \citenamefont {Xie}, \citenamefont {Taniguchi}, \citenamefont {Watanabe},\
  and\ \citenamefont {Young}}]{zhou2021superconductivity}%
  \BibitemOpen
  \bibfield  {author} {\bibinfo {author} {\bibfnamefont {H.}~\bibnamefont
  {Zhou}}, \bibinfo {author} {\bibfnamefont {T.}~\bibnamefont {Xie}}, \bibinfo
  {author} {\bibfnamefont {T.}~\bibnamefont {Taniguchi}}, \bibinfo {author}
  {\bibfnamefont {K.}~\bibnamefont {Watanabe}},\ and\ \bibinfo {author}
  {\bibfnamefont {A.~F.}\ \bibnamefont {Young}},\ }\href
  {https://www.nature.com/articles/s41586-021-03926-0} {\bibfield  {journal}
  {\bibinfo  {journal} {Nature}\ }\textbf {\bibinfo {volume} {598}},\ \bibinfo
  {pages} {434} (\bibinfo {year} {2021})}\BibitemShut {NoStop}%
\bibitem [{\citenamefont {Zou}\ \emph {et~al.}(2013)\citenamefont {Zou},
  \citenamefont {Zhang}, \citenamefont {Clapp}, \citenamefont {MacDonald},\
  and\ \citenamefont {Zhu}}]{zou2013transport}%
  \BibitemOpen
  \bibfield  {author} {\bibinfo {author} {\bibfnamefont {K.}~\bibnamefont
  {Zou}}, \bibinfo {author} {\bibfnamefont {F.}~\bibnamefont {Zhang}}, \bibinfo
  {author} {\bibfnamefont {C.}~\bibnamefont {Clapp}}, \bibinfo {author}
  {\bibfnamefont {A.}~\bibnamefont {MacDonald}},\ and\ \bibinfo {author}
  {\bibfnamefont {J.}~\bibnamefont {Zhu}},\ }\href
  {https://pubs.acs.org/doi/10.1021/nl303375a} {\bibfield  {journal} {\bibinfo
  {journal} {Nano letters}\ }\textbf {\bibinfo {volume} {13}},\ \bibinfo
  {pages} {369} (\bibinfo {year} {2013})}\BibitemShut {NoStop}%
\bibitem [{\citenamefont {Jhang}\ \emph {et~al.}(2011)\citenamefont {Jhang},
  \citenamefont {Craciun}, \citenamefont {Schmidmeier}, \citenamefont
  {Tokumitsu}, \citenamefont {Russo}, \citenamefont {Yamamoto}, \citenamefont
  {Skourski}, \citenamefont {Wosnitza}, \citenamefont {Tarucha}, \citenamefont
  {Eroms},\ and\ \citenamefont {Strunk}}]{jhang2011stacking}%
  \BibitemOpen
  \bibfield  {author} {\bibinfo {author} {\bibfnamefont {S.~H.}\ \bibnamefont
  {Jhang}}, \bibinfo {author} {\bibfnamefont {M.~F.}\ \bibnamefont {Craciun}},
  \bibinfo {author} {\bibfnamefont {S.}~\bibnamefont {Schmidmeier}}, \bibinfo
  {author} {\bibfnamefont {S.}~\bibnamefont {Tokumitsu}}, \bibinfo {author}
  {\bibfnamefont {S.}~\bibnamefont {Russo}}, \bibinfo {author} {\bibfnamefont
  {M.}~\bibnamefont {Yamamoto}}, \bibinfo {author} {\bibfnamefont
  {Y.}~\bibnamefont {Skourski}}, \bibinfo {author} {\bibfnamefont
  {J.}~\bibnamefont {Wosnitza}}, \bibinfo {author} {\bibfnamefont
  {S.}~\bibnamefont {Tarucha}}, \bibinfo {author} {\bibfnamefont
  {J.}~\bibnamefont {Eroms}},\ and\ \bibinfo {author} {\bibfnamefont
  {C.}~\bibnamefont {Strunk}},\ }\href
  {https://link.aps.org/doi/10.1103/PhysRevB.84.161408} {\bibfield  {journal}
  {\bibinfo  {journal} {Physical Review B}\ }\textbf {\bibinfo {volume} {84}},\
  \bibinfo {pages} {161408} (\bibinfo {year} {2011})}\BibitemShut {NoStop}%
\bibitem [{\citenamefont {Polyakov}\ and\ \citenamefont
  {Shklovskii}(1993{\natexlab{b}})}]{PhysRevLett.70.3796}%
  \BibitemOpen
  \bibfield  {author} {\bibinfo {author} {\bibfnamefont {D.~G.}\ \bibnamefont
  {Polyakov}}\ and\ \bibinfo {author} {\bibfnamefont {B.~I.}\ \bibnamefont
  {Shklovskii}},\ }\href {https://doi.org/10.1103/PhysRevLett.70.3796}
  {\bibfield  {journal} {\bibinfo  {journal} {Phys. Rev. Lett.}\ }\textbf
  {\bibinfo {volume} {70}},\ \bibinfo {pages} {3796} (\bibinfo {year}
  {1993}{\natexlab{b}})}\BibitemShut {NoStop}%
\bibitem [{\citenamefont {Venugopal}\ \emph {et~al.}(2011)\citenamefont
  {Venugopal}, \citenamefont {Chan}, \citenamefont {Li}, \citenamefont
  {Magnuson}, \citenamefont {Kirk}, \citenamefont {Colombo}, \citenamefont
  {Ruoff},\ and\ \citenamefont {Vogel}}]{10.1063/1.3592338}%
  \BibitemOpen
  \bibfield  {author} {\bibinfo {author} {\bibfnamefont {A.}~\bibnamefont
  {Venugopal}}, \bibinfo {author} {\bibfnamefont {J.}~\bibnamefont {Chan}},
  \bibinfo {author} {\bibfnamefont {X.}~\bibnamefont {Li}}, \bibinfo {author}
  {\bibfnamefont {C.~W.}\ \bibnamefont {Magnuson}}, \bibinfo {author}
  {\bibfnamefont {W.~P.}\ \bibnamefont {Kirk}}, \bibinfo {author}
  {\bibfnamefont {L.}~\bibnamefont {Colombo}}, \bibinfo {author} {\bibfnamefont
  {R.~S.}\ \bibnamefont {Ruoff}},\ and\ \bibinfo {author} {\bibfnamefont
  {E.~M.}\ \bibnamefont {Vogel}},\ }\href {https://doi.org/10.1063/1.3592338}
  {\bibfield  {journal} {\bibinfo  {journal} {Journal of Applied Physics}\
  }\textbf {\bibinfo {volume} {109}},\ \bibinfo {pages} {104511} (\bibinfo
  {year} {2011})},\ \Eprint
  {https://arxiv.org/abs/https://pubs.aip.org/aip/jap/article-pdf/doi/10.1063/1.3592338/15082620/104511\_1\_online.pdf}
  {https://pubs.aip.org/aip/jap/article-pdf/doi/10.1063/1.3592338/15082620/104511\_1\_online.pdf}
  \BibitemShut {NoStop}%
\bibitem [{\citenamefont {Stepanov}\ \emph {et~al.}(2016)\citenamefont
  {Stepanov}, \citenamefont {Barlas}, \citenamefont {Espiritu}, \citenamefont
  {Che}, \citenamefont {Watanabe}, \citenamefont {Taniguchi}, \citenamefont
  {Smirnov},\ and\ \citenamefont {Lau}}]{PhysRevLett.117.076807}%
  \BibitemOpen
  \bibfield  {author} {\bibinfo {author} {\bibfnamefont {P.}~\bibnamefont
  {Stepanov}}, \bibinfo {author} {\bibfnamefont {Y.}~\bibnamefont {Barlas}},
  \bibinfo {author} {\bibfnamefont {T.}~\bibnamefont {Espiritu}}, \bibinfo
  {author} {\bibfnamefont {S.}~\bibnamefont {Che}}, \bibinfo {author}
  {\bibfnamefont {K.}~\bibnamefont {Watanabe}}, \bibinfo {author}
  {\bibfnamefont {T.}~\bibnamefont {Taniguchi}}, \bibinfo {author}
  {\bibfnamefont {D.}~\bibnamefont {Smirnov}},\ and\ \bibinfo {author}
  {\bibfnamefont {C.~N.}\ \bibnamefont {Lau}},\ }\href
  {https://doi.org/10.1103/PhysRevLett.117.076807} {\bibfield  {journal}
  {\bibinfo  {journal} {Phys. Rev. Lett.}\ }\textbf {\bibinfo {volume} {117}},\
  \bibinfo {pages} {076807} (\bibinfo {year} {2016})}\BibitemShut {NoStop}%
\bibitem [{\citenamefont {Taychatanapat}\ \emph
  {et~al.}(2011{\natexlab{a}})\citenamefont {Taychatanapat}, \citenamefont
  {Watanabe}, \citenamefont {Taniguchi},\ and\ \citenamefont
  {Jarillo-Herrero}}]{taychatanapat2011quantum}%
  \BibitemOpen
  \bibfield  {author} {\bibinfo {author} {\bibfnamefont {T.}~\bibnamefont
  {Taychatanapat}}, \bibinfo {author} {\bibfnamefont {K.}~\bibnamefont
  {Watanabe}}, \bibinfo {author} {\bibfnamefont {T.}~\bibnamefont
  {Taniguchi}},\ and\ \bibinfo {author} {\bibfnamefont {P.}~\bibnamefont
  {Jarillo-Herrero}},\ }\href {https://www.nature.com/articles/nphys2008}
  {\bibfield  {journal} {\bibinfo  {journal} {Nature Physics}\ }\textbf
  {\bibinfo {volume} {7}},\ \bibinfo {pages} {621} (\bibinfo {year}
  {2011}{\natexlab{a}})}\BibitemShut {NoStop}%
\bibitem [{\citenamefont {Goldman}\ \emph
  {et~al.}(1990{\natexlab{b}})\citenamefont {Goldman}, \citenamefont {Jain},\
  and\ \citenamefont {Shayegan}}]{goldman1990nature}%
  \BibitemOpen
  \bibfield  {author} {\bibinfo {author} {\bibfnamefont {V.}~\bibnamefont
  {Goldman}}, \bibinfo {author} {\bibfnamefont {J.~K.}\ \bibnamefont {Jain}},\
  and\ \bibinfo {author} {\bibfnamefont {M.}~\bibnamefont {Shayegan}},\ }\href
  {https://link.aps.org/doi/10.1103/PhysRevLett.65.907} {\bibfield  {journal}
  {\bibinfo  {journal} {Phys. Rev. Lett.}\ }\textbf {\bibinfo {volume} {65}},\
  \bibinfo {pages} {907} (\bibinfo {year} {1990}{\natexlab{b}})}\BibitemShut
  {NoStop}%
\bibitem [{\citenamefont {Martin}\ \emph {et~al.}(2008)\citenamefont {Martin},
  \citenamefont {Akerman}, \citenamefont {Ulbricht}, \citenamefont {Lohmann},
  \citenamefont {Smet}, \citenamefont {von Klitzing},\ and\ \citenamefont
  {Yacoby}}]{Martin2008}%
  \BibitemOpen
  \bibfield  {author} {\bibinfo {author} {\bibfnamefont {J.}~\bibnamefont
  {Martin}}, \bibinfo {author} {\bibfnamefont {N.}~\bibnamefont {Akerman}},
  \bibinfo {author} {\bibfnamefont {G.}~\bibnamefont {Ulbricht}}, \bibinfo
  {author} {\bibfnamefont {T.}~\bibnamefont {Lohmann}}, \bibinfo {author}
  {\bibfnamefont {J.~H.}\ \bibnamefont {Smet}}, \bibinfo {author}
  {\bibfnamefont {K.}~\bibnamefont {von Klitzing}},\ and\ \bibinfo {author}
  {\bibfnamefont {A.}~\bibnamefont {Yacoby}},\ }\href
  {https://doi.org/10.1038/nphys781} {\bibfield  {journal} {\bibinfo  {journal}
  {Nature Physics}\ }\textbf {\bibinfo {volume} {4}},\ \bibinfo {pages} {144}
  (\bibinfo {year} {2008})}\BibitemShut {NoStop}%
\bibitem [{\citenamefont
  {Pruisken}(1988{\natexlab{b}})}]{pruisken1988universal}%
  \BibitemOpen
  \bibfield  {author} {\bibinfo {author} {\bibfnamefont {A.}~\bibnamefont
  {Pruisken}},\ }\href@noop {} {\bibfield  {journal} {\bibinfo  {journal}
  {Physical review letters}\ }\textbf {\bibinfo {volume} {61}},\ \bibinfo
  {pages} {1297} (\bibinfo {year} {1988}{\natexlab{b}})}\BibitemShut {NoStop}%
\bibitem [{\citenamefont {Wei}\ \emph {et~al.}(1990)\citenamefont {Wei},
  \citenamefont {Hwang}, \citenamefont {Tsui},\ and\ \citenamefont
  {Pruisken}}]{WEI199034}%
  \BibitemOpen
  \bibfield  {author} {\bibinfo {author} {\bibfnamefont {H.}~\bibnamefont
  {Wei}}, \bibinfo {author} {\bibfnamefont {S.}~\bibnamefont {Hwang}}, \bibinfo
  {author} {\bibfnamefont {D.}~\bibnamefont {Tsui}},\ and\ \bibinfo {author}
  {\bibfnamefont {A.}~\bibnamefont {Pruisken}},\ }\href
  {https://doi.org/https://doi.org/10.1016/0039-6028(90)90825-S} {\bibfield
  {journal} {\bibinfo  {journal} {Surface Science}\ }\textbf {\bibinfo {volume}
  {229}},\ \bibinfo {pages} {34} (\bibinfo {year} {1990})}\BibitemShut
  {NoStop}%
\bibitem [{\citenamefont {Koch}\ \emph {et~al.}(1992)\citenamefont {Koch},
  \citenamefont {Haug}, \citenamefont {Klitzing},\ and\ \citenamefont
  {Ploog}}]{PhysRevB.46.1596}%
  \BibitemOpen
  \bibfield  {author} {\bibinfo {author} {\bibfnamefont {S.}~\bibnamefont
  {Koch}}, \bibinfo {author} {\bibfnamefont {R.~J.}\ \bibnamefont {Haug}},
  \bibinfo {author} {\bibfnamefont {K.~v.}\ \bibnamefont {Klitzing}},\ and\
  \bibinfo {author} {\bibfnamefont {K.}~\bibnamefont {Ploog}},\ }\href
  {https://doi.org/10.1103/PhysRevB.46.1596} {\bibfield  {journal} {\bibinfo
  {journal} {Phys. Rev. B}\ }\textbf {\bibinfo {volume} {46}},\ \bibinfo
  {pages} {1596} (\bibinfo {year} {1992})}\BibitemShut {NoStop}%
\bibitem [{\citenamefont {Koch}\ \emph
  {et~al.}(1991{\natexlab{a}})\citenamefont {Koch}, \citenamefont {Haug},
  \citenamefont {Klitzing},\ and\ \citenamefont {Ploog}}]{PhysRevLett.67.883}%
  \BibitemOpen
  \bibfield  {author} {\bibinfo {author} {\bibfnamefont {S.}~\bibnamefont
  {Koch}}, \bibinfo {author} {\bibfnamefont {R.~J.}\ \bibnamefont {Haug}},
  \bibinfo {author} {\bibfnamefont {K.~v.}\ \bibnamefont {Klitzing}},\ and\
  \bibinfo {author} {\bibfnamefont {K.}~\bibnamefont {Ploog}},\ }\href
  {https://doi.org/10.1103/PhysRevLett.67.883} {\bibfield  {journal} {\bibinfo
  {journal} {Phys. Rev. Lett.}\ }\textbf {\bibinfo {volume} {67}},\ \bibinfo
  {pages} {883} (\bibinfo {year} {1991}{\natexlab{a}})}\BibitemShut {NoStop}%
\bibitem [{\citenamefont {Cobaleda}\ \emph {et~al.}(2014)\citenamefont
  {Cobaleda}, \citenamefont {Pezzini}, \citenamefont {Rodriguez}, \citenamefont
  {Diez},\ and\ \citenamefont {Bellani}}]{PhysRevB.90.161408}%
  \BibitemOpen
  \bibfield  {author} {\bibinfo {author} {\bibfnamefont {C.}~\bibnamefont
  {Cobaleda}}, \bibinfo {author} {\bibfnamefont {S.}~\bibnamefont {Pezzini}},
  \bibinfo {author} {\bibfnamefont {A.}~\bibnamefont {Rodriguez}}, \bibinfo
  {author} {\bibfnamefont {E.}~\bibnamefont {Diez}},\ and\ \bibinfo {author}
  {\bibfnamefont {V.}~\bibnamefont {Bellani}},\ }\href
  {https://doi.org/10.1103/PhysRevB.90.161408} {\bibfield  {journal} {\bibinfo
  {journal} {Phys. Rev. B}\ }\textbf {\bibinfo {volume} {90}},\ \bibinfo
  {pages} {161408} (\bibinfo {year} {2014})}\BibitemShut {NoStop}%
\bibitem [{\citenamefont {Amin}\ \emph {et~al.}(2018)\citenamefont {Amin},
  \citenamefont {Ray}, \citenamefont {Pal}, \citenamefont {Pandit},\ and\
  \citenamefont {Bid}}]{Amin2018}%
  \BibitemOpen
  \bibfield  {author} {\bibinfo {author} {\bibfnamefont {K.~R.}\ \bibnamefont
  {Amin}}, \bibinfo {author} {\bibfnamefont {S.~S.}\ \bibnamefont {Ray}},
  \bibinfo {author} {\bibfnamefont {N.}~\bibnamefont {Pal}}, \bibinfo {author}
  {\bibfnamefont {R.}~\bibnamefont {Pandit}},\ and\ \bibinfo {author}
  {\bibfnamefont {A.}~\bibnamefont {Bid}},\ }\href
  {https://doi.org/10.1038/s42005-017-0001-4} {\bibfield  {journal} {\bibinfo
  {journal} {Communications Physics}\ }\textbf {\bibinfo {volume} {1}},\
  \bibinfo {pages} {1} (\bibinfo {year} {2018})}\BibitemShut {NoStop}%
\bibitem [{\citenamefont {Wei}\ \emph {et~al.}(1988)\citenamefont {Wei},
  \citenamefont {Tsui}, \citenamefont {Paalanen},\ and\ \citenamefont
  {Pruisken}}]{PhysRevLett.61.1294}%
  \BibitemOpen
  \bibfield  {author} {\bibinfo {author} {\bibfnamefont {H.~P.}\ \bibnamefont
  {Wei}}, \bibinfo {author} {\bibfnamefont {D.~C.}\ \bibnamefont {Tsui}},
  \bibinfo {author} {\bibfnamefont {M.~A.}\ \bibnamefont {Paalanen}},\ and\
  \bibinfo {author} {\bibfnamefont {A.~M.~M.}\ \bibnamefont {Pruisken}},\
  }\href {https://doi.org/10.1103/PhysRevLett.61.1294} {\bibfield  {journal}
  {\bibinfo  {journal} {Phys. Rev. Lett.}\ }\textbf {\bibinfo {volume} {61}},\
  \bibinfo {pages} {1294} (\bibinfo {year} {1988})}\BibitemShut {NoStop}%
\bibitem [{\citenamefont {Koch}\ \emph
  {et~al.}(1991{\natexlab{b}})\citenamefont {Koch}, \citenamefont {Haug},
  \citenamefont {Klitzing},\ and\ \citenamefont {Ploog}}]{PhysRevB.43.6828}%
  \BibitemOpen
  \bibfield  {author} {\bibinfo {author} {\bibfnamefont {S.}~\bibnamefont
  {Koch}}, \bibinfo {author} {\bibfnamefont {R.~J.}\ \bibnamefont {Haug}},
  \bibinfo {author} {\bibfnamefont {K.~v.}\ \bibnamefont {Klitzing}},\ and\
  \bibinfo {author} {\bibfnamefont {K.}~\bibnamefont {Ploog}},\ }\href
  {https://doi.org/10.1103/PhysRevB.43.6828} {\bibfield  {journal} {\bibinfo
  {journal} {Phys. Rev. B}\ }\textbf {\bibinfo {volume} {43}},\ \bibinfo
  {pages} {6828} (\bibinfo {year} {1991}{\natexlab{b}})}\BibitemShut {NoStop}%
\bibitem [{\citenamefont {Giesbers}\ \emph {et~al.}(2009)\citenamefont
  {Giesbers}, \citenamefont {Zeitler}, \citenamefont {Ponomarenko},
  \citenamefont {Yang}, \citenamefont {Novoselov}, \citenamefont {Geim},\ and\
  \citenamefont {Maan}}]{PhysRevB.80.241411}%
  \BibitemOpen
  \bibfield  {author} {\bibinfo {author} {\bibfnamefont {A.~J.~M.}\
  \bibnamefont {Giesbers}}, \bibinfo {author} {\bibfnamefont {U.}~\bibnamefont
  {Zeitler}}, \bibinfo {author} {\bibfnamefont {L.~A.}\ \bibnamefont
  {Ponomarenko}}, \bibinfo {author} {\bibfnamefont {R.}~\bibnamefont {Yang}},
  \bibinfo {author} {\bibfnamefont {K.~S.}\ \bibnamefont {Novoselov}}, \bibinfo
  {author} {\bibfnamefont {A.~K.}\ \bibnamefont {Geim}},\ and\ \bibinfo
  {author} {\bibfnamefont {J.~C.}\ \bibnamefont {Maan}},\ }\href
  {https://doi.org/10.1103/PhysRevB.80.241411} {\bibfield  {journal} {\bibinfo
  {journal} {Phys. Rev. B}\ }\textbf {\bibinfo {volume} {80}},\ \bibinfo
  {pages} {241411} (\bibinfo {year} {2009})}\BibitemShut {NoStop}%
\bibitem [{\citenamefont {Shen}\ \emph {et~al.}(2012)\citenamefont {Shen},
  \citenamefont {Neal}, \citenamefont {Bolen}, \citenamefont {Gu},
  \citenamefont {Engel}, \citenamefont {Capano},\ and\ \citenamefont
  {Ye}}]{Shen2012}%
  \BibitemOpen
  \bibfield  {author} {\bibinfo {author} {\bibfnamefont {T.}~\bibnamefont
  {Shen}}, \bibinfo {author} {\bibfnamefont {A.~T.}\ \bibnamefont {Neal}},
  \bibinfo {author} {\bibfnamefont {M.~L.}\ \bibnamefont {Bolen}}, \bibinfo
  {author} {\bibfnamefont {J.~J.}\ \bibnamefont {Gu}}, \bibinfo {author}
  {\bibfnamefont {L.~W.}\ \bibnamefont {Engel}}, \bibinfo {author}
  {\bibfnamefont {M.~A.}\ \bibnamefont {Capano}},\ and\ \bibinfo {author}
  {\bibfnamefont {P.~D.}\ \bibnamefont {Ye}},\ }\href
  {https://doi.org/10.1063/1.3675464} {\bibfield  {journal} {\bibinfo
  {journal} {Journal of Applied Physics}\ }\textbf {\bibinfo {volume} {111}},\
  \bibinfo {pages} {013716} (\bibinfo {year} {2012})}\BibitemShut {NoStop}%
\bibitem [{\citenamefont {Pallecchi}\ \emph {et~al.}(2013)\citenamefont
  {Pallecchi}, \citenamefont {Ridene}, \citenamefont {Kazazis}, \citenamefont
  {Lafont}, \citenamefont {Schopfer}, \citenamefont {Poirier}, \citenamefont
  {Goerbig}, \citenamefont {Mailly},\ and\ \citenamefont
  {Ouerghi}}]{Pallecchi2013}%
  \BibitemOpen
  \bibfield  {author} {\bibinfo {author} {\bibfnamefont {E.}~\bibnamefont
  {Pallecchi}}, \bibinfo {author} {\bibfnamefont {M.}~\bibnamefont {Ridene}},
  \bibinfo {author} {\bibfnamefont {D.}~\bibnamefont {Kazazis}}, \bibinfo
  {author} {\bibfnamefont {F.}~\bibnamefont {Lafont}}, \bibinfo {author}
  {\bibfnamefont {F.}~\bibnamefont {Schopfer}}, \bibinfo {author}
  {\bibfnamefont {W.}~\bibnamefont {Poirier}}, \bibinfo {author} {\bibfnamefont
  {M.~O.}\ \bibnamefont {Goerbig}}, \bibinfo {author} {\bibfnamefont
  {D.}~\bibnamefont {Mailly}},\ and\ \bibinfo {author} {\bibfnamefont
  {A.}~\bibnamefont {Ouerghi}},\ }\href
  {http://www.ncbi.nlm.nih.gov/pmc/articles/PMC3646355/} {\bibfield  {journal}
  {\bibinfo  {journal} {Scientific Reports}\ }\textbf {\bibinfo {volume} {3}},\
  \bibinfo {pages} {1791} (\bibinfo {year} {2013})}\BibitemShut {NoStop}%
\bibitem [{\citenamefont {Peters}\ \emph {et~al.}(2014)\citenamefont {Peters},
  \citenamefont {Giesbers}, \citenamefont {Burghard},\ and\ \citenamefont
  {Kern}}]{Peters2014}%
  \BibitemOpen
  \bibfield  {author} {\bibinfo {author} {\bibfnamefont {E.~C.}\ \bibnamefont
  {Peters}}, \bibinfo {author} {\bibfnamefont {A.~J.~M.}\ \bibnamefont
  {Giesbers}}, \bibinfo {author} {\bibfnamefont {M.}~\bibnamefont {Burghard}},\
  and\ \bibinfo {author} {\bibfnamefont {K.}~\bibnamefont {Kern}},\ }\href
  {https://pubs.aip.org/aip/apl/article/104/20/203109/131097/Scaling-in-the-quantum-Hall-regime-of-graphene}
  {\bibfield  {journal} {\bibinfo  {journal} {Appl. Phys. Lett.}\ }\textbf
  {\bibinfo {volume} {104}},\ \bibinfo {pages} {203109} (\bibinfo {year}
  {2014})}\BibitemShut {NoStop}%
\bibitem [{\citenamefont {Liu}\ \emph {et~al.}(2016)\citenamefont {Liu},
  \citenamefont {Wang}, \citenamefont {Woo}, \citenamefont {Shih},
  \citenamefont {Liou}, \citenamefont {Ho}, \citenamefont {Chen}, \citenamefont
  {Liang},\ and\ \citenamefont {Wang}}]{PhysRevB.93.041421}%
  \BibitemOpen
  \bibfield  {author} {\bibinfo {author} {\bibfnamefont {C.-H.}\ \bibnamefont
  {Liu}}, \bibinfo {author} {\bibfnamefont {P.-H.}\ \bibnamefont {Wang}},
  \bibinfo {author} {\bibfnamefont {T.-P.}\ \bibnamefont {Woo}}, \bibinfo
  {author} {\bibfnamefont {F.-Y.}\ \bibnamefont {Shih}}, \bibinfo {author}
  {\bibfnamefont {S.-C.}\ \bibnamefont {Liou}}, \bibinfo {author}
  {\bibfnamefont {P.-H.}\ \bibnamefont {Ho}}, \bibinfo {author} {\bibfnamefont
  {C.-W.}\ \bibnamefont {Chen}}, \bibinfo {author} {\bibfnamefont {C.-T.}\
  \bibnamefont {Liang}},\ and\ \bibinfo {author} {\bibfnamefont {W.-H.}\
  \bibnamefont {Wang}},\ }\href {https://doi.org/10.1103/PhysRevB.93.041421}
  {\bibfield  {journal} {\bibinfo  {journal} {Phys. Rev. B}\ }\textbf {\bibinfo
  {volume} {93}},\ \bibinfo {pages} {041421} (\bibinfo {year}
  {2016})}\BibitemShut {NoStop}%
\bibitem [{\citenamefont {Amado}\ \emph {et~al.}(2012)\citenamefont {Amado},
  \citenamefont {Diez}, \citenamefont {Rossella}, \citenamefont {Bellani},
  \citenamefont {Lopez-Romero},\ and\ \citenamefont {Maude}}]{Amado2012}%
  \BibitemOpen
  \bibfield  {author} {\bibinfo {author} {\bibfnamefont {M.}~\bibnamefont
  {Amado}}, \bibinfo {author} {\bibfnamefont {E.}~\bibnamefont {Diez}},
  \bibinfo {author} {\bibfnamefont {F.}~\bibnamefont {Rossella}}, \bibinfo
  {author} {\bibfnamefont {V.}~\bibnamefont {Bellani}}, \bibinfo {author}
  {\bibfnamefont {D.}~\bibnamefont {Lopez-Romero}},\ and\ \bibinfo {author}
  {\bibfnamefont {D.~K.}\ \bibnamefont {Maude}},\ }\href
  {http://stacks.iop.org/0953-8984/24/i=30/a=305302} {\bibfield  {journal}
  {\bibinfo  {journal} {Journal of Physics: Condensed Matter}\ }\textbf
  {\bibinfo {volume} {24}},\ \bibinfo {pages} {305302} (\bibinfo {year}
  {2012})}\BibitemShut {NoStop}%
\bibitem [{\citenamefont {Jabakhanji}\ \emph {et~al.}(2014)\citenamefont
  {Jabakhanji}, \citenamefont {Michon}, \citenamefont {Consejo}, \citenamefont
  {Desrat}, \citenamefont {Portail}, \citenamefont {Tiberj}, \citenamefont
  {Paillet}, \citenamefont {Zahab}, \citenamefont {Cheynis}, \citenamefont
  {Lafont}, \citenamefont {Schopfer}, \citenamefont {Poirier}, \citenamefont
  {Bertran}, \citenamefont {Le~F\'evre}, \citenamefont {Taleb-Ibrahimi},
  \citenamefont {Kazazis}, \citenamefont {Escoffier}, \citenamefont {Camargo},
  \citenamefont {Kopelevich}, \citenamefont {Camassel},\ and\ \citenamefont
  {Jouault}}]{PhysRevB.89.085422}%
  \BibitemOpen
  \bibfield  {author} {\bibinfo {author} {\bibfnamefont {B.}~\bibnamefont
  {Jabakhanji}}, \bibinfo {author} {\bibfnamefont {A.}~\bibnamefont {Michon}},
  \bibinfo {author} {\bibfnamefont {C.}~\bibnamefont {Consejo}}, \bibinfo
  {author} {\bibfnamefont {W.}~\bibnamefont {Desrat}}, \bibinfo {author}
  {\bibfnamefont {M.}~\bibnamefont {Portail}}, \bibinfo {author} {\bibfnamefont
  {A.}~\bibnamefont {Tiberj}}, \bibinfo {author} {\bibfnamefont
  {M.}~\bibnamefont {Paillet}}, \bibinfo {author} {\bibfnamefont
  {A.}~\bibnamefont {Zahab}}, \bibinfo {author} {\bibfnamefont
  {F.}~\bibnamefont {Cheynis}}, \bibinfo {author} {\bibfnamefont
  {F.}~\bibnamefont {Lafont}}, \bibinfo {author} {\bibfnamefont
  {F.}~\bibnamefont {Schopfer}}, \bibinfo {author} {\bibfnamefont
  {W.}~\bibnamefont {Poirier}}, \bibinfo {author} {\bibfnamefont
  {F.}~\bibnamefont {Bertran}}, \bibinfo {author} {\bibfnamefont
  {P.}~\bibnamefont {Le~F\'evre}}, \bibinfo {author} {\bibfnamefont
  {A.}~\bibnamefont {Taleb-Ibrahimi}}, \bibinfo {author} {\bibfnamefont
  {D.}~\bibnamefont {Kazazis}}, \bibinfo {author} {\bibfnamefont
  {W.}~\bibnamefont {Escoffier}}, \bibinfo {author} {\bibfnamefont {B.~C.}\
  \bibnamefont {Camargo}}, \bibinfo {author} {\bibfnamefont {Y.}~\bibnamefont
  {Kopelevich}}, \bibinfo {author} {\bibfnamefont {J.}~\bibnamefont
  {Camassel}},\ and\ \bibinfo {author} {\bibfnamefont {B.}~\bibnamefont
  {Jouault}},\ }\href {https://doi.org/10.1103/PhysRevB.89.085422} {\bibfield
  {journal} {\bibinfo  {journal} {Phys. Rev. B}\ }\textbf {\bibinfo {volume}
  {89}},\ \bibinfo {pages} {085422} (\bibinfo {year} {2014})}\BibitemShut
  {NoStop}%
\bibitem [{\citenamefont {Dresselhaus}\ \emph {et~al.}(2021)\citenamefont
  {Dresselhaus}, \citenamefont {Sbierski},\ and\ \citenamefont
  {Gruzberg}}]{DRESSELHAUS2021168676}%
  \BibitemOpen
  \bibfield  {author} {\bibinfo {author} {\bibfnamefont {E.~J.}\ \bibnamefont
  {Dresselhaus}}, \bibinfo {author} {\bibfnamefont {B.}~\bibnamefont
  {Sbierski}},\ and\ \bibinfo {author} {\bibfnamefont {I.~A.}\ \bibnamefont
  {Gruzberg}},\ }\href
  {https://doi.org/https://doi.org/10.1016/j.aop.2021.168676} {\bibfield
  {journal} {\bibinfo  {journal} {Annals of Physics}\ }\textbf {\bibinfo
  {volume} {435}},\ \bibinfo {pages} {168676} (\bibinfo {year} {2021})},\
  \bibinfo {note} {special Issue on Localisation 2020}\BibitemShut {NoStop}%
\bibitem [{\citenamefont {Zirnbauer}(2019)}]{ZIRNBAUER2019458}%
  \BibitemOpen
  \bibfield  {author} {\bibinfo {author} {\bibfnamefont {M.~R.}\ \bibnamefont
  {Zirnbauer}},\ }\href
  {https://doi.org/https://doi.org/10.1016/j.nuclphysb.2019.02.017} {\bibfield
  {journal} {\bibinfo  {journal} {Nuclear Physics B}\ }\textbf {\bibinfo
  {volume} {941}},\ \bibinfo {pages} {458} (\bibinfo {year}
  {2019})}\BibitemShut {NoStop}%
\bibitem [{\citenamefont {Zhao}\ \emph {et~al.}(2008)\citenamefont {Zhao},
  \citenamefont {Tu}, \citenamefont {Hao}, \citenamefont {Guo}, \citenamefont
  {Jiang},\ and\ \citenamefont {Guo}}]{PhysRevB.78.233301}%
  \BibitemOpen
  \bibfield  {author} {\bibinfo {author} {\bibfnamefont {Y.~J.}\ \bibnamefont
  {Zhao}}, \bibinfo {author} {\bibfnamefont {T.}~\bibnamefont {Tu}}, \bibinfo
  {author} {\bibfnamefont {X.~J.}\ \bibnamefont {Hao}}, \bibinfo {author}
  {\bibfnamefont {G.~C.}\ \bibnamefont {Guo}}, \bibinfo {author} {\bibfnamefont
  {H.~W.}\ \bibnamefont {Jiang}},\ and\ \bibinfo {author} {\bibfnamefont
  {G.~P.}\ \bibnamefont {Guo}},\ }\href
  {https://doi.org/10.1103/PhysRevB.78.233301} {\bibfield  {journal} {\bibinfo
  {journal} {Phys. Rev. B}\ }\textbf {\bibinfo {volume} {78}},\ \bibinfo
  {pages} {233301} (\bibinfo {year} {2008})}\BibitemShut {NoStop}%
\bibitem [{\citenamefont {Hohls}\ \emph
  {et~al.}(2002{\natexlab{c}})\citenamefont {Hohls}, \citenamefont {Zeitler},\
  and\ \citenamefont {Haug}}]{PhysRevLett.88.036802}%
  \BibitemOpen
  \bibfield  {author} {\bibinfo {author} {\bibfnamefont {F.}~\bibnamefont
  {Hohls}}, \bibinfo {author} {\bibfnamefont {U.}~\bibnamefont {Zeitler}},\
  and\ \bibinfo {author} {\bibfnamefont {R.~J.}\ \bibnamefont {Haug}},\ }\href
  {https://doi.org/10.1103/PhysRevLett.88.036802} {\bibfield  {journal}
  {\bibinfo  {journal} {Phys. Rev. Lett.}\ }\textbf {\bibinfo {volume} {88}},\
  \bibinfo {pages} {036802} (\bibinfo {year} {2002}{\natexlab{c}})}\BibitemShut
  {NoStop}%
\bibitem [{\citenamefont {Wei}\ \emph {et~al.}(1992{\natexlab{b}})\citenamefont
  {Wei}, \citenamefont {Lin}, \citenamefont {Tsui},\ and\ \citenamefont
  {Pruisken}}]{PhysRevB.45.3926}%
  \BibitemOpen
  \bibfield  {author} {\bibinfo {author} {\bibfnamefont {H.~P.}\ \bibnamefont
  {Wei}}, \bibinfo {author} {\bibfnamefont {S.~Y.}\ \bibnamefont {Lin}},
  \bibinfo {author} {\bibfnamefont {D.~C.}\ \bibnamefont {Tsui}},\ and\
  \bibinfo {author} {\bibfnamefont {A.~M.~M.}\ \bibnamefont {Pruisken}},\
  }\href {https://doi.org/10.1103/PhysRevB.45.3926} {\bibfield  {journal}
  {\bibinfo  {journal} {Phys. Rev. B}\ }\textbf {\bibinfo {volume} {45}},\
  \bibinfo {pages} {3926} (\bibinfo {year} {1992}{\natexlab{b}})}\BibitemShut
  {NoStop}%
\bibitem [{\citenamefont {Yoo}\ \emph {et~al.}(1994)\citenamefont {Yoo},
  \citenamefont {Kwon},\ and\ \citenamefont {Park}}]{YOO1994821}%
  \BibitemOpen
  \bibfield  {author} {\bibinfo {author} {\bibfnamefont {K.-H.}\ \bibnamefont
  {Yoo}}, \bibinfo {author} {\bibfnamefont {H.}~\bibnamefont {Kwon}},\ and\
  \bibinfo {author} {\bibfnamefont {J.}~\bibnamefont {Park}},\ }\href
  {https://doi.org/https://doi.org/10.1016/0038-1098(94)90320-4} {\bibfield
  {journal} {\bibinfo  {journal} {Solid State Communications}\ }\textbf
  {\bibinfo {volume} {92}},\ \bibinfo {pages} {821} (\bibinfo {year}
  {1994})}\BibitemShut {NoStop}%
\bibitem [{\citenamefont {Koch}\ \emph {et~al.}(1995)\citenamefont {Koch},
  \citenamefont {Haug}, \citenamefont {von Klitzing},\ and\ \citenamefont
  {Ploog}}]{SKoch1995}%
  \BibitemOpen
  \bibfield  {author} {\bibinfo {author} {\bibfnamefont {S.}~\bibnamefont
  {Koch}}, \bibinfo {author} {\bibfnamefont {R.~J.}\ \bibnamefont {Haug}},
  \bibinfo {author} {\bibfnamefont {K.}~\bibnamefont {von Klitzing}},\ and\
  \bibinfo {author} {\bibfnamefont {K.}~\bibnamefont {Ploog}},\ }\href
  {https://doi.org/10.1088/0268-1242/10/2/015} {\bibfield  {journal} {\bibinfo
  {journal} {Semiconductor Science and Technology}\ }\textbf {\bibinfo {volume}
  {10}},\ \bibinfo {pages} {209} (\bibinfo {year} {1995})}\BibitemShut
  {NoStop}%
\bibitem [{\citenamefont {Huang}\ \emph {et~al.}(2004)\citenamefont {Huang},
  \citenamefont {Chang}, \citenamefont {Cheng}, \citenamefont {Liang},\ and\
  \citenamefont {Hwang}}]{HUANG2004232}%
  \BibitemOpen
  \bibfield  {author} {\bibinfo {author} {\bibfnamefont {C.}~\bibnamefont
  {Huang}}, \bibinfo {author} {\bibfnamefont {Y.}~\bibnamefont {Chang}},
  \bibinfo {author} {\bibfnamefont {H.}~\bibnamefont {Cheng}}, \bibinfo
  {author} {\bibfnamefont {C.-T.}\ \bibnamefont {Liang}},\ and\ \bibinfo
  {author} {\bibfnamefont {G.}~\bibnamefont {Hwang}},\ }\href
  {https://doi.org/https://doi.org/10.1016/j.physe.2003.11.256} {\bibfield
  {journal} {\bibinfo  {journal} {Physica E: Low-dimensional Systems and
  Nanostructures}\ }\textbf {\bibinfo {volume} {22}},\ \bibinfo {pages} {232}
  (\bibinfo {year} {2004})},\ \bibinfo {note} {15th International Conference on
  Electronic Propreties of Two-Dimensional Systems (EP2DS-15)}\BibitemShut
  {NoStop}%
\bibitem [{\citenamefont {Tu}\ \emph {et~al.}(2007)\citenamefont {Tu},
  \citenamefont {Zhao}, \citenamefont {Guo}, \citenamefont {Hao},\ and\
  \citenamefont {Guo}}]{TU2007108}%
  \BibitemOpen
  \bibfield  {author} {\bibinfo {author} {\bibfnamefont {T.}~\bibnamefont
  {Tu}}, \bibinfo {author} {\bibfnamefont {Y.-J.}\ \bibnamefont {Zhao}},
  \bibinfo {author} {\bibfnamefont {G.-P.}\ \bibnamefont {Guo}}, \bibinfo
  {author} {\bibfnamefont {X.-J.}\ \bibnamefont {Hao}},\ and\ \bibinfo {author}
  {\bibfnamefont {G.-C.}\ \bibnamefont {Guo}},\ }\href
  {https://doi.org/https://doi.org/10.1016/j.physleta.2007.03.059} {\bibfield
  {journal} {\bibinfo  {journal} {Physics Letters A}\ }\textbf {\bibinfo
  {volume} {368}},\ \bibinfo {pages} {108} (\bibinfo {year}
  {2007})}\BibitemShut {NoStop}%
\bibitem [{\citenamefont {Nakajima}\ \emph {et~al.}(2007)\citenamefont
  {Nakajima}, \citenamefont {Ueda},\ and\ \citenamefont
  {Komiyama}}]{doi:10.1143/JPSJ.76.094703}%
  \BibitemOpen
  \bibfield  {author} {\bibinfo {author} {\bibfnamefont {T.}~\bibnamefont
  {Nakajima}}, \bibinfo {author} {\bibfnamefont {T.}~\bibnamefont {Ueda}},\
  and\ \bibinfo {author} {\bibfnamefont {S.}~\bibnamefont {Komiyama}},\ }\href
  {https://doi.org/10.1143/JPSJ.76.094703} {\bibfield  {journal} {\bibinfo
  {journal} {Journal of the Physical Society of Japan}\ }\textbf {\bibinfo
  {volume} {76}},\ \bibinfo {pages} {094703} (\bibinfo {year} {2007})},\
  \Eprint {https://arxiv.org/abs/https://doi.org/10.1143/JPSJ.76.094703}
  {https://doi.org/10.1143/JPSJ.76.094703} \BibitemShut {NoStop}%
\bibitem [{\citenamefont {Li}\ \emph {et~al.}(2010)\citenamefont {Li},
  \citenamefont {Xia}, \citenamefont {Vicente}, \citenamefont {Sullivan},
  \citenamefont {Pan}, \citenamefont {Tsui}, \citenamefont {Pfeiffer},\ and\
  \citenamefont {West}}]{PhysRevB.81.033305}%
  \BibitemOpen
  \bibfield  {author} {\bibinfo {author} {\bibfnamefont {W.}~\bibnamefont
  {Li}}, \bibinfo {author} {\bibfnamefont {J.~S.}\ \bibnamefont {Xia}},
  \bibinfo {author} {\bibfnamefont {C.}~\bibnamefont {Vicente}}, \bibinfo
  {author} {\bibfnamefont {N.~S.}\ \bibnamefont {Sullivan}}, \bibinfo {author}
  {\bibfnamefont {W.}~\bibnamefont {Pan}}, \bibinfo {author} {\bibfnamefont
  {D.~C.}\ \bibnamefont {Tsui}}, \bibinfo {author} {\bibfnamefont {L.~N.}\
  \bibnamefont {Pfeiffer}},\ and\ \bibinfo {author} {\bibfnamefont {K.~W.}\
  \bibnamefont {West}},\ }\href {https://doi.org/10.1103/PhysRevB.81.033305}
  {\bibfield  {journal} {\bibinfo  {journal} {Phys. Rev. B}\ }\textbf {\bibinfo
  {volume} {81}},\ \bibinfo {pages} {033305} (\bibinfo {year}
  {2010})}\BibitemShut {NoStop}%
\bibitem [{\citenamefont {Wang}\ \emph
  {et~al.}(2016{\natexlab{b}})\citenamefont {Wang}, \citenamefont {Liu},
  \citenamefont {Zhu}, \citenamefont {Shan}, \citenamefont {Wang},
  \citenamefont {Fu}, \citenamefont {Du}, \citenamefont {Pfeiffer},
  \citenamefont {West}, \citenamefont {Xie}, \citenamefont {Du},\ and\
  \citenamefont {Lin}}]{PhysRevB.93.075307}%
  \BibitemOpen
  \bibfield  {author} {\bibinfo {author} {\bibfnamefont {X.}~\bibnamefont
  {Wang}}, \bibinfo {author} {\bibfnamefont {H.}~\bibnamefont {Liu}}, \bibinfo
  {author} {\bibfnamefont {J.}~\bibnamefont {Zhu}}, \bibinfo {author}
  {\bibfnamefont {P.}~\bibnamefont {Shan}}, \bibinfo {author} {\bibfnamefont
  {P.}~\bibnamefont {Wang}}, \bibinfo {author} {\bibfnamefont {H.}~\bibnamefont
  {Fu}}, \bibinfo {author} {\bibfnamefont {L.}~\bibnamefont {Du}}, \bibinfo
  {author} {\bibfnamefont {L.~N.}\ \bibnamefont {Pfeiffer}}, \bibinfo {author}
  {\bibfnamefont {K.~W.}\ \bibnamefont {West}}, \bibinfo {author}
  {\bibfnamefont {X.~C.}\ \bibnamefont {Xie}}, \bibinfo {author} {\bibfnamefont
  {R.-R.}\ \bibnamefont {Du}},\ and\ \bibinfo {author} {\bibfnamefont
  {X.}~\bibnamefont {Lin}},\ }\href
  {https://doi.org/10.1103/PhysRevB.93.075307} {\bibfield  {journal} {\bibinfo
  {journal} {Phys. Rev. B}\ }\textbf {\bibinfo {volume} {93}},\ \bibinfo
  {pages} {075307} (\bibinfo {year} {2016}{\natexlab{b}})}\BibitemShut
  {NoStop}%
\bibitem [{\citenamefont {Khouri}\ \emph {et~al.}(2016)\citenamefont {Khouri},
  \citenamefont {Bendias}, \citenamefont {Leubner}, \citenamefont {Br\"une},
  \citenamefont {Buhmann}, \citenamefont {Molenkamp}, \citenamefont {Zeitler},
  \citenamefont {Hussey},\ and\ \citenamefont {Wiedmann}}]{PhysRevB.93.125308}%
  \BibitemOpen
  \bibfield  {author} {\bibinfo {author} {\bibfnamefont {T.}~\bibnamefont
  {Khouri}}, \bibinfo {author} {\bibfnamefont {M.}~\bibnamefont {Bendias}},
  \bibinfo {author} {\bibfnamefont {P.}~\bibnamefont {Leubner}}, \bibinfo
  {author} {\bibfnamefont {C.}~\bibnamefont {Br\"une}}, \bibinfo {author}
  {\bibfnamefont {H.}~\bibnamefont {Buhmann}}, \bibinfo {author} {\bibfnamefont
  {L.~W.}\ \bibnamefont {Molenkamp}}, \bibinfo {author} {\bibfnamefont
  {U.}~\bibnamefont {Zeitler}}, \bibinfo {author} {\bibfnamefont {N.~E.}\
  \bibnamefont {Hussey}},\ and\ \bibinfo {author} {\bibfnamefont
  {S.}~\bibnamefont {Wiedmann}},\ }\href
  {https://doi.org/10.1103/PhysRevB.93.125308} {\bibfield  {journal} {\bibinfo
  {journal} {Phys. Rev. B}\ }\textbf {\bibinfo {volume} {93}},\ \bibinfo
  {pages} {125308} (\bibinfo {year} {2016})}\BibitemShut {NoStop}%
\bibitem [{\citenamefont {Bennaceur}\ \emph {et~al.}(2012)\citenamefont
  {Bennaceur}, \citenamefont {Jacques}, \citenamefont {Portier}, \citenamefont
  {Roche},\ and\ \citenamefont {Glattli}}]{PhysRevB.86.085433}%
  \BibitemOpen
  \bibfield  {author} {\bibinfo {author} {\bibfnamefont {K.}~\bibnamefont
  {Bennaceur}}, \bibinfo {author} {\bibfnamefont {P.}~\bibnamefont {Jacques}},
  \bibinfo {author} {\bibfnamefont {F.}~\bibnamefont {Portier}}, \bibinfo
  {author} {\bibfnamefont {P.}~\bibnamefont {Roche}},\ and\ \bibinfo {author}
  {\bibfnamefont {D.~C.}\ \bibnamefont {Glattli}},\ }\href
  {https://doi.org/10.1103/PhysRevB.86.085433} {\bibfield  {journal} {\bibinfo
  {journal} {Phys. Rev. B}\ }\textbf {\bibinfo {volume} {86}},\ \bibinfo
  {pages} {085433} (\bibinfo {year} {2012})}\BibitemShut {NoStop}%
\bibitem [{\citenamefont {Amado}\ \emph {et~al.}(2010)\citenamefont {Amado},
  \citenamefont {Diez}, \citenamefont {L{\'o}pez-Romero}, \citenamefont
  {Rossella}, \citenamefont {Caridad}, \citenamefont {Dionigi}, \citenamefont
  {Bellani},\ and\ \citenamefont {Maude}}]{amado2010plateau}%
  \BibitemOpen
  \bibfield  {author} {\bibinfo {author} {\bibfnamefont {M.}~\bibnamefont
  {Amado}}, \bibinfo {author} {\bibfnamefont {E.}~\bibnamefont {Diez}},
  \bibinfo {author} {\bibfnamefont {D.}~\bibnamefont {L{\'o}pez-Romero}},
  \bibinfo {author} {\bibfnamefont {F.}~\bibnamefont {Rossella}}, \bibinfo
  {author} {\bibfnamefont {J.}~\bibnamefont {Caridad}}, \bibinfo {author}
  {\bibfnamefont {F.}~\bibnamefont {Dionigi}}, \bibinfo {author} {\bibfnamefont
  {V.}~\bibnamefont {Bellani}},\ and\ \bibinfo {author} {\bibfnamefont
  {D.}~\bibnamefont {Maude}},\ }\href
  {https://doi.org/10.1088/1367-2630/12/5/053004} {\bibfield  {journal}
  {\bibinfo  {journal} {New Journal of Physics}\ }\textbf {\bibinfo {volume}
  {12}},\ \bibinfo {pages} {053004} (\bibinfo {year} {2010})}\BibitemShut
  {NoStop}%
\bibitem [{\citenamefont {Taychatanapat}\ \emph
  {et~al.}(2011{\natexlab{b}})\citenamefont {Taychatanapat}, \citenamefont
  {Watanabe}, \citenamefont {Taniguchi},\ and\ \citenamefont
  {Jarillo-Herrero}}]{Taychatanapat2011}%
  \BibitemOpen
  \bibfield  {author} {\bibinfo {author} {\bibfnamefont {T.}~\bibnamefont
  {Taychatanapat}}, \bibinfo {author} {\bibfnamefont {K.}~\bibnamefont
  {Watanabe}}, \bibinfo {author} {\bibfnamefont {T.}~\bibnamefont
  {Taniguchi}},\ and\ \bibinfo {author} {\bibfnamefont {P.}~\bibnamefont
  {Jarillo-Herrero}},\ }\href {https://doi.org/10.1038/nphys2008} {\bibfield
  {journal} {\bibinfo  {journal} {Nature Physics}\ }\textbf {\bibinfo {volume}
  {7}},\ \bibinfo {pages} {621} (\bibinfo {year}
  {2011}{\natexlab{b}})}\BibitemShut {NoStop}%
\bibitem [{\citenamefont {Datta}\ \emph {et~al.}(2017)\citenamefont {Datta},
  \citenamefont {Dey}, \citenamefont {Samanta}, \citenamefont {Agarwal},
  \citenamefont {Borah}, \citenamefont {Watanabe}, \citenamefont {Taniguchi},
  \citenamefont {Sensarma},\ and\ \citenamefont {Deshmukh}}]{Datta2017}%
  \BibitemOpen
  \bibfield  {author} {\bibinfo {author} {\bibfnamefont {B.}~\bibnamefont
  {Datta}}, \bibinfo {author} {\bibfnamefont {S.}~\bibnamefont {Dey}}, \bibinfo
  {author} {\bibfnamefont {A.}~\bibnamefont {Samanta}}, \bibinfo {author}
  {\bibfnamefont {H.}~\bibnamefont {Agarwal}}, \bibinfo {author} {\bibfnamefont
  {A.}~\bibnamefont {Borah}}, \bibinfo {author} {\bibfnamefont
  {K.}~\bibnamefont {Watanabe}}, \bibinfo {author} {\bibfnamefont
  {T.}~\bibnamefont {Taniguchi}}, \bibinfo {author} {\bibfnamefont
  {R.}~\bibnamefont {Sensarma}},\ and\ \bibinfo {author} {\bibfnamefont
  {M.~M.}\ \bibnamefont {Deshmukh}},\ }\href
  {https://doi.org/10.1038/ncomms14518} {\bibfield  {journal} {\bibinfo
  {journal} {Nature Communications}\ }\textbf {\bibinfo {volume} {8}},\
  \bibinfo {pages} {14518} (\bibinfo {year} {2017})}\BibitemShut {NoStop}%
\bibitem [{\citenamefont {Datta}\ \emph {et~al.}(2019)\citenamefont {Datta},
  \citenamefont {Adak}, \citenamefont {kun Shi}, \citenamefont {Watanabe},
  \citenamefont {Taniguchi}, \citenamefont {Song},\ and\ \citenamefont
  {Deshmukh}}]{doi:10.1126/sciadv.aax6550}%
  \BibitemOpen
  \bibfield  {author} {\bibinfo {author} {\bibfnamefont {B.}~\bibnamefont
  {Datta}}, \bibinfo {author} {\bibfnamefont {P.~C.}\ \bibnamefont {Adak}},
  \bibinfo {author} {\bibfnamefont {L.}~\bibnamefont {kun Shi}}, \bibinfo
  {author} {\bibfnamefont {K.}~\bibnamefont {Watanabe}}, \bibinfo {author}
  {\bibfnamefont {T.}~\bibnamefont {Taniguchi}}, \bibinfo {author}
  {\bibfnamefont {J.~C.~W.}\ \bibnamefont {Song}},\ and\ \bibinfo {author}
  {\bibfnamefont {M.~M.}\ \bibnamefont {Deshmukh}},\ }\href
  {https://doi.org/10.1126/sciadv.aax6550} {\bibfield  {journal} {\bibinfo
  {journal} {Science Advances}\ }\textbf {\bibinfo {volume} {5}},\ \bibinfo
  {pages} {eaax6550} (\bibinfo {year} {2019})},\ \Eprint
  {https://arxiv.org/abs/https://www.science.org/doi/pdf/10.1126/sciadv.aax6550}
  {https://www.science.org/doi/pdf/10.1126/sciadv.aax6550} \BibitemShut
  {NoStop}%
\bibitem [{\citenamefont {Winterer}\ \emph
  {et~al.}(2022{\natexlab{b}})\citenamefont {Winterer}, \citenamefont {Seiler},
  \citenamefont {Ghazaryan}, \citenamefont {Geisenhof}, \citenamefont
  {Watanabe}, \citenamefont {Taniguchi}, \citenamefont {Serbyn},\ and\
  \citenamefont {Weitz}}]{Winterer2022}%
  \BibitemOpen
  \bibfield  {author} {\bibinfo {author} {\bibfnamefont {F.}~\bibnamefont
  {Winterer}}, \bibinfo {author} {\bibfnamefont {A.~M.}\ \bibnamefont
  {Seiler}}, \bibinfo {author} {\bibfnamefont {A.}~\bibnamefont {Ghazaryan}},
  \bibinfo {author} {\bibfnamefont {F.~R.}\ \bibnamefont {Geisenhof}}, \bibinfo
  {author} {\bibfnamefont {K.}~\bibnamefont {Watanabe}}, \bibinfo {author}
  {\bibfnamefont {T.}~\bibnamefont {Taniguchi}}, \bibinfo {author}
  {\bibfnamefont {M.}~\bibnamefont {Serbyn}},\ and\ \bibinfo {author}
  {\bibfnamefont {R.~T.}\ \bibnamefont {Weitz}},\ }\href
  {https://doi.org/10.1021/acs.nanolett.2c00435} {\bibfield  {journal}
  {\bibinfo  {journal} {Nano Letters}\ }\textbf {\bibinfo {volume} {22}},\
  \bibinfo {pages} {3317} (\bibinfo {year} {2022}{\natexlab{b}})}\BibitemShut
  {NoStop}%
\bibitem [{\citenamefont {Datta}\ \emph
  {et~al.}(2018{\natexlab{b}})\citenamefont {Datta}, \citenamefont {Agarwal},
  \citenamefont {Samanta}, \citenamefont {Ratnakar}, \citenamefont {Watanabe},
  \citenamefont {Taniguchi}, \citenamefont {Sensarma},\ and\ \citenamefont
  {Deshmukh}}]{PhysRevLett.121.056801}%
  \BibitemOpen
  \bibfield  {author} {\bibinfo {author} {\bibfnamefont {B.}~\bibnamefont
  {Datta}}, \bibinfo {author} {\bibfnamefont {H.}~\bibnamefont {Agarwal}},
  \bibinfo {author} {\bibfnamefont {A.}~\bibnamefont {Samanta}}, \bibinfo
  {author} {\bibfnamefont {A.}~\bibnamefont {Ratnakar}}, \bibinfo {author}
  {\bibfnamefont {K.}~\bibnamefont {Watanabe}}, \bibinfo {author}
  {\bibfnamefont {T.}~\bibnamefont {Taniguchi}}, \bibinfo {author}
  {\bibfnamefont {R.}~\bibnamefont {Sensarma}},\ and\ \bibinfo {author}
  {\bibfnamefont {M.~M.}\ \bibnamefont {Deshmukh}},\ }\href
  {https://doi.org/10.1103/PhysRevLett.121.056801} {\bibfield  {journal}
  {\bibinfo  {journal} {Phys. Rev. Lett.}\ }\textbf {\bibinfo {volume} {121}},\
  \bibinfo {pages} {056801} (\bibinfo {year} {2018}{\natexlab{b}})}\BibitemShut
  {NoStop}%
\bibitem [{\citenamefont {Campos}\ \emph {et~al.}(2016)\citenamefont {Campos},
  \citenamefont {Taychatanapat}, \citenamefont {Serbyn}, \citenamefont
  {Surakitbovorn}, \citenamefont {Watanabe}, \citenamefont {Taniguchi},
  \citenamefont {Abanin},\ and\ \citenamefont
  {Jarillo-Herrero}}]{PhysRevLett.117.066601}%
  \BibitemOpen
  \bibfield  {author} {\bibinfo {author} {\bibfnamefont {L.~C.}\ \bibnamefont
  {Campos}}, \bibinfo {author} {\bibfnamefont {T.}~\bibnamefont
  {Taychatanapat}}, \bibinfo {author} {\bibfnamefont {M.}~\bibnamefont
  {Serbyn}}, \bibinfo {author} {\bibfnamefont {K.}~\bibnamefont
  {Surakitbovorn}}, \bibinfo {author} {\bibfnamefont {K.}~\bibnamefont
  {Watanabe}}, \bibinfo {author} {\bibfnamefont {T.}~\bibnamefont {Taniguchi}},
  \bibinfo {author} {\bibfnamefont {D.~A.}\ \bibnamefont {Abanin}},\ and\
  \bibinfo {author} {\bibfnamefont {P.}~\bibnamefont {Jarillo-Herrero}},\
  }\href {https://doi.org/10.1103/PhysRevLett.117.066601} {\bibfield  {journal}
  {\bibinfo  {journal} {Phys. Rev. Lett.}\ }\textbf {\bibinfo {volume} {117}},\
  \bibinfo {pages} {066601} (\bibinfo {year} {2016})}\BibitemShut {NoStop}%
\end{thebibliography}%

\end{document}